\documentclass[11pt, a4paper]{article}
\usepackage[absolute]{textpos}
\usepackage[top=3cm,bottom=3cm,left=2cm,right=2cm]{geometry}
\usepackage{amsmath,amssymb,mathtools,dsfont,emptypage,graphicx,dcolumn,bm,booktabs,colortbl,collcell,enumerate,float,verbatim,multirow,xparse,slashed}
\usepackage{cite}
\usepackage{tabularx}
\usepackage{multirow}
\usepackage{graphicx}
\usepackage{subcaption}
\usepackage{appendix}
\usepackage{booktabs} 
\usepackage{colortbl}
\usepackage{collcell}
\usepackage{soul}
\usepackage{xparse}
\usepackage{relsize}
\usepackage[labelfont=bf, labelsep=period]{caption}

\usepackage[compat=1.1.0]{tikz-feynman}
\usepackage{contour}

\newcommand{\virg}[1]{``#1''}
\newcommand{\bra}[1]{\langle #1 |}
\newcommand{\ket}[1]{| #1 \rangle}

\def\kin{\mathsmaller{\mathrm{kin}}}
\def\min{\mathsmaller{\mathrm{min}}}
\def\max{\mathsmaller{\mathrm{max}}}
\def\upper{\mathsmaller{\mathrm{upper}}}
\def\DM{\mathsmaller{\mathrm{DM}}}
\def\BH{\mathsmaller{\mathrm{BH}}}
\def\LOS{\mathsmaller{\mathrm{LOS}}}
\def\DIS{\mathsmaller{\mathrm{DIS}}}
\def\EL{\mathsmaller{\mathrm{EL}}}
\def\rel{\mathsmaller{\mathrm{rel}}}
\def\NR{\mathsmaller{\mathrm{NR}}}
\def\SI{\mathsmaller{\mathrm{SI}}}
\def\SD{\mathsmaller{\mathrm{SD}}}
\def\EM{\mathsmaller{\mathrm{EM}}}
\def\quark{\mathsmaller{\mathrm{quark}}}

\def\zero{\mathsmaller{\mathrm{(0)}}}
\def\det{\mathsmaller{\rm det}}

\definecolor{myred}{cmyk}{0,1,1,0.8}
\definecolor{mygreen}{rgb}{0.27, 0.64, 0.48}
\definecolor{myblue}{cmyk}{0.8, 0.4, 0, 0.2}
\definecolor{mygray}{gray}{.95}
\definecolor{nicered}{rgb}{.85, 0.1, 0.2}
\definecolor{niceorange}{rgb}{0.9, 0.3, 0.2}
\definecolor{nicepurple}{rgb}{0.7, 0.0, 0.4}
\definecolor{niceblue}{rgb}{0.0, 0.14, 0.8}
\usepackage[linktoc=page,bookmarks=false,colorlinks=true,
linkbordercolor=niceblue, linkcolor = nicered,
citebordercolor=niceblue,citecolor=niceblue,urlbordercolor=niceblue, urlcolor = niceblue]{hyperref}

\begin{document}


\begin{center}
{\bf\LARGE Boosted dark matter versus dark matter-induced\\
\vspace{.2em} neutrinos from single and stacked blazars} \\
[5mm]
\renewcommand*{\thefootnote}{\fnsymbol{footnote}}
Andrea Giovanni~De Marchi$^{a,b}$ \footnote{\href{mailto:andreagiovanni.demarchi@gmail.com}{andreagiovanni.demarchi@unibo.it}},
Alessandro Granelli$^{a,b}$ \footnote{\href{mailto:alessandro.granelli@unibo.it}{alessandro.granelli@unibo.it}},
Jacopo Nava$^{a,b}$
\footnote{\href{mailto:jacopo.nava2@unibo.it}{jacopo.nava2@unibo.it}}
and
Filippo Sala$^{a,b}$
\footnote{\href{mailto:f.sala@unibo.it}{f.sala@unibo.it}, FS is on leave from LPTHE, CNRS \&
Sorbonne Universit\'e, Paris, France} 
\\
$^{a}$\,{\it \small Dipartimento di Fisica e Astronomia, Università di Bologna, via Irnerio 46, 40126 Bologna, Italy; and} \\
$^{b}$\,{\it \small INFN, Sezione di Bologna, viale Berti Pichat 6/2, 40127, Bologna, Italy.}
\end{center}

\begin{center}
    {\bf Abstract}
\end{center}
    \noindent
    The physics responsible for the production of observed high-energy neutrinos have not been established so far, neither for the diffuse astrophysical ones nor for those detected from single blazars.
    We recently proposed that both could be explained by deep inelastic scatterings between sub-GeV dark matter (DM) around blazars and protons within their jets. Here, we compute the proton-recoil signals at the neutrino detectors Super-Kamiokande, KamLAND, Borexino, JUNO, Hyper-Kamiokande and DUNE induced by DM that is itself boosted by the scatterings with protons in blazar jets. We do it for the four cases of vector, axial, scalar and pseudoscalar mediators of DM-quark interactions.
    We perform the analysis for the single blazar TXS 0506+056 and for a sample of more than 300 stacked blazars. 
    We find that searches for such blazar-boosted 
    DM leave room for a variety of DM models to explain observations of high-energy neutrinos. We check that the depletion of the DM spike induced by DM-proton and DM-DM interactions does not compromise the DM interpretation for high-energy neutrinos, but challenges other blazar-DM signals.
    


\renewcommand*{\thefootnote}{\arabic{footnote}}
\setcounter{footnote}{0}

\par\noindent\rule{\textwidth}{0.5pt}
\setcounter{tocdepth}{2}
\tableofcontents
\par\noindent\rule{\textwidth}{0.5pt}

\section{Introduction} 

Dark matter (DM) is the only known hypothesis that explains, at once, several independent observations at the scales of the entire Universe, clusters and galaxies \cite{Cirelli:2024ssz}.  The mass and the non-gravitational interactions of DM constituents are unknown and the object of enormous theoretical and experimental investigation.
Over the past decades, significant effort has been dedicated to searches for the DM that composes the Milky Way halo via the recoil signal it induces in fixed target nuclei. These experimental efforts have not yet resulted in any unambiguous DM detection.
However, most of these \virg{standard} (in the sense that they look for halo DM) direct detection experiments are sensitive to DM masses above the GeV scale.
Indeed, the nuclear recoil energy induced by their scatterings with sub-GeV DM falls below the detection threshold of typical experiments,  leading to a sharp loss of their sensitivity.

Improved performance can be achieved in this mass regime with novel low-threshold techniques, some already looking for DM and others under investigation, see e.g.~\cite{Essig:2022dfa} for an overview.
In addition, one can leverage existing huge detectors, with higher thresholds, to look for subdominant populations of DM that are more energetic than the halo one, like DM upscattered by cosmic-rays~\cite{Bringmann:2018cvk,Ema:2018bih}, produced in atmospheric showers~\cite{Alvey:2019zaa} or boosted in the jets of blazars (i.e.~supermassive black holes with two jets, one pointing towards us)~\cite{Wang:2021jic}.
This strategy has already been successfully used to establish stringent constraints on sub-GeV DM both by standard DD experiments~\cite{Alvey:2019zaa,PandaX:2023tfq} and by large neutrino ones~\cite{Ema:2020ulo,Super-Kamiokande:2022ncz}. 

The same DM-nucleon interactions, that are responsible for DM DD and for energetic DM sub-populations, also induce inelastic DM-proton scatterings that break the proton apart, triggering hadronic cascades and neutrino emission. The high-energy neutrinos generated in this way, by DM around blazars, could be the origin of the event observed at IceCube in 2017~\cite{IceCube:2018dnn} from the blazar TXS 0506+056, as we have proven in \cite{DeMarchi:2024riu}, and also of the diffuse astrophysical neutrino flux, as we have demonstrated in~\cite{DeMarchi:2025xag}.
Since the same DM-proton collisions in blazar jets accelerate DM in our direction, a compelling test of these intriguing results is provided by direct searches for blazar-boosted DM (BBDM) on Earth. Indeed, the properties of BBDM are determined by the same parameters, both astrophysical and DM ones, that control the DM-induced neutrino fluxes.

The study of BBDM interacting with nuclei, however, necessitates critical improvements since the first proposal~\cite{Wang:2021jic}.
These include the exploration of large neutrino detectors, as well as the consideration of explicit mediators instead of constant cross sections.  While the first improvement regards the strength of these searches, the second one regards their consistency, because we are not aware of any DM model that results in a cross section which is constant over the energies involved in blazars and, at the same time, so large as to possibly result in a signal.
In addition, the use of a constant cross section does not allow one to compare BBDM searches with other DM signals, like the neutrino ones mentioned above, because it does not take into account the different energies involved in different searches. This in turn either impacts their comparison by orders of magnitude (e.g.,~with standard DD searches) or makes that comparison impossible (e.g.,~accelerator searches).\footnote{The necessity of each of these improvements in the study of BBDM has instead been demonstrated for DM interacting with electrons~\cite{Granelli:2022ysi,Bhowmick:2022zkj,Jeesun:2025gzt}.} Furthermore, the studies on BBDM have focused so far on single sources, whereas thousands of blazars are known and potentially many of them could contribute to a diffuse BBDM flux. Improving these studies by considering multiple sources is crucial, as searches involving a population of emitters are potentially more robust and statistically significant than those based on individual sources.

\medskip 
After implementing the aforementioned improvements in the BBDM scenario, we aim to address the following questions.
What lessons can we learn about DM-nucleon interactions from BBDM, once neutrino detectors and explicit DM models are taken into account?
Do BBDM searches constrain or exclude sub-GeV DM explanations of high-energy neutrinos, both from TXS 0506+056 and diffuse? What complementary information do they provide on DM and blazar jets?  

\medskip

Our work 
is organised as follows.
In Sec.~\ref{sec:intro_astro} we review the properties of blazars and of DM around them.
In Sec.~\ref{sec:intro_particle} we define the DM interactions object of our study and compute the relevant cross sections, including terms usually omitted in the literature but important in our context. We defer details on elastic and inelastic ones, respectively, to  App.~\ref{app:elastic} and~\ref{app:DIS}.
Then, in Sec.~\ref{sec:BBDM} we compute BBDM fluxes and their signals from TXS 0506+056, as well as from a population of blazars, at a variety of large neutrino detectors, including Earth attenuation (for which we provide more details in App.~\ref{app:EarthAtt}). We then derive  
novel limits and sensitivities on DM-nucleon interactions.
In Sec.~\ref{sec:limits} we compare them with a variety of other tests of the same DM dynamics, along with the DM-around-blazar explanation of high-energy neutrinos, both diffuse and from TXS 0506+056.  In the same section, we also discuss the potential depletion of the DM spike due to DM-proton collisions. The reader can find further details on the results for a different benchmark choice of the parameters that determine the DM profile, as well as on the effects of DM $4\to2$ annihilations and DM self-interactions ($2 \to 2$ processes), 
respectively in App.~\ref{app:BMCII} and \ref{app:Spikedepletion}. Finally, in Sec.~\ref{sec:summary} we conclude and discuss other directions worth future exploration.

\section{Blazars and dark matter around them}
\label{sec:intro_astro}

\subsection{Blazars and their jet models}
\label{subsec:blazar_jet}
Active galactic nuclei (AGN) are compact regions at the centre of certain galaxies that can outshine their hosts, and are among the brightest objects in the Universe. According to current understanding, they are powered by the accretion of matter onto supermassive black holes (BHs) with typical masses of $M_{\BH} \approx 10^{8-9}M_\odot$, $M_\odot$ denoting the solar mass. 
They can feature a pair of back-to-back jets, made of relativistic plasma, launched along the polar axis of the hosted supermassive BH and perpendicularly to the accretion disk. Jetted AGN are called \textit{blazars} when one of the two jets forms a small angle $\theta_\LOS \lesssim 15^\circ-20^\circ$ with our line-of-sight (LOS) \cite{Urry:1995mg, Giommi:2011sn, Giommi:2013mck, Padovani:2017zpf}.

Although blazars are relatively rare -- with more than 3500 identified to date (see, e.g., \cite{Massaro:2015nia}) -- they dominate the $\gamma$-ray sky within the $50\,\text{MeV}$ to $1\,\text{TeV}$ energy range, comprising $\sim 56\%$ (up to $\lesssim 88\%$) of the fourth \textit{Fermi}-LAT catalogue \cite{Fermi-LAT:2022byn, Ballet:2023qzs}. Their electromagnetic activity is mainly non-thermal, with a spectral energy distribution (SED) of photons that covers the entire electromagnetic spectrum and exhibits two distinct peaks: one in the infrared/X-ray band and the other at $\gamma$-ray frequencies \cite{Padovani:2017zpf}.
Understanding the origin of this broad and doubly-peaked SED requires detailed modelling of the blazar jet physics.

The main structures in blazar jet models are the so-called \textit{blobs}: spherical regions of plasma which move at a bulk relativistic speed $\beta_B$ along the jet axis. The accelerated charged particles that are confined within the blobs radiate non-thermally while propagating through magnetic fields and ambient radiation. A particularly compelling category of blazar jet models is that of the hybrid lepto-hadronic ones, according to which both electrons and protons are present in the blob with ultra-relativistic velocities, and the SED arises from a combination of leptonic and hadronic processes (see, e.g., \cite{Cerruti:2020lfj} for a review). A distinctive prediction of this class of models is the production of 
neutrino fluxes from hadronic cascades initiated when protons collide with photons or other protons in the blazar environment. After IceCube's first $\gtrsim 3\sigma$ spatial association between multi-TeV neutrino events and a blazar, TXS 0506+056 \cite{IceCube:2018dnn, IceCube:2018cha, Padovani:2018acg} \footnote{Two significant associations of neutrino signals and TXS 0506+056 have been reported: one single neutrino event in 2017 in coincidence with a six-month $\gamma$-ray flaring episode \cite{IceCube:2018dnn}, and $\sim 13$ events in 2014/2015 \cite{IceCube:2018cha}. The latter, however, were not accompanied by an enhanced electromagnetic activity of the same blazar \cite{Fermi-LAT:2019hte}.} an intense campaign to explain both the neutrino emissions and the electromagnetic activities of TXS 0506+056 with lepto-hadronic modelling has been carried out \cite{Gao:2018mnu, MAGIC:2018sak, Cerruti:2018tmc,Cerruti:2018tmc_erratum, Keivani:2018rnh} (see also \cite{ Banik:2019jlm, Banik:2019twt, Cao:2019fnn, Das:2022nyp, MAGIC:2022gyf}), supporting the hypothesis that highly-accelerated protons are present in its jets. Other tentative associations of different neutrino events with other blazars, albeit with lower significance, have also been reported \cite{Kadler:2016ygj, Paliya:2020mqm, Rodrigues:2020fbu, Oikonomou:2021akf, Liao:2022csg, Sahakyan:2022nbz, Fermi-LAT:2019hte, Jiang:2024nwa, Ji:2024dgn, Ji:2024zbv} (see also \cite{Giommi:2021bar, Boettcher:2022dci}) \footnote{Also, the active galaxy NGC 1068 has shown evidence of neutrino emission
    at 4.2$\sigma$ \cite{IceCube:2022der}, but this object is not a blazar.
    Furthermore, the study~\cite{IceCube:2019cia} suggests a correlation at the level of 3.3$\sigma$ of
    IceCube neutrinos with a catalogue composed of 110 known $\gamma$-ray emitters, of which TXS 0506+056 and NGC 1068 are two main contributors.}, and the same class of models has been applied to several other sources (as, e.g., 
    in \cite{Rodrigues:2023vbv}), advancing the idea that the lepto-hadronic framework may constitute a comprehensive description of blazar jet emission.

 Motivated by these studies and observations, in this work we concentrate on lepto-hadronic models to describe the proton content in blazar jets. Specifically, we model the proton energy spectrum in blazar jets with a homogeneous and isotropic single power-law in the blob's frame:
 \begin{equation}\label{eq:JetSpectrumBlob}
  \frac{d\Gamma'_p}{d\gamma'_p \, d\Omega'} = \frac{\kappa_p}{4\pi} {\gamma'_p}^{-\alpha_p}\,e^{-\gamma_p'/\gamma'_{\max_p}} \,.
  \end{equation}
  Here, $d\Gamma_p'$ is the infinitesimal rate of particles ejected in the blobs along the direction $d\Omega'$ and with $\gamma_p'\equiv E_p'/m_p$ lying in the range $[\gamma'_p, \;\gamma'_p + d\gamma'_p]$, $E'_p$ being the energy of the proton and $m_p= 0.938\,\text{GeV}$ its mass; $\alpha_p\geq 0$ is the slope of the power-law; $\kappa_p$ is an overall normalisation constant. We have also included an exponential cut-off at $\gamma'_{\max_p}$.
  In the rest of this section, primed (unprimed) symbols denote non-invariant quantities computed in the blob’s (observer’s) rest frame. 
The spectrum in Eq.~\eqref{eq:JetSpectrumBlob} is related to that in the observer's frame by a Lorentz transformation of boost factor $\Gamma_B \equiv (1-\beta_B^2)^{-1/2}$ along the jet axis, which gives  \cite{Wang:2021jic, Granelli:2022ysi} (see also \cite{Gorchtein:2010xa}):
  \begin{equation}
 \label{eq:CRSpectrum}
     \frac{d\Gamma_p}{d\gamma_pd\Omega} 
     = \frac{\kappa_p}{4\pi}\,\gamma_p^{-\alpha_p} \frac{\beta_p(1-\beta_p\beta_B  \mu)^{-\alpha_p} \Gamma_B^{-\alpha_p} e^{-\gamma_p/\gamma_{\max_p}}}{\sqrt{(1-\beta_p \beta_B \mu)^2 - (1-\beta_p^2)(1-\beta_B^2)}}\,,
 \end{equation}
 where $\beta_p =[1-1/\gamma_p^2]^{1/2}$ is the proton's speed and $\mu$ is the cosine of the angle between its direction of motion and the jet axis.

We perform our analysis focusing on the single source TXS 0506+056, as well as a sample of blazars. 
For TXS 0506+056, we fix the parameters of the proton spectrum according to the 
lepto-hadronic model of \cite{Keivani:2018rnh} fitted to the observed SED during the six-month flaring activity of 2017.
For the blazar sample, we consider the 324 blazars and their corresponding lepto-hadronic fit of their steady activity as presented in \cite{Rodrigues:2023vbv}. The relevant parameters that we extrapolate from \cite{Keivani:2018rnh, Rodrigues:2023vbv} are the minimal and maximal Lorentz boost factors of the protons $\gamma'_{\min_p}$, $\gamma'_{\max_p}$; the blob Lorentz factor $\Gamma_B$; the LOS angle of the jet $\theta_\LOS$ assumed to be $\theta_\LOS=1/\Gamma_B$; \footnote{In \cite{Keivani:2018rnh}, the LOS angle was taken to be zero, given the assumed relation $\Gamma_B = \mathcal{D}/2$, where $\mathcal{D} \equiv [\Gamma_{\rm B}\left(1-\beta_{\rm B}\cos\theta_{\LOS}\right)]^{-1}$ is the Doppler factor determined by the fit. Here, we prefer to consider a non-vanishing angle and use the condition $\Gamma_B = \mathcal{D}$, which implies $\theta_{\LOS} = 1/\Gamma_B$. Under this assumption, the proton luminosity given in \cite{Keivani:2018rnh} should be multiplied by a factor of 4 for consistency. However, for the sake of conservativeness, we refrain from applying this correction.}
the proton luminosity $L_p$, which is related to the normalisation constant $\kappa_p$ via \cite{Granelli:2022ysi}
\begin{equation}
L_p= \kappa_p m_p \Gamma_B^2 \int_{\gamma'_{\min_p}}^{\gamma'_{\max_p}} dx\,x^{1- \alpha_p}\,.
\label{eq: luminosity_normalisation}
\end{equation}
We summarise in Table~\ref{tab:AGNparameters} the relevant 
jet 
parameters 
of TXS 0506+056 and
of the sample of blazars~\cite{Rodrigues:2023vbv}, together with the redshift $z$, the luminosity distance $d_L \equiv (1+z)\,c \int_0^z dz'/H(z')$ (c being the speed of light, $H(z) = H_0 \sqrt{\Omega_m(1+z)^3 + \Omega_\Lambda}$ the Hubble expansion rate, $H_0$ its present value, $\Omega_m$ the matter density parameter and $\Omega_\Lambda$ the dark energy one), the BH mass $M_{\BH}$ and the corresponding Schwarzschild radius $R_S$.

\newcommand{\thickhline}{\noalign{\hrule height 3pt}} 

\setlength{\arrayrulewidth}{.8pt}

\setlength{\tabcolsep}{0pt}
\newcolumntype{C}{@{}>{\centering\arraybackslash}X}
\begin{table}
\centering
\begin{tabularx}{\linewidth}{|@{}C||C|C@{}|}
    \hline
    \multicolumn{3}{|c|}{\rule{0pt}{2.5ex} \cellcolor{gray!20}\bf Lepto-Hadronic Model Parameters}\\
    \thickhline
    \rule{0pt}{2.5ex}
{\bf Parameter (units)} & {\bf TXS 0506+056 \cite{Keivani:2018rnh}} & {\bf Blazar sample \cite{Rodrigues:2023vbv}}\\ 
\hline
    \rule{0pt}{2.5ex}
$z$ & 0.337 & $\left[ 0.04, 3.41 \right]$\\
$d_L$ (Mpc) & 1765 & $\left[ 171, 30.2\times 10^3 \right]$\\
$M_\BH (M_\odot)$  & $3 \times 10^8$ & $\left[ 10^8, 10^9 \right]$\\
$R_S$ (pc) & $3\times 10^{-5}$  & $[10^{-5}, 10^{-4}]$\\
$\Gamma_B$ & 24.2 & $\left[ 3.4, 31.5 \right]$\\
${\theta_\LOS}(^{\circ})$ & 2.37 & $\left[ 1.81, 16.95 \right]$\\
$\alpha_p$ & 2 & 1\\
$\gamma'_{\min_p}$ & 1 & 100\\
$\gamma'_{\max_p}$ & $1.6 \times 10^{7}$ & $\left[ 10^6, 3.1 \times 10^7 \right]$\\
$L_p$ (erg/s)  & $1.85 \times 10^{50}$ & $\left[ 2.2 \times 10^{44}, 3.1 \times 10^{50}\right]$\\
$\kappa_p\,(\text{s}^{-1}\text{sr}^{-1})$ & $1.27 \times 10^{49}$ &  $\left[ 3.1 \times 10^{37}, 1.3 \times 10^{46}\right]$\\
\hline
\end{tabularx}
\caption{The relevant 
jet 
parameters from the 
lepto-hadronic fit~\cite{Keivani:2018rnh} for the blazar TXS 0506+056 
(during its 2017 flare) 
and the sample of blazars~\cite{Rodrigues:2023vbv} 
(average steady flux) 
used in our calculations. Values in square brackets denote the ranges for the sources in \cite{Rodrigues:2023vbv}. Also listed are the redshift $z$; the luminosity distance $d_L$, which we compute from $z$ assuming a standard cosmological model and taking the value of the Hubble constant today as $H_0 = 70.2 \,\text{km}\,\text{s}^{-1}\,\text{Mpc}^{-1}$, the matter density parameter $\Omega_m = 0.315$ and the dark energy one $\Omega_\Lambda = 0.685$ \cite{ParticleDataGroup:2024cfk};
the BH mass in solar mass units $M_\odot$, and the corresponding Schwarzschild radius $R_S$. The values of $z$ and $M_{\BH}$ are taken from~\cite{Paiano:2018qeq,Padovani:2019xcv} for TXS 0506+056 and~\cite{Rodrigues:2023vbv} for the sample of blazars.}\label{tab:AGNparameters}
\end{table}

\subsection{Dark matter around blazars}

DM is expected to form dense spikes in the vicinity of supermassive BHs, like those that power blazars. As demonstrated in \cite{Gondolo:1999ef}, the adiabatic growth of BH at the centre of a spherical DM halo with a power-law density $\rho^\text{halo}_{\DM}(r)=\mathcal{N} r^{-\gamma}$ reshapes the halo profile into a steeper distribution
 \begin{equation}
 \label{eq:Rsp}
     \rho^\text{spike}_{\DM}(r) = \mathcal{N} R_\text{sp}^{-\gamma} \left(\frac{R_\text{sp}}{r}\right)^{\alpha(\gamma)}, \quad\text{with}\quad \alpha(\gamma)=\frac{9-2\gamma}{4-\gamma},
     \end{equation}
where $\mathcal{N}$ is a normalisation constant, $r$ the
radial distance from the central BH and $R_{\text{sp}}$ the radial extension of the spike. The latter is given as a function of $\gamma$, the BH mass and the normalisation constant by~\cite{Gondolo:1999ef}
\begin{equation}
R_\text{sp} \simeq \epsilon(\gamma) \left(\frac{M_{\BH}}{\mathcal{N}}\right)^{1/(3-\gamma)},
\end{equation}
where $\epsilon(\gamma)\approx 0.1$ for $0.5\leq \gamma \leq 1.5$~\cite{Merritt:2003qc}.
For the blazars under consideration, we model the total DM profile $\rho_{\DM}(r)$ as 
\begin{equation}\label{eq:rhoDM}
\rho_{\DM}(r) = \mathcal{N}R_{\text{sp}}^{-\gamma}\begin{cases}
0&r\leq 2R_S\,;\\
g(r)(R_{\text{sp}}/r)^{\alpha(\gamma)}&2R_S \leq r < R_\text{sp}\,;\\
(R_{\text{sp}}/r)^\gamma& r\geq R_{\text{sp}}\,;
\end{cases}
\end{equation}
where $g(r)=(1-2R_S/r)^{3/2} $ accounts for 
the inevitable capture of DM onto the BH after including relativistic effects \cite{Sadeghian:2013laa}. In the analysis that follows, we fix $\gamma=1$ for the initial DM profile, inspired by a Navarro-Frenk-White (NFW) distribution \cite{Navarro:1995iw, Navarro:1996gj}, so that  $\alpha(\gamma = 1) = 7/3$.

Due to the blazar jet emission outshining the dynamics of the host galaxy, information on the DM distribution around blazars is limited. Consequently, the normalisation of the DM profile remains somewhat arbitrary.
Following the same procedure as in \cite{DeMarchi:2024riu}, we fix $R_\text{sp} = R_\star$,  $R_\star\approx 10^6R_S$ being the typical radius of influence of a BH on stars \cite{Kormendy:2013dxa}.
This normalisation, for $\gamma = 1$, results in \begin{equation}\label{eq:mathcalN}
\mathcal{N} \simeq 10^{-6} M_\odot/R_S^2 \left(\frac{M_\BH}{10^8 M_\odot}\right).
\end{equation} 
As can be easily checked, the total 
DM mass contained within $R_{\text{sp}}$ is $M_{\DM}^{\text{spike}}\simeq 18.5\%\,M_{\BH}$. This means that our normalisation does not interfere with typical BH mass estimates, which are performed within the sphere of BH influence \cite{Labita:2006jg, Pei:2021qay}, at least up to corrections of order $\mathcal{O}(10\%)$ (the uncertainty in the BH mass estimate in these objects is in any case much larger).
Furthermore, our normalisation is more conservative than that adopted in similar studies on DM around AGN~\cite{Gorchtein:2010xa, Wang:2021jic, Granelli:2022ysi, Cline:2022qld, Ferrer:2022kei, Bhowmick:2022zkj, Cline:2023tkp, Herrera:2023nww, CDEX:2024qzq, Gustafson:2024aom, Mishra:2025juk, Wang:2025ztb}.
Observational evidence of DM spikes around SMBHs has been tentatively reported recently, from friction in a binary system~\cite{chan2024first} and from reverberation mapping of AGNs~\cite{Sharma:2025ynw}.

All the information on the DM distribution that is relevant to our  calculation ultimately reduces to the following parameter \cite{Wang:2021jic, Granelli:2022ysi} (see also \cite{Gorchtein:2010xa}):
\begin{equation}\label{eq:DMColumnDensity}
    \Sigma_{\DM}(r)\equiv \int_{R_{\min}}^{r} \rho_{\DM}(r') dr',
\end{equation}
where $R_{\min}$ is the minimal radial extension of the blazar jet. The above integral is quickly saturated by the spike contribution, which reads
\begin{equation}\label{eq:SigmaDMspike_analytical}
    \Sigma_{\DM}^\text{spike}   \equiv \Sigma_\DM(R_{\text{sp}}) \simeq 
        \dfrac{\epsilon(\gamma)^{3-\gamma}}{\alpha(\gamma)-1}\frac{M_{\BH}}{R_{\min}^2}\left(\dfrac{R_{\min}}{R_\text{sp}}\right)^{3-\alpha(\gamma)}\,,
\end{equation}
where we have used the DM profile in Eq.~\eqref{eq:rhoDM} without fixing $\gamma$, after inverting the relation in Eq.~\eqref{eq:Rsp} to get $\mathcal{N}$, and integrated Eq.~\eqref{eq:DMColumnDensity} up to $R_{\text{sp}}$. For $\gamma = 1$ and $R_{\text{sp}} = 10^6 R_S$, the above relation gives \begin{equation}
\Sigma_{\DM}^{\text{spike}} 
\simeq 3\times 10^{-7}\,\frac{M_\BH}{R_S^2}\left(\frac{2R_S}{R_\min}\right)^{4/3}
 \simeq 1.28\times 10^{31}\,\text{GeV}\,\text{cm}^{-2}\left(\frac{3\times 10^8 M_\odot}{M_\BH}\right)\left(\frac{2R_S}{R_\min}\right)^{4/3}.
\end{equation}

 Various effects can alter the formation and evolution of the DM spike. For instance, DM annihilations over the BH lifetime $t_{\BH}$~\cite{Gondolo:1999ef} flatten the DM profile at inner radii reducing the slope of the spike to $\leq 0.5$ for $r<r_\text{ann}$~\cite{Shapiro:2016ypb}, $r_\text{ann}$ being defined by the condition $ \rho_{\DM}(r_\text{ann}) = \rho_\text{core}\equiv m_\chi/(\langle\sigma_\text{ann} v_\text{rel}\rangle t_{\BH})$, which gives
\begin{equation}
r_\text{ann} = R_\text{sp}
    \left[\dfrac{M_{\BH}\epsilon(\gamma)^{3-\gamma}}{R_\text{sp}^3\rho_\text{core}}\right]^{1/\alpha(\gamma)} 
    \quad \text{for} \quad 2R_S \leq r_\text{ann} \leq R_\text{sp},
\end{equation} 
and $\langle\sigma_\text{ann} v_\text{rel}\rangle$ is the DM averaged annihilation cross section times relative velocity. The DM profile including the effects of DM annihilation can be approximated as $\tilde{\rho}_\DM(r) \equiv \rho_\DM(r) \rho_\text{core}/(\rho_\DM(r) + \rho_\text{core})$ \cite{Gondolo:1999ef}. Other effects can influence the DM spike, such as merging of galaxies~\cite{Ullio:2001fb, Merritt:2002vj} and
the gravitational interaction of stars close to the BH~\cite{Gnedin:2003rj, Bertone:2005hw}.
Sizeable uncertainties also reside in $R_{\min}$, the minimal radius of the jet, and they have a substantial impact on $\Sigma_{\DM}$. 
Therefore, to effectively account for 
all these  
astrophysical effects on the DM spike, instead of considering different scenarios of softenings or depletions, we find it more practical to adopt the density profile in Eq.~\eqref{eq:rhoDM} and consider different benchmark cases (BMCs) for $R_{\min}$.
In particular, as in \cite{DeMarchi:2024riu}, we choose
$R_{\min} = 10^2 R_S$ (BMCI) and $R_{\min} = 10^4 R_S$ (BMCII), based on results from blazar studies indicating where the jet is likely well-accelerated \cite{Rodrigues:2023vbv} (see also \cite{Okino:2021iuj}).

\begin{figure}[t!]
    \centering
    \includegraphics[width=0.8\linewidth]{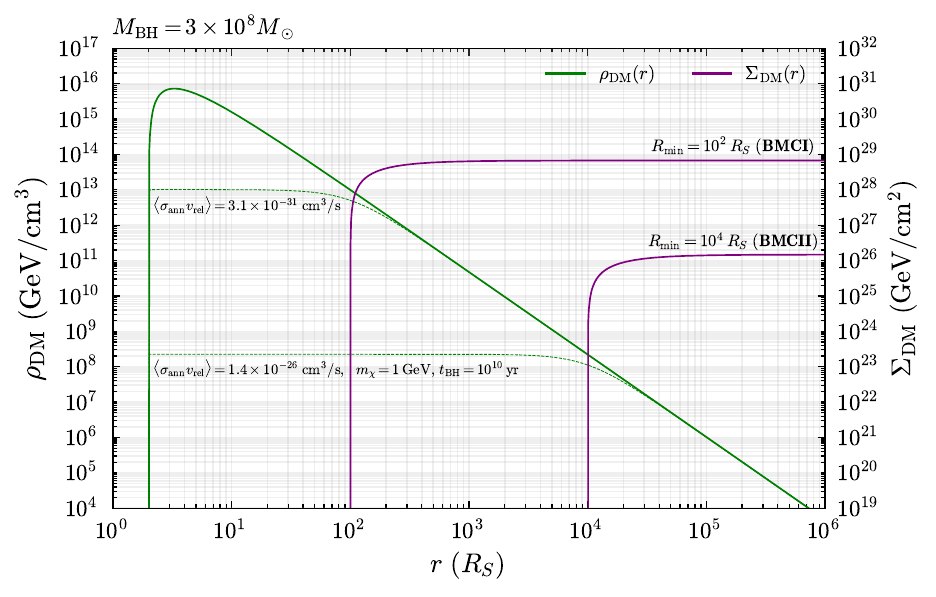}
    \caption{The DM density profile $\rho_\DM(r)$ (solid green, left axis) and its column density $\Sigma_{\DM}(r)$ (solid purple, right axis) for BH mass $M_{\BH}=3 \times 10^8 M_\odot$. The DM column density is plotted for the two BMCs: $R_{\min} = 10^2 R_S$ (BMCI, upper curve) and $R_{\min} = 10^4 R_S$ (BMCII, lower curve). We also plot the DM profile $\tilde{\rho}_\DM(r)$ (green dashed) which includes DM annihilations for the maximal values of $\left\langle \sigma_\text{ann} v_\text{rel}\right\rangle$ which yield $\Sigma_{\DM}$ consistent with BMCI and BMCII, taking the BH lifetime $t_{\BH}=10^{10}$ yr and DM mass $m_\chi=1$ GeV.}
    \label{fig:DMDensitiyPlot}
\end{figure}

For the adopted density profile, $M_\BH = 3\times 10^8\,M_\odot$ and $\gamma = 1$, we find 
\begin{equation}
    \Sigma_{\DM}^\text{spike} \simeq 
    \begin{array}{ll}
    6.9\times 10^{28} \,\text{GeV}\,\text{cm}^{-2}&\text{for BMCI} 
    \\
    1.5\times 10^{26}\,\text{GeV}\,\text{cm}^{-2} &\text{for BMCII}
    \end{array}
\end{equation}
We note that taking $\left\langle \sigma_\text{ann} v_\text{rel}\right\rangle$ such that $r_\text{ann} \leq R_\min$, and calculating the DM column density for $\tilde{\rho}_\DM(r)$, would practically lead to values of $\Sigma_{\DM}^\text{spike}$ between the two BMCs considered. 
Imposing $r_\text{ann} \leq R_\min$ gives 
\begin{equation}
    \begin{array}{ll}
    \left\langle \sigma_\text{ann} v_\text{rel}\right\rangle \lesssim \sigma_\text{BMCI}~\equiv
    3.1\times 10^{-31}\,\,\text{cm}^3\,\text{s}^{-1}\,(m_{\chi}/\text{GeV})(10^{10}\,\text{yr}/t_\BH)&\text{for BMCI},\\
    \left\langle \sigma_\text{ann} v_\text{rel}\right\rangle \lesssim \sigma_\text{BMCII}\equiv1.4\times 10^{-26}\,\,\text{cm}^3\,\text{s}^{-1}\,(m_{\chi}/\text{GeV})(10^{10}\,\text{yr}/t_\BH)&\text{for BMCII}.
    \end{array}
    \end{equation} Then, values of $\left\langle \sigma_\text{ann} v_\text{rel}\right\rangle \lesssim \sigma_{\text{BMCI}}$ would yield a DM column density consistent with the BMCs considered. Intermediate values $\sigma_{\text{BMCI}} \lesssim \left\langle \sigma_\text{ann} v_\text{rel}\right\rangle \lesssim \sigma_{\text{BMCII}}$ would correspond to situations between BMCI and BMCII. Larger values $\left\langle \sigma_\text{ann} v_\text{rel}\right\rangle \gtrsim \sigma_{\text{BMCII}}$ would lead to DM column densities smaller than that of our BMCII. However, values of $\left\langle \sigma_\text{ann} v_\text{rel}\right\rangle$ larger than 
 about $10^{-28}$~cm$^3$/s, for sub-GeV DM, are very hard to realise because of stringent indirect detection constraints -- see, e.g., Sec.~6.13 of ~\cite{Cirelli:2024ssz} for a recent summary.
 Our BMCII can then be considered as very conservative with respect to possible DM annihilations.
    To clarify this discussion, we depict in Fig.~\ref{fig:DMDensitiyPlot} the DM profile (green curves) adopted in our analysis (thick solid), as well as the core plus spike profile in case $r_\text{ann} = R_\min$ (thin dashed). In the same plot, we show the corresponding DM column density for BMCI and BMCII.

\section{Dark matter-nucleon interactions}
\label{sec:intro_particle}

\subsection{Toy models}
\label{sec:ToyModels}
We consider a DM particle consisting, for definiteness, of a Dirac fermion $\chi$ with mass $m_\chi$.
Modelling the DM-nucleon interactions with a constant cross section would be inconsistent for our purposes, which involve processes (upscatterings in blazar jets, scatterings on nuclei on Earth) that happen in a range of energies that spans orders of magnitude. We are not aware of any particle DM model, within reach of the searches object of this paper, that results in a cross section with the same value over that range of energies.
We therefore model the DM-nucleon interactions by specifying their mediator. We consider, in particular, vector, scalar, axial and pseudoscalar mediators.
We report the associated Lagrangians for DM-quark interactions below.
They induce non-relativistic DM-nucleon interactions of different kinds, relevant for experiments looking for the halo DM components. Our toy-models thus allow to explore the complementarity of those \virg{standard} DM searches with the ones proposed here.
To make this manifest, we report below also the non-relativistic operator induced by each mediator.

\begin{itemize}
\item {\bf Scalar mediator}: DM-quark interactions are mediated by a scalar $\phi$ with mass $m_\phi$ as
\begin{equation}
    \mathcal{L}_{\chi q \phi} = g_{\chi \phi} \bar{\chi}\chi \phi + g_{q \phi} \bar{q}q \phi 
    \quad \rightarrow \quad
    \mathcal{O}_{\chi N \phi} = 1\!\!1\,, \quad (\text{SI})
    \label{eq:L_scalar}
\end{equation}
where $q$ runs over light quarks and $\mathcal{O}_{\chi N \phi}$ is the spin-independent (SI) DM interaction with nucleons $N$ induced by $\mathcal{L}_{\chi q \phi}$.

\item{\bf Vector mediator}:  DM-quark interactions are mediated by a vector $V$ with mass $m_{V}$ as
\begin{equation}
\mathcal{L}_{\chi q V} = g_{\chi V} \bar{\chi}\gamma^\mu \chi V_\mu +  g_{q V} \bar{q} \gamma^\mu q V_\mu
\quad \rightarrow \quad
    \mathcal{O}_{\chi N V} = 1\!\!1\,, \quad (\text{SI})
\label{eq:L_vector}
\end{equation}
where $q$ runs over light quarks and $\mathcal{O}_{\chi N V}$ is the SI DM interaction with nucleons induced by $\mathcal{L}_{\chi q V}$.
\item{\bf Axial mediator}:  We consider the interaction with an axial mediator $V'$ with mass $m_{V'}$ as follows
\begin{equation}
    \mathcal{L}_{\chi q V'} = g_{\chi V'} \bar{\chi}\gamma^\mu\gamma_5 \chi V'_\mu +  g_{q V'} \bar{q} \gamma^\mu \gamma_5 q V'_\mu
    \quad \rightarrow \quad
    \mathcal{O}_{\chi N V'} = \vec{S}_\chi \cdot \vec{S}_N\,, \quad (\text{SD})
    \label{eq:L_axial}
    \end{equation}
where $q$ runs over light quarks, $\vec{S}_\chi$ and $\vec{S}_N$ denote respectively the DM and nucleon spin vectors, and $\mathcal{O}_{\chi N V'}$ is the spin-dependent (SD) DM interaction with nucleons induced by $\mathcal{L}_{\chi q V'}$.

\item
{\bf Pseudoscalar (ALP) mediator}: DM-quark interactions are mediated by a pseudoscalar $a$ with mass $m_a$ (i.e. an axion-like particle, or ALP) as
\begin{equation}
\mathcal{L}_{\chi q a} = i g_{\chi a}\bar{\chi}\gamma_5\chi a + i g_{q a} \bar{q} \gamma_5 q a,
\quad \rightarrow \quad
    \mathcal{O}_{\chi N a} = \left(\vec{S}_\chi \cdot \frac{\vec{Q}}{m_N}\right)\left(\vec{S}_N \cdot \frac{\vec{Q}}{m_N}\right), \quad (\text{SMD})
\label{eq:L_pseudoscalar}
\end{equation}
where $q$ runs over light quarks, $\mathcal{O}_{\chi N a}$ is the spin- and momentum-dependent (SMD) DM-nucleon interaction induced by $\mathcal{L}_{\chi q a}$, and $\vec{Q}$ is the 3-momentum transfer.

\end{itemize}

\noindent The DM-quark interactions above do not respect the electroweak (EW) symmetry and so are valid below its breaking scale. We will briefly comment on possible EW-invariant UV completions for each of them in Sec.~\ref{sec:constraints_modeldep}, together with the constraints that they induce on the couplings $g_{\chi Y}$ and $g_{q Y}$ of Eqs.~(\ref{eq:L_scalar}-\ref{eq:L_pseudoscalar}),  where $Y$ is the mediator under consideration.

 In order to ease the comparison of our results with the literature, it is convenient to recast them in terms of the non-relativistic SI ($\sigma^N_\SI$) and SD ($\sigma^N_\SD$) DM–nucleon scattering cross sections, $N=p,n$. These are obtained by integrating out the mediator $Y$ from $\mathcal{L}_{\chi q Y}$ and by computing the cross sections in the limit of vanishing momentum transfer and  $s \to (m_\chi + m_N)^2$, being 
 $s$ the centre-of-mass energy squared.

For $Y=\phi, V$ ($\sigma^{\NR}_{Y \chi N} =\sigma^N_\SI$) and $Y=V'$ ($\sigma^{\NR}_{Y \chi N} =\sigma^N_\SD$), with DM spin $s_\chi=1/2$, one obtains
\begin{equation}\label{eq:NRlimit}
    \sigma^{\NR}_{Y \chi N} =\frac{g_{\chi Y}^2 g_{N Y}^2}{\pi}\frac {\mu_{\chi N}^2}{m_Y^4}\,,
    \quad Y=\phi,V,V'\,,
\end{equation}
where $\mu_{\chi N}$ is the DM-nucleon reduced mass and $g_{NY}$ the DM-nucleon coupling, which we derive for each model in App.~\ref{app:elastic} and report synthetically here 
\begin{align}
&g_{p\phi} \simeq 8.4 \, g_{u\phi} + 7.6 \, g_{d \phi} ,
&& g_{n\phi} \simeq 7.6 \,g_{u \phi} + 8.5\, g_{d\phi},
\\
&g_{pV} = 2\, g_{uV} + g_{dV},
&& g_{nV} = g_{uV} + 2 \,g_{dV},
\\
&g_{p V'} \simeq 1.5 g_{uV'}-0.7 g_{dV'}, &&  g_{n V'} \simeq -0.7 g_{uV'}+1.5 g_{dV'}\,.
\end{align}
For $Y=a$ instead, the tree-level SMD contribution to the non-relativistic  cross section is suppressed by the exchanged momentum, as shown by $\mathcal{O}_{\chi N a}$ in Eq.~\eqref{eq:L_pseudoscalar}.
In this pseudoscalar-mediated model a spin-independent cross section $\sigma^N_\SI$ is generated at higher orders, by box diagrams with $a$'s running in the loop, whose contributions have been computed in~\cite{Abe:2018emu}.
This one-loop SI cross section dominates over the tree-level SMD one in determining the direct detection signals of halo DM, which are anyway considerably weaker than those induced by the other mediators. 

The neutrino and BBDM signals object of this work require relativistic DM-quark interactions. To understand their interplay with the standard searches for non-relativistic halo DM, it is necessary to perform one extra step down the ladder of effective theories, and determine the DM scatterings with nuclei induced by the relativistic interactions of Eqs.~(\ref{eq:L_scalar}-\ref{eq:L_pseudoscalar}). 
For the SI case, DM couples coherently to all the nucleons inside a given nucleus, and the DM-nucleus cross section is enhanced as 
\begin{equation}\label{eq:SInucleus}
\sigma^A_\SI=A^2 \frac{\mu_{\chi A}^2}{\mu_{\chi N}^2} \sigma_\SI\,,
\quad\text{where} \quad 
\sigma_\SI=\frac{\mu_{\chi N}^2}{\pi A^2}\left[Z\frac{g_{\chi Y}g_{p Y}}{m_Y^2}+(A-Z)\frac{g_{\chi Y}g_{n Y}}{m_Y^2}\right]^2\,,
\quad Y=\phi,V\,,
\end{equation} 
being $Z (A)$ the atomic (mass) number of the target nucleus.
For the pseudoscalar mediator $a$, $\sigma^A_\SI$ is given by the same expression of  Eq.~(\ref{eq:SInucleus}), where instead $\sigma_\SI$ is given by a loop-suppressed contribution~\cite{Abe:2018emu}. 
Finally, the SD scattering is not coherently enhanced, 
\begin{equation}
\sigma^A_\SD
= S(0) \frac{\mu_{\chi A}^2}{\mu_{\chi N}^2}\sigma_\SD\,,\quad \text{where} \quad \sigma_\SD = 
\frac{g_{\chi V'}^2 g_{N V'}^2}{\pi}\frac {\mu_{\chi N}^2}{m_{V'}^4}\,,
\end{equation}
i.e.~$\sigma_\SD = \sigma^{\NR}_{Y \chi N}$ of Eq.~(\ref{eq:NRlimit}) for the axial mediator $Y=V'$, and where $S(0)$ is 
a combination of nuclear response functions in the limit of vanishing momentum transfer, which is not $A^2$-enhanced and depends on the magnitude of the spin of the nucleus and on the spatial distribution of the individual nucleons -- expressions for it can be found, e.g., in \cite{Cirelli:2024ssz}.

\subsection{Elastic scattering}
We now move beyond the fully non-relativistic limit and present cross sections that include the dependence on $Q^2 \neq 0$.
For sufficiently small momentum transfer, the collision of a proton $p$ and
a DM particle can be regarded as elastic. We define the 4-momentum of the incoming proton in the DM rest frame as $p_p = (E_p, \vec{p}_p)$ while for the DM particle
$p_\chi = (m_\chi, \vec{0})$. Concerning the final states, we denote the 4-momenta of the scattered DM particle and the outgoing proton, respectively, as $k_p = (E_p, \vec{k}_p)$ and $k_\chi = (E'_\chi, \vec{k}_\chi)$. The Feynman diagram for the process is sketched below using the package Ti\textit{k}Z-Feynman~\cite{Ellis:2016jkw}
\begin{center}
\begin{tabular}{cc}
\begin{tikzpicture}[baseline=(current bounding box.center)]
\begin{feynman}
\vertex(V1);
\vertex(pxi)[left=2cm of V1];
\vertex(pxf)[right=2cm of V1];
\node[blob, below = 1.5cm of V1](V2);
\path (V2.+180) ++ (0:-1.58) node[vertex] (ppi);
\path (V2.+145) ++ (0:-1.65) node[vertex] (ppiup);
\path (V2.+220) ++ (0:-1.68) node[vertex] (ppidown);
\vertex(ppf)[right=1.95cm of V2];
\path (V2.-40) ++ (00:1.68) node[vertex] (ppfdown);
\path (V2.+35) ++ (00:1.68) node[vertex] (ppfup);
\diagram*{
(pxi)--[fermion, edge label=\(\chi\)](V1),
(ppi)--[fermion](V2),
(ppiup)--[with arrow=0.48](V2.+145),
(ppidown)--[with arrow=0.476](V2.+220),
(V1)--[edge label' = \(Y\), color = nicepurple](V2),
(V1) --[fermion, edge label=\(\chi\)](pxf),
(V2.+35) --[with arrow = 0.51](ppfup),
(V2) --[fermion](ppf),
(V2.-40) --[with arrow=0.52](ppfdown)};
\vertex(X1) [above left=.2cm of ppiup];
\vertex(X2) [below left=.2cm of ppidown];
\draw [decoration={brace, mirror}, decorate] (X1) -- (X2) node [pos=0.5, left = .2em] {\(p\)};
\vertex(Y1) [above right=.2cm of ppfup];
\vertex(Y2) [below right=.2cm of ppfdown];
\draw [decoration={brace}, decorate] (Y1) -- (Y2) node [pos=0.5, right = .2em] {\(p\)};
\end{feynman}
\end{tikzpicture}
\end{tabular}
\end{center}
where $Y=\phi,\,a,\,V,\,V'$ denotes either the scalar, pseudoscalar,
vector or axial mediator. For the $2\to2$ elastic scattering under consideration, the differential cross section can be computed as $d\sigma^{\EL}_{Y\chi p}/d\mu_s^* = |\overline{\mathcal{M}_{Y\chi p}}|^2/(32\pi s)$,
with $-1\leq\mu_s^*\leq 1$ being the cosine of the scattering angle of emission in the center-of-mass rest frame, $s = (p_\chi + p_p)^2$ and
$|\overline{\mathcal{M}_{Y\chi p}}|^2$ the squared matrix element averaged over the initial spins and summed over the final spins.
The momentum transfer squared in the rest frame of $\chi$ reads $Q^2 = -(p_\chi - k_\chi)^2 = 2m_\chi T_\chi$, with $T_\chi \equiv E_\chi-m_\chi$ the kinetic energy of the outgoing $\chi$.
The latter is given as a function of $\mu_s^*$ by
\begin{equation}
T_\chi(\mu_s^*) = T_\chi^{\max}(T_p) \frac{1+\mu_s^*}{2}\quad\text{with}\quad T_\chi^{\max}(T_p) 
= \frac{2m_\chi T_p\left(T_p + 2m_p \right)}{2m_\chi T_p + (m_p+m_\chi)^2},
\end{equation}
where $T_p = E_p-m_p$ is the kinetic energy of the incoming proton.
Consider now the following \virg{trick}:
\begin{equation}\label{eq:trick}
    \frac{d\sigma^{\EL}_{Y\chi p}}{dT_\chi} = \int d\mu_s \frac{d\sigma^{\EL}_{Y\chi p}}{dT_\chi d\mu_s} = \int d\mu_s \frac{d\sigma^{\EL}_{Y\chi p}}{d\mu_s} \delta(T_\chi - T_\chi(\mu_s^*)) 
    = 2\int d\mu_s \,\frac{d\sigma^{\EL}_{Y\chi p}}{d\mu_s^*}\frac{\delta(\mu_s-
    \mu_s^{\EL}(T_p,T_\chi))}{T_\chi^{\max}(T_p)},
\end{equation}
with 
\begin{equation}\label{eq:musEL}
   \mu_s^{\EL}(T_p, T_\chi)\equiv \frac{(T_p+m_p+m_\chi)T_\chi}{\sqrt{(T_p^2+2m_pT_p)(T_\chi^2+2m_\chi T_\chi)}} \,.
        \end{equation}
From the relation in Eq.~\eqref{eq:trick} we obtain
\begin{equation}
    \frac{d\sigma^{\EL}_{Y\chi p}}{dT_\chi d\mu_s} =
    \frac{|\overline{\mathcal{M}_{Y\chi p}}|^2}{16\pi s}\frac{\delta(\mu_s-\mu_s^{\EL}(T_p,T_\chi))}{T_\chi^{\max}(T_p)}\,.
\end{equation}
By computing the averaged Feynman amplitude squared $|\overline{\mathcal{M}_{Y\chi p}}|^2$ we obtain the following differential elastic cross sections for $Y=\phi,a,V,V'$:

\begin{eqnarray}\label{eq:scalarelastic}
    \frac{d\sigma^{\EL}_{\phi \chi p}}{dT_\chi d\mu_s} &=&
    \frac{\sigma^{\NR}_{\phi \chi p}}{s}\frac{m_\phi^4}{16 \mu_{\chi p}^2}\frac{(4m_\chi^2+Q^2)(4m_p^2+Q^2)}{(m_\phi^2+Q^2)^2}G_S^2(Q^2)\frac{\delta(\mu_s-\mu_s^{\EL}(T_p,T_\chi))}{T_\chi^{\max}(T_p)},\\
\label{eq:pseudoscalarelastic}
   \frac{d\sigma^{\EL}_{a \chi p}}{dT_\chi d\mu_s} &=&
   \frac{
   g_{\chi a}^2g_{pa}^2}
   {16 \pi s}
   \frac{(Q^2)^2}{(m_a^2+Q^2)^2}\widehat{G}_{5,\,p}^2(Q^2)\frac{\delta(\mu_s-\mu_s^{\EL}(T_p,T_\chi))}{T_\chi^{\max}(T_p)},
   \end{eqnarray}
   \begin{equation}
\label{eq:vectorelastic}
\frac{d\sigma^{\EL}_{V \chi p}}{dT_\chi d\mu_s} =
\frac{\sigma^{\NR}_{V \chi p}}{s} \frac{m_p^4}{4\mu_{\chi p}^2} \left[ A^{V,\,p}(Q^2) +  \frac{(s-u)^2}{m_p^4} C^{V,\,p}(Q^2)+ \frac{m_\chi^2}{m_p^2}D^{V,\,p}(Q^2)\right]\frac{\delta(\mu_s-\mu_s^{\EL}(T_p,T_\chi))}{T_\chi^{\max}(T_p)(1+Q^2/m_{V}^2)^2},
\end{equation}
\begin{equation}
\label{eq:axialelastic}
\hspace{-.2em}\frac{d\sigma^{\EL}_{V' \chi p}}{dT_\chi d\mu_s} =
\frac{\sigma^{\NR}_{V' \chi p}}{s} \frac{m_p^4}{4\mu_{\chi p}^2}\left[ A^{V',\,p}(Q^2) +  \frac{(s-u)^2}{m_p^4} C^{V',\,p}(Q^2)+ \frac{m_\chi^2}{m_p^2}D^{V',\,p}(Q^2)\right]\frac{\delta(\mu_s-\mu_s^{\EL}(T_p,T_\chi))}{T_\chi^{\max}(T_p)(1+Q^2/m_{V'}^2)^2},
\end{equation}
where 
the functions $G_S, \widehat{G}_{5,\,p}, A^{V,\,p}, C^{V,\,p}, D^{V,\,p}$, $A^{V',\,p}$, $C^{V',\,p}$ and $D^{V',\,p}$ are nuclear form factors for the proton, which at $Q^2=0$ take the values $G_S = 1$, $\widehat{G}_{5,\,p} =1 
$, $A^{V,\,p} = D^{V,\,p} = A^{V',\,p} = 0$, $C^{V,\,p} = C^{V',\,p} = 1/4$, $D^{V',\,p} = 8$.
Details on the derivation
and on the expressions of the form factors are given in App.~\ref{app:elastic}.

\subsection{Inelastic scattering}
\label{sec:DIS}
When the momentum transfer becomes $\mathcal O (\text{GeV}^2)$, the contribution of the \textit{deep inelastic scattering} (DIS) starts to be relevant.
The process is represented diagrammatically below \cite{Ellis:2016jkw}
\begin{center}
\begin{tabular}{cc}
\begin{tikzpicture}[baseline=(current bounding box.center)]
\begin{feynman}
\vertex(V1);
\vertex(p1)[left=2cm of V1];
\vertex(k3)[right=2cm of V1];
\node[blob, below = 1.5cm of V1](V2);
\path (V2.+180) ++ (00:-1.58) node[vertex] (p2);
\path (V2.+145) ++ (00:-1.65) node[vertex] (p2up);
\path (V2.+220) ++ (00:-1.68) node[vertex] (p2down);
\vertex(kXup)[right=1.95cm of V2];
\path (V2.-45) ++ (00:1.68) node[vertex] (kXdown);
\path (V2.+45) ++ (45:1) node[vertex] (p2q);
\path (V2.+45) ++ (45:1) node[vertex] (p2q);
\vertex[right=1cm of p2q](k4q);
\diagram*{
(p1)--[fermion, edge label=\(\chi\), pos = .6](V1),
(p2)--[fermion](V2),
(p2up)--[with arrow=0.48](V2.+145),
(p2down)--[with arrow=0.47](V2.+220),
(V1)--[edge label' = \(Y\), color = nicepurple](p2q),
(V1) --[fermion, edge label=\(\chi\), pos = .6](k3),
(V2.+45) --[fermion, edge label' = \(q\,\text{, }\bar{q}\)](p2q)--[fermion, edge label = \(q\,\text{, }\bar{q}\)](k4q),
(V2) --[fermion](kXup),
(V2.-45) --[with arrow=0.53](kXdown),
};
\vertex(X1) [right=.2cm of k4q];
\vertex(X2) [right=.2cm of kXdown];
\draw [decoration={brace}, decorate] (X1) -- (X2) node [pos=0.5, right = .2em] {\(X\)};
\vertex(N1) [left=.1cm of p2up];
\vertex(N2) [left=.1cm of p2down];
\draw [decoration={brace, mirror}, decorate] (N1) -- (N2) node [pos=0.5, left = .2em] {\(p\)};
\end{feynman}
\end{tikzpicture}
\end{tabular}
\end{center}
where $Y$ denotes the mediator, $p$ the initial proton, $q\,(\bar{q})$ the (anti)quark participating the scattering, and $X$ the outgoing hadronic state.
We fix the momenta of the initial DM and proton respectively as $p_\chi =(E_\chi, \vec{p}_\chi)$ and $p_p= (E_p, \vec{p}_p)$,  with $k_\chi = (E'_\chi, \vec{k}_\chi) $ and $k_q = (E'_q, \vec{k}_q)$ that of the outgoing DM and scattered quark momenta.
Then, the squared center-of-mass energy and momentum transfer are given in terms of the momenta, respectively, by $s = (p_p + p_\chi)^2$ and $Q^2 = (p_\chi-k_\chi)^2 $. 
The differential DIS cross section for DM-proton scattering can be expressed as (see App.~\ref{app:DIS} for details):
\begin{equation}\label{eq:Dismaster}
\frac{d\sigma^{\DIS}_{Y\chi p}}{dxdy} =\frac{y}{16 \pi}\frac{Q^2 \sum_{\kappa=q\,,\bar{q}} f_\kappa^p(x,\,Q^2)|\overline{\mathcal{M}_{Y\chi}^{\quark}}|^2}{(Q^2)^2-4m_p^2m_\chi^2 x^2 y^2}\,,
\end{equation}
being $f_{q(\bar{q})}^{p}$ the parton distribution functions (PDFs) for the (anti)quark in the proton, $\overline{\mathcal{M}_{Y\chi}^{\quark}}$ is the spin-averaged Feynman amplitude at the quark level, $ y \equiv p_p\cdot q/(p_p\cdot p_\chi) 
$  the \textit{inelasticity} parameter and $ x \equiv Q^2/(2 p_p\cdot q)= Q^2/[(s-m_p^2-m_\chi^2)\, y]$ the \textit{Bjorken scaling variable}. 
After expanding the matrix element for the various toy models considered, we obtain the following final expressions for the DIS differential cross section:
\begin{eqnarray} 
\frac{d\sigma^{\DIS}_{\phi \chi p}}{dxdy} &=&   \frac{\sigma_{\phi \chi p}^{\NR}}{16\mu_{\chi p}^2 g_{p\phi}^2}\frac{(Q^2)^2(Q^2+4m_\chi^2)\,y \sum_{\kappa=q,\bar{q}}g_{\kappa\phi}^2f^p_\kappa(x,Q^2)}{[(Q^2)^2-4m_N^2m_\chi^2 x^2 y^2] (1+Q^2/m_{\phi}^2)^2}\,,
\label{eq: analytic scalar mediator cross section}
\\
    \frac{d\sigma^{\DIS}_{a \chi p}}{dxdy} &=&  \frac{
    g_{\chi a}^2}{16\pi m_a^4
    }\frac{(Q^2)^3 y\sum_{\kappa=q,\bar{q}}g_{\kappa a}^2 f^p_\kappa(x,Q^2)}{[(Q^2)^2-4m_N^2m_\chi^2 x^2 y^2] (1+Q^2/m_{a}^2)^2}\,,
    \label{eq: analytic pseudoscalar mediator cross section}
    \\ 
\frac{d\sigma^{\DIS}_{V \chi p}}{dxdy} &=& \frac{\sigma_{V \chi p}^{\NR}}{8\mu_{\chi p}^2 g_{pV}^2}\frac{(Q^2)^3 y \sum_{\kappa=q,\bar{q}}g_{\kappa V}^2 f^p_\kappa(x,Q^2)}{[(Q^2)^2-4m_N^2m_\chi^2 x^2 y^2] (1+Q^2/m_{V}^2)^2} \left(1+\frac{2}{y^2}-\frac{2}{y}-\frac{2m_p^2 x^2}{Q^2}\right)\,,
\label{eq: analytic vector mediator cross section}
\end{eqnarray}

The DIS cross section in the case of an axial mediator, $\sigma_{V'}^\DIS$, coincide with $\sigma_V^\DIS$ provided that the formal substitutions $g_{\chi V}\to g_{\chi V'}$, $g_{pV} \to g_{pV'}$, $\sigma_{V \chi p}^{\NR}\to \sigma_{V' \chi p}^{\NR}$, $g_{\kappa V} \to g_{\kappa V'}$, $\kappa = q,\bar{q}$, and $m_V \to m_{V'}$ are made. To align the notation with the elastic case, we perform the change of variables $(x,y)\to (T_\chi, \mu_s)$, which yields \begin{equation}
  \frac{d\sigma^{\DIS}_{Y \chi p}}{dT_\chi d \mu_s}=\frac{\sqrt{(T_p^2+2m_pT_p)(T_\chi^2+2m_\chi T_\chi)}}{\sqrt{(T_p^2+2m_pT_p)(T_\chi^2+2m_\chi T_\chi)}\,\mu_s-T_\chi (T_p+m_p)}\frac{1}{T_p+m_p}\frac{d\sigma^{\DIS}_{Y \chi p}}{dxdy}\,,
\end{equation}
where the DIS quantities can be written in terms of the new variables as 
\begin{eqnarray}
x&=&\frac{T_\chi m_\chi}{\sqrt{(T_p^2+2m_pT_p)(T_\chi^2+2m_\chi T_\chi)}\,\mu_s-T_\chi (T_p+m_p)}\,,\\
y&=&\frac{\sqrt{(T_p^2+2m_pT_p)(T_\chi^2+2m_\chi T_\chi)}}{{m_\chi(T_p+m_p)}}\,\mu_s-\frac{T_\chi}{m_\chi}.
\end{eqnarray}

\section{Blazar-boosted dark matter fluxes and their signals}
\label{sec:BBDM}

\subsection{Boosted-dark matter flux from blazars}
The same DM-hadron inelastic collisions that disintegrate protons in blazar jets, as well as the elastic ones at smaller momentum transfer, can boost DM particles in the vicinity of a blazar towards the Earth. For a single blazar, we estimate the BBDM flux  exiting its jet and going in our direction as
\begin{equation}\label{eq:spectrumDM}
    \frac{d\Phi_\chi}{dT_\chi} =\frac{\Sigma^\text{spike}_\text{DM}}{2\pi m_\chi d_L^2}\int_{0}^{2\pi}d\phi_s \int_{\gamma_p^\kin(T_\chi)}^{\gamma_p^\upper}d\gamma_p\int d \mu_s \frac{d\Gamma_p}{d\gamma_p d\Omega}\frac{d\sigma^{\EL+\DIS}_{Y \chi p}}{dT_\chi d \mu_s}\, ,
\end{equation}
where the differential cross section includes both the elastic and inelastic contributions, while $\mu_s$ is the cosine of the DM-proton scattering angle in the frame where DM is at rest. The proton spectrum is expressed in terms of $\mu$, which is related to $\mu_s$ and $\phi_s$ by a rotation of an angle $\theta_{\text{LOS}}$ \cite{Wang:2021jic}
\begin{equation}
\mu (\mu_s, \phi_s)=\mu_s \cos\theta_{\text{LOS}}+ \sin \phi_s \sin\theta_{\text{LOS}}\sqrt{1-\mu_s^2}\,.
    \end{equation}
    In the elastic case, $\mu_s$ is fixed by kinematics to be $\mu_s = \mu_s^{\EL}(T_p, T_\chi)$, as defined in Eq.~\eqref{eq:musEL}. In the inelastic case, the integration over $\mu_s$ is non-trivial and has to be performed in the range $\mu_s^{\EL}(T_p, T_\chi)\leq \mu_s \leq 1$, where the lower extremum represents the elastic scattering limit of the DIS ($x=1$). \footnote{A similar calculation of the BBDM flux from TXS 0506+056 for the vector mediator case, also including inelastic scatterings, has been carried out in \cite{Wang:2025ztb}. In \cite{Wang:2025ztb}, however, the integration over $\mu_s$ has been performed only for the cross section, while the proton spectrum was calculated at a fixed angle.
    This approach results in an approximation that, when $m_\chi \gtrsim \mathcal{O}(100)\,\text{MeV}$, is too conservative 
    for the BBDM flux at the relevant energies, which comes out considerably smaller than that obtained by us using Eq.~\eqref{eq:spectrumDM}. \label{footnote:2503.22105}}

The  DM flux shows mild dependence on the upper extreme of integration over $\gamma_p$, which we set to $\gamma_p^\upper = 1+10^8\,\text{GeV}/m_p$ for all the considered blazars.   
 The lower extremum is a function of the DM kinetic energy $T_\chi$, reading
\begin{equation}\label{eq:gamma_p_min}
     \gamma_p^\kin(T_\chi) = \frac{1}{2m_p}\left[\left(T_\chi-2m_p\right)+\sqrt{\left(T_\chi+2m_p\right)^2 + (m_p-m_\chi)^2\frac{2T_\chi}{m_\chi}}\,\right]+1.
 \end{equation}
 We also define $T_p^\kin(T_\chi) \equiv m_p[\gamma_p^\kin(T_\chi) -1]\geq T_\chi$, and note that the equality only holds if $m_\chi = m_p$.

The expression of the differential BBDM flux in Eq.~\eqref{eq:spectrumDM} is written in terms of the DM kinetic energy $T_\chi$ at the source. However,  the DM momentum at Earth $|\vec{p}_\chi^{\,(0)}|$ is related to that at the source $|\vec{p}_\chi|$ by $|\vec{p}_\chi^{\,(0)}|=|\vec{p}_\chi|/(1+z)$ -- we remind that $z$ is the redshift of the source -- and the DM kinetic energy at Earth $T_\chi^{\zero}$ can be expressed in terms of $T_\chi$ at production as
\begin{equation}\label{eq:kinredshift}
    T_\chi^{\zero}=m_\chi\left[\sqrt{\frac{T_\chi(T_\chi+2 m_\chi)}{(1+z)^2m_\chi^2}+1}-1\right]\,.
\end{equation}
Therefore, upon inverting Eq.~\eqref{eq:kinredshift}, the BBDM flux at Earth is given by
\begin{equation}
\frac{d\Phi_\chi}{dT_\chi^{\zero}}=\frac{dT_\chi}{dT_\chi^{\zero}}\frac{d\Phi_\chi}{dT_\chi}\Bigg|_{T_\chi = T_\chi(T_\chi^{\zero})}\,\,,
\end{equation}
where 
\begin{equation}
   \frac{dT_\chi}{dT_\chi^{\zero}} = 
(1+z) \frac{ T_\chi^{\zero} + m_\chi }{ \sqrt{ T_\chi^{\zero} (T_\chi^{\zero}+ 2m_\chi) } }\,.
\end{equation}

To ensure that DM particles are not boosted before the formation of the blazar, leading to an inconsistency, the DM travel time from the source to Earth ($t_\DM$) must not exceed the estimated age of the central BH. That is, $t_\DM \leq t_\BH$, where $t_\DM$ is given by

\begin{equation}\label{eq:redshift}
 t_\DM=\int_0^z \frac{dz'}{(1+z')H(z')}\sqrt{1+\left(\frac{1+z}{1+z'}\right)^2\frac{m_\chi^2}{T_\chi(T_\chi+2m_\chi)}}\,.
\end{equation}

\begin{figure}
    \centering
    \includegraphics[width=0.8\linewidth]{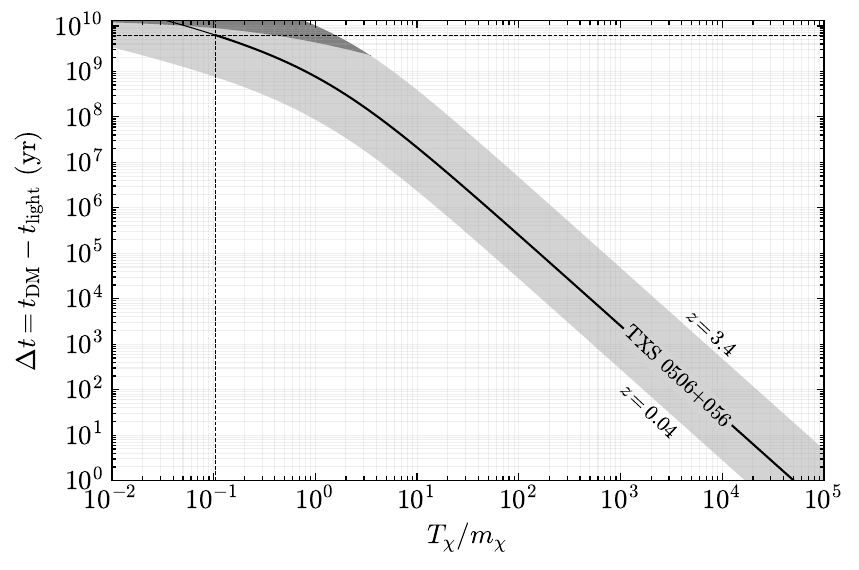}
    \caption{Time delay of the BBDM flux at Earth with respect to the photons emitted by TXS 0506+056 (black solid) and the blazar sample (gray area) as a function of $T_\chi/m_\chi$. Also shown in dashed is the cut $T_\chi/m_\chi\geq 0.104$ (vertical line) and the relative $\Delta t=t_\DM-t_\text{light}$ (horizontal line) for TXS 0506+056 obtained by imposing $t_\DM \leq t_\BH\simeq 10^{10}$ yr. In the darker gray region we find $t_\DM(z) \geq t_{\rm Universe} \simeq 1.379\times 10^{10}\,\text{yr}$ for the considered sample of blazars.}
    \label{fig:DMdelay}
\end{figure}

\begin{figure}
    \centering
        \includegraphics[width=.43\textwidth]{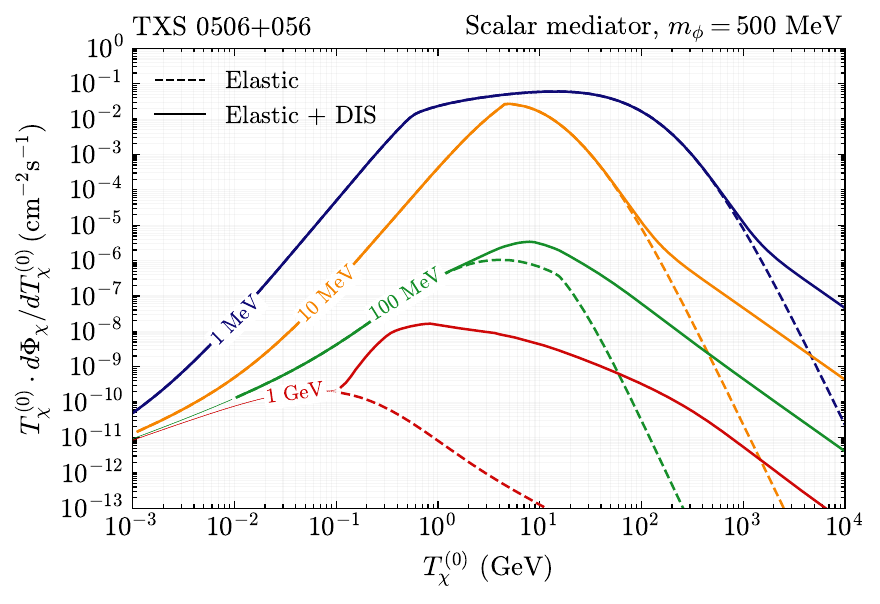}                \includegraphics[width=.43\textwidth]{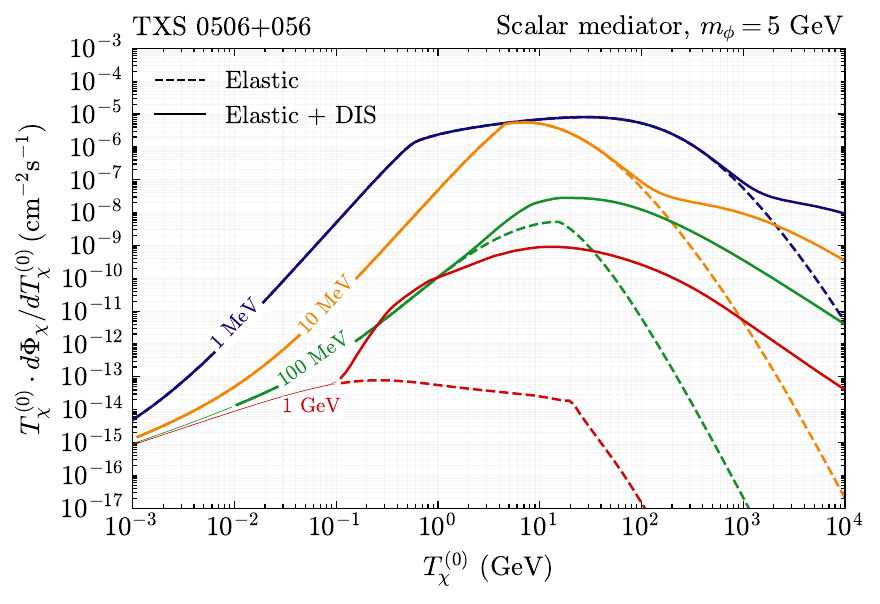}       \includegraphics[width=.43\textwidth]{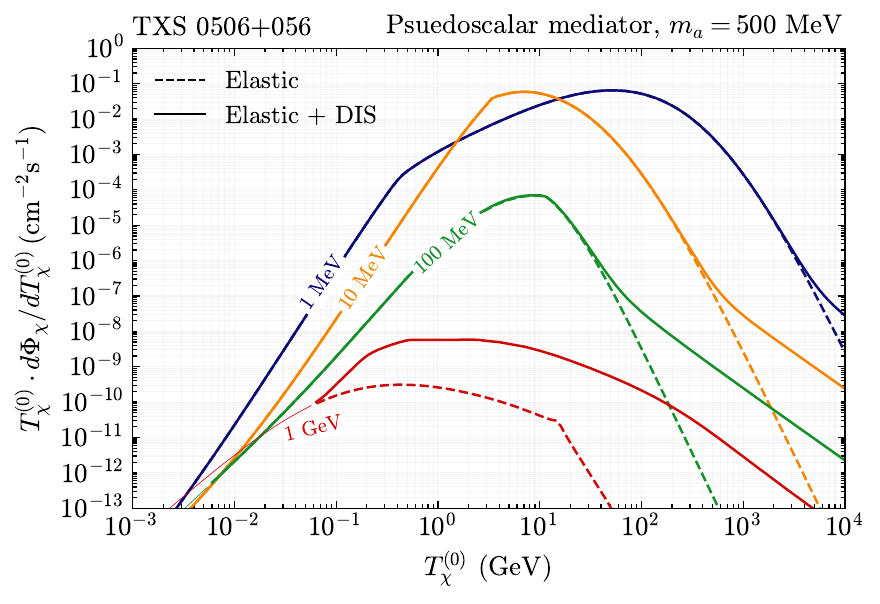}\includegraphics[width=.43\textwidth]{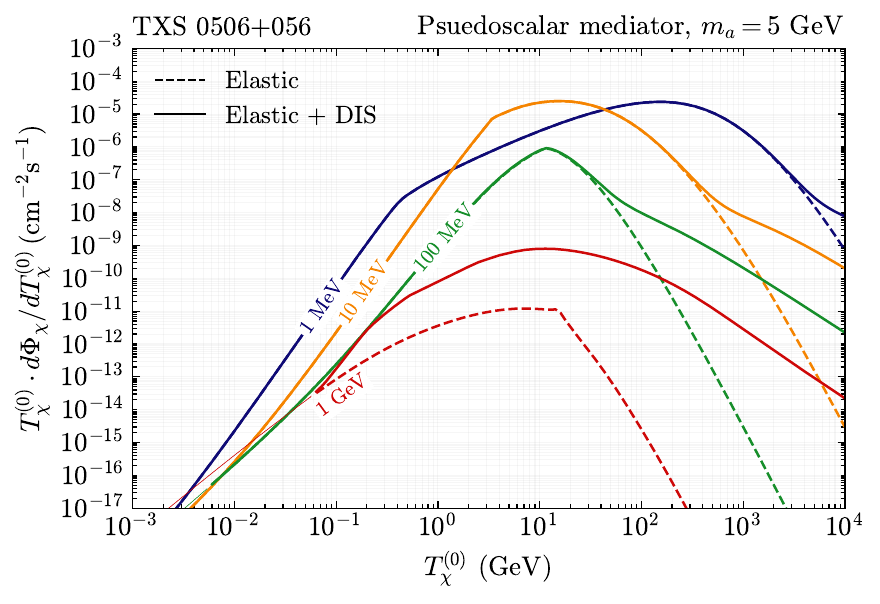}
        \includegraphics[width=.43\textwidth]{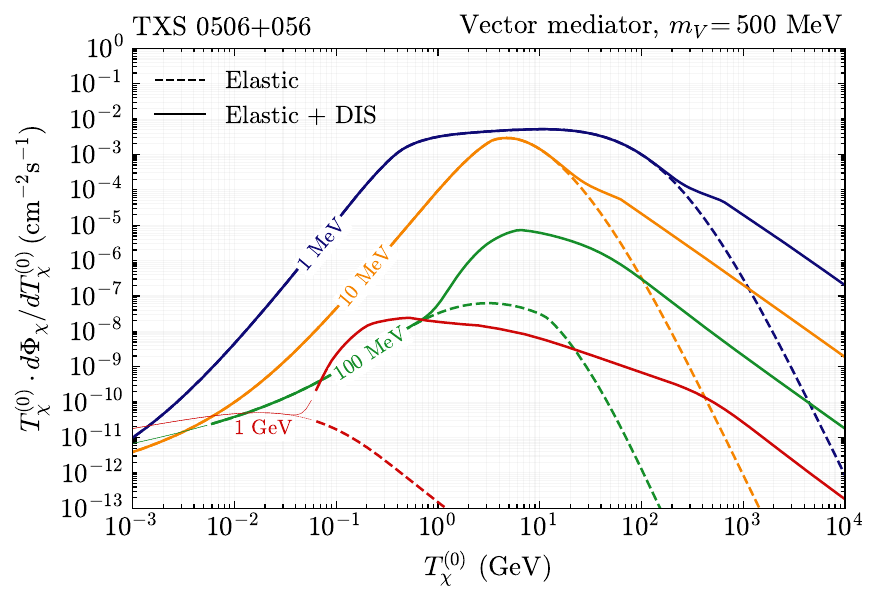} 
        \includegraphics[width=.43\textwidth]{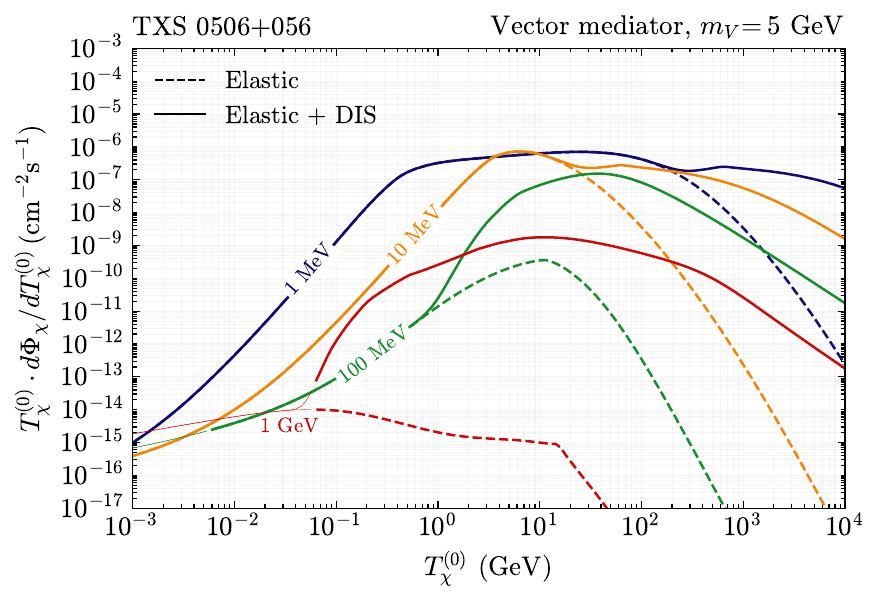} 
        \includegraphics[width=.43\textwidth]{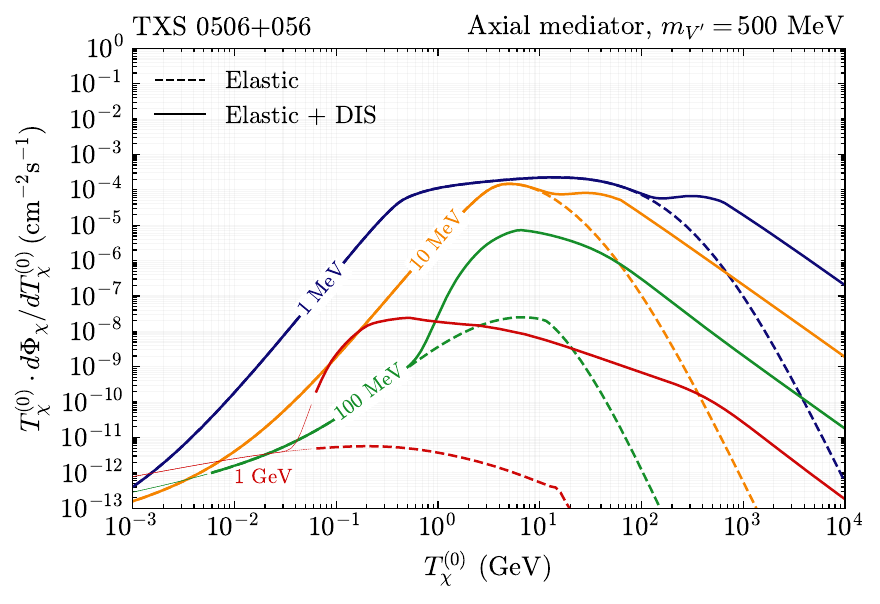} 
        \includegraphics[width=.43\textwidth]{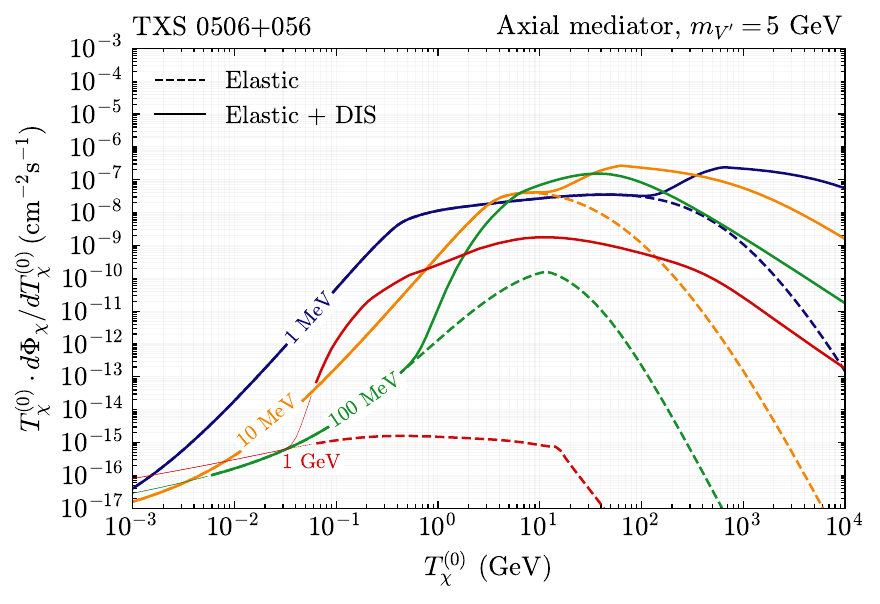} 
    \caption{BBDM flux at Earth surface from TXS 0506+056  for $m_\chi = 1\,\text{MeV}$ (blue), $10\,\text{MeV}$ (orange), $100\,\text{ MeV}$ (green), $1\,\text{GeV}$ (red). Dashed (solid) curves correspond to the elastic (elastic + DIS) contribution, for a scalar (first row), pseudoscalar (second row), vector (third row) and axial (last row) mediator with mass $m_Y=500\,\text{MeV}$ (left panels) and $5\,\text{GeV}$ (right panels). The plots are obtained for $g_{\chi Y} g_{u Y} =  g_{\chi Y} g_{dY} = 0.1$ and $\Sigma_\DM = 6.9\times10^{28}\,\text{GeV}\,\text{cm}^{-2}$. The thick (thin) lines are obtained after imposing the cut $T_\chi/m_\chi\geq 0.104$ ($< 0.104$). Note: the range of the vertical axes differ between left and right panels.}
    \label{fig:BBDM_Fluxes}
\end{figure}

\begin{figure}
    \centering
        \includegraphics[width=.43\textwidth]{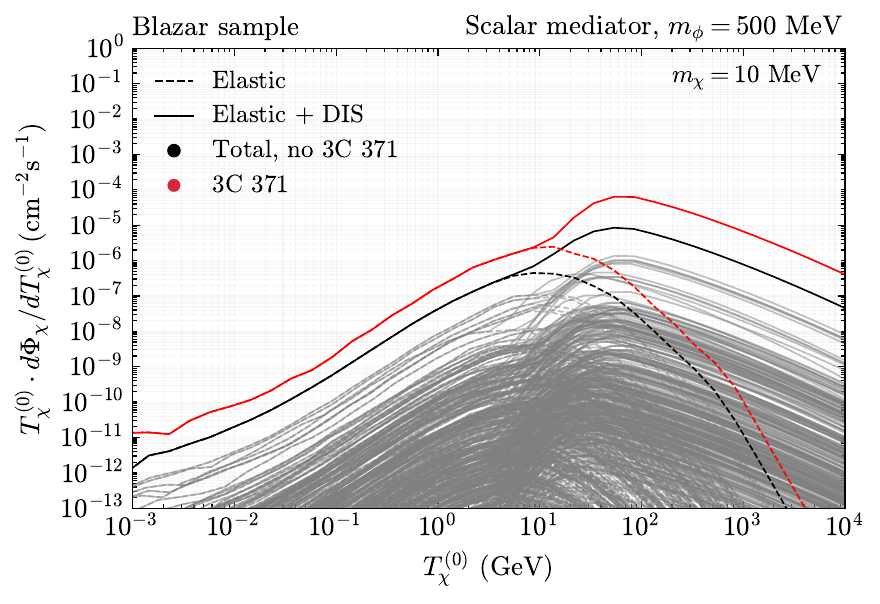}       \includegraphics[width=.43\textwidth]{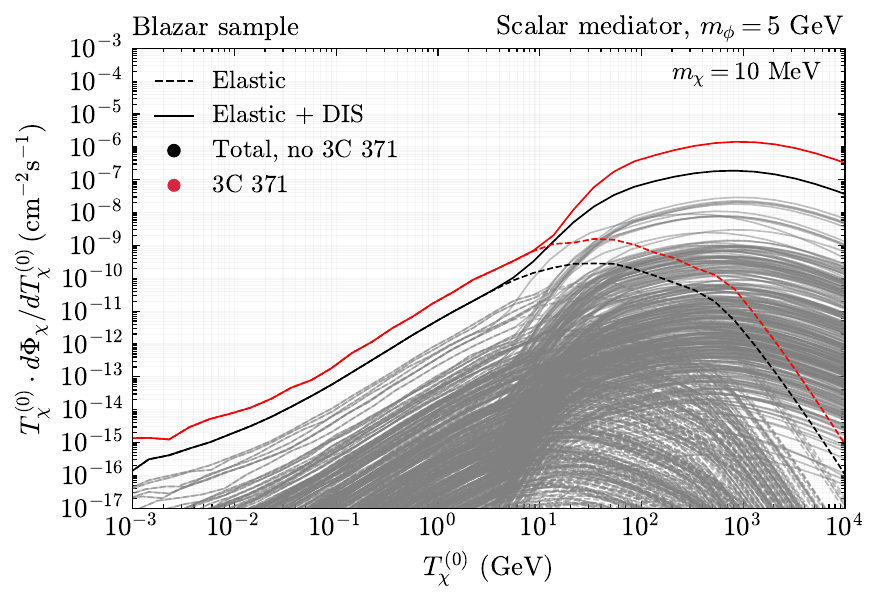}  
        \includegraphics[width=.43\textwidth]{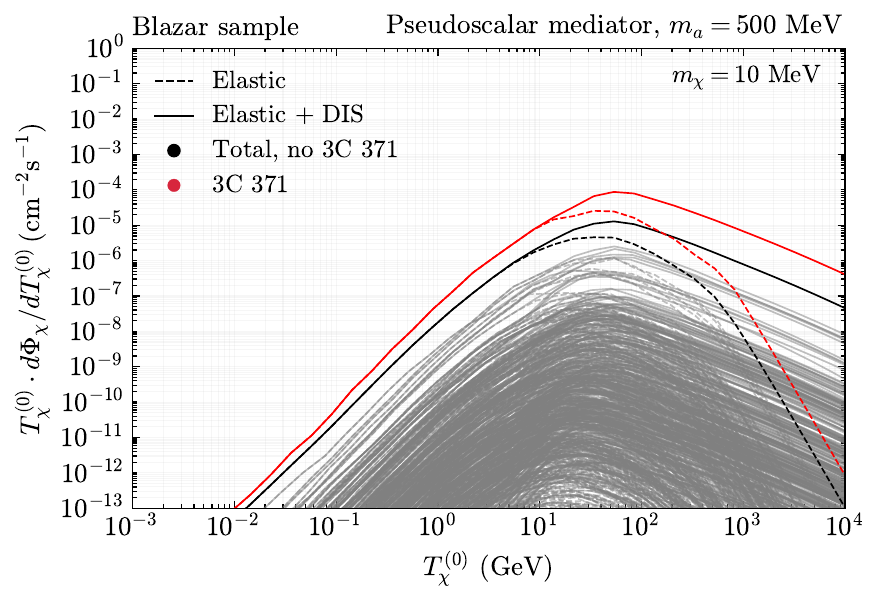}
        \includegraphics[width=.43\textwidth]{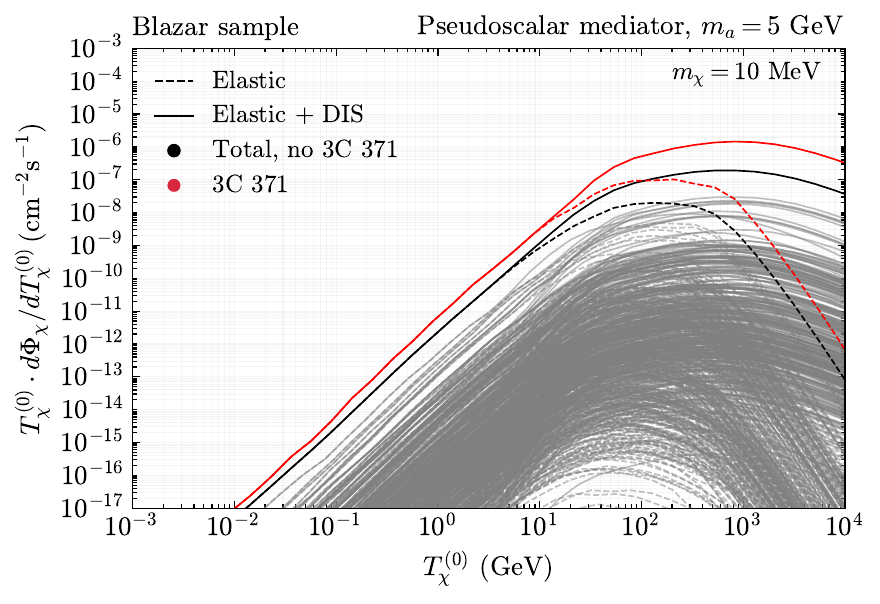}
\includegraphics[width=.43\textwidth]{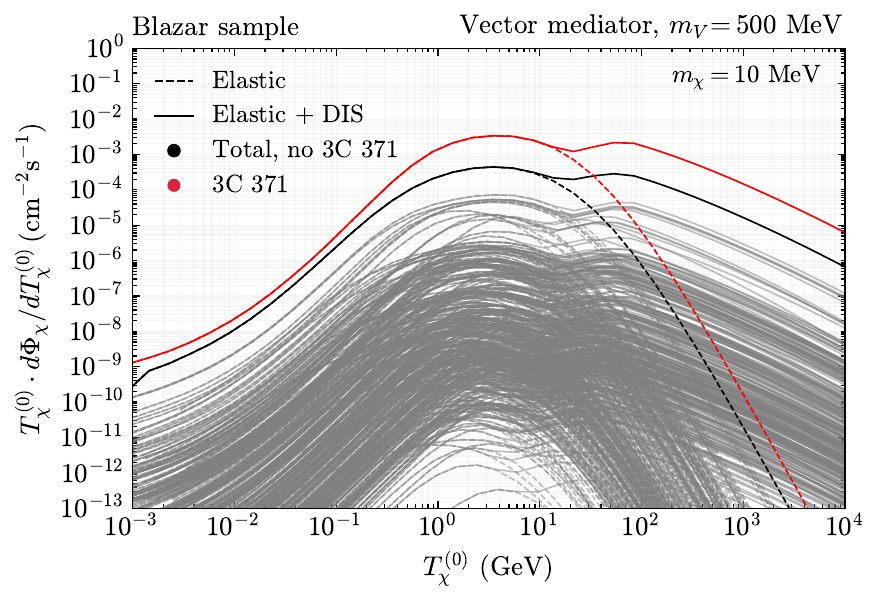}
        \includegraphics[width=.43\textwidth]{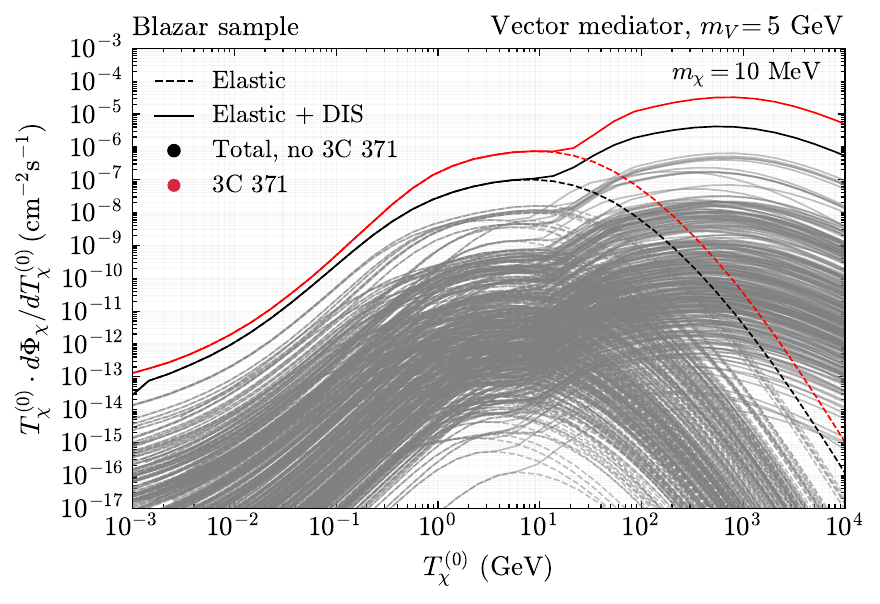} 
        \includegraphics[width=.43\textwidth]{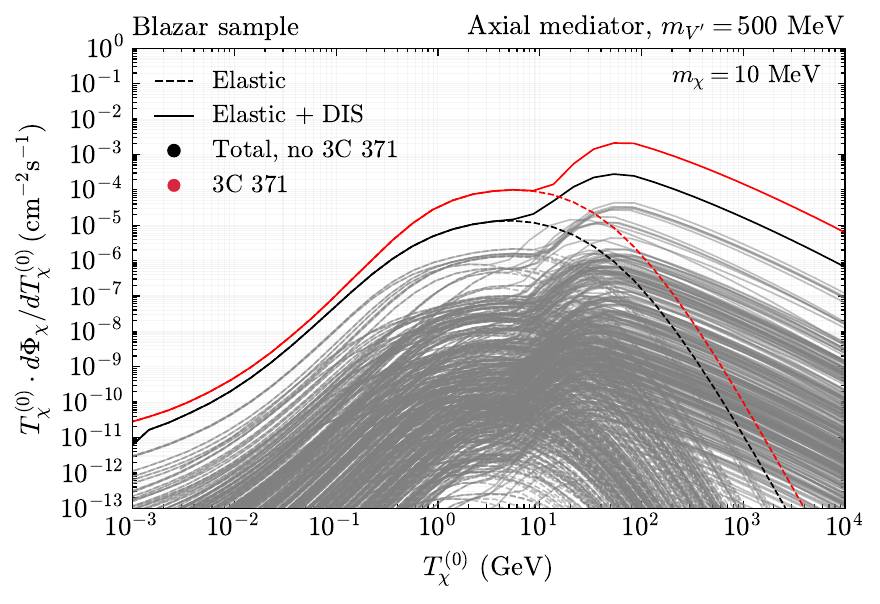} 
        \includegraphics[width=.43\textwidth]{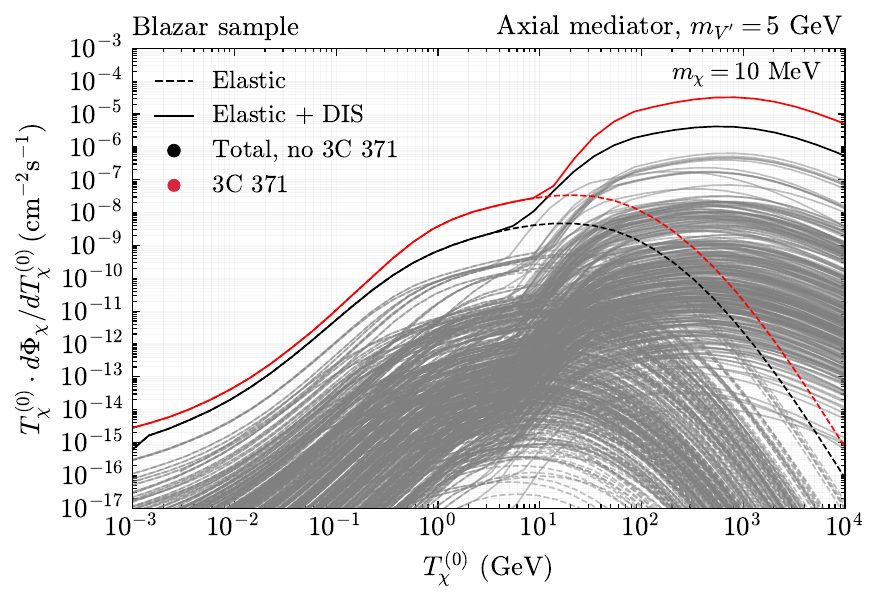} 
    \caption{BBDM flux from 324 blazars for scalar (top), pseudoscalar (second row), vector (third row), and axial (bottom) mediators. Gray lines show individual blazars; red highlights 3C 371, which dominates the flux; black is the total excluding 3C 371. Dashed and solid lines represent elastic and elastic + DIS contributions, respectively. We fix $m_Y=500\,\text{MeV}$ ($m_Y = 5\,\text{GeV}$) in the left (right) panels, with $m_\chi = 10\,\text{MeV}$, and the overall normalisation with $g_{\chi Y} g_{u Y} =  g_{\chi Y} g_{dY} = 0.1$ and $\Sigma^{\rm spike}_\DM$ as in BMCI. For each blazar we have imposed the cut on $T_\chi/m_\chi$ that results from the condition $t_{\DM} \leq t_{\rm Universe} \simeq 1.379\times 10^{10}\,\text{yr}$. Note: the range of the vertical axes differ between left and right panels.}
    \label{fig:BBDM_Fluxes_SAMPLE}
\end{figure}

We further note that, since DM travels slower than light, it arrives at Earth with a delay with respect to the photons emitted at the same time.
In particular, a BBDM-induced signal would lag behind any $\gamma$-ray and/or neutrino flaring activity. We show in Fig.~\ref{fig:DMdelay} the time delay $\Delta t=t_\DM-t_\text{light}$ of the BBDM flux at Earth with respect to the photons emitted by the same source for TXS 0506+056 and for the sample of blazars in~\cite{Rodrigues:2023vbv} as a function of $T_\chi/m_\chi$, where $t_{\rm light}(z) = \int_0^z dz'/[(1+z')H(z')]$.
For TXS 0506+056, we require that $t_\DM \leq t_\BH=10^{10}$ yr, which yields $T_\chi/m_\chi\geq 0.104$ using $H_0 = 70.2\, \text{km}\,\text{s}^{-1}\,\text{Mpc}^{-1}$, $\Omega_\Lambda = 0.685$, $\Omega_m = 0.315$ \cite{ParticleDataGroup:2024cfk}. 
In what follows, we artificially set the BBDM flux to zero when this condition is not satisfied.
In Fig.~\ref{fig:DMdelay}, the black solid curve corresponds to TXS 0506+056, the vertical dashed line to the cut on $T_\chi/m_\chi$, while the horizontal line corresponds to $t_\DM(z_{\mathsmaller{\rm TXS}} = 0.337)= t_\BH$. For all the sources in the sample, corresponding to the gray area in the plot, we impose instead a 
cut on $T_\chi/m_\chi$ by requiring that $t_\DM(z)\leq t_\text{Universe}=1.38\times10^{10}\,\text{yr}$. The cut on the sample applies to the darker gray region of the figure.  One can observe that $\Delta t$ can exceed by many orders of magnitude the typical duration of the flaring activity, which is $\mathcal{O}(\text{yr})$. Therefore, as our BBDM estimates for TXS 0506+056 are based on the prominent six-month flare observed in 2017, they should be considered optimistic, as they implicitly assume that similar flaring activity has persisted throughout the blazar's past emission history. Our results on the sample of blazars are instead based on steady activities, and are thus more solid in this regard.

We show in Fig.~\ref{fig:BBDM_Fluxes} the redshifted BBDM fluxes from TXS 0506+056 at Earth for our DM models, where we set the mediator masses to $m_Y= 500\,\text{MeV}$ and $5\,\text{GeV}$, the couplings to $g_{\chi Y} g_{uY} = g_{\chi Y} g_{d Y} = 0.1$, and choose $m_\chi=1\text{ MeV}, 10\text{ MeV}, 100\text{ MeV}, 1\text{ GeV}$ as benchmark values for the DM mass. We separate the contribution of the elastic scatterings only (dashed) from the total elastic + DIS (solid). The figure highlights the role of the inelastic contributions as the DM mass approaches the GeV scale.  Indeed, DIS requires $m_\chi T_\chi\gtrsim  \mathcal O(\text{GeV}^2)$, and the scattering imposes $T_p \geq T_p^{\kin}(T_\chi)\geq T_\chi$: the lower the DM mass, the higher the required proton energy to contribute to the DIS. Due to the power-law spectrum of protons in blazar jets, the contribution of inelastic scatterings to the flux becomes increasingly suppressed as the DM mass decreases. Also, the DIS has a milder impact for the pseudoscalar mediator case. This is because, for a fixed value of the quark couplings to the mediator, the pseudoscalar scenario enhances the associated coupling to the nucleons, leading to an increased contribution to the flux from elastic scatterings, see App.~\ref{app:elastic} for details.
Analogously, due to a similar conspiracy of couplings, the elastic contribution in the vector mediator case is larger than in the axial scenario, while the respective DIS contributions are equivalent. The scalar mediator case is intermediate between the vector/axial and pseudoscalar scenarios.
The mediator mass $m_Y$ also controls the relative importance of elastic and inelastic contributions to the flux. Increasing $m_Y$ tends to suppress elastic scattering more strongly than DIS, due to the lower momentum transfer involved.
Coming finally to low $T_\chi$'s, the different behaviour of the DM flux for the various mediators is induced by the different low-$Q^2$ behaviour of the associated elastic cross sections. This qualitatively explains why the lowest low-energy DM flux is suppressed 
in the pseudoscalar mediator case.

 We show in Fig.~\ref{fig:BBDM_Fluxes_SAMPLE} the diffuse BBDM flux originating from the sample of 324 blazars analysed in~\cite{Rodrigues:2023vbv}, choosing the benchmark parameters $m_\chi = 10\,\text{MeV}$, couplings $g_{\chi Y} g_{uY} = g_{\chi Y} g_{dY} = 0.1$, mediator masses $m_Y = 500\,\text{MeV}$ and $5\,\text{GeV}$.
 Since our BBDM flux at Earth would be dominated by the blazar 3C 371, the contribution of which is shown in red colour in the Fig.~\ref{fig:BBDM_Fluxes_SAMPLE}, we decide not to include it in the cumulative flux, as its uniqueness would demand a more specific and careful fit than the global one performed in~\cite{Rodrigues:2023vbv} (for the same reason, we had excluded the same blazar from the computation of the cumulative neutrino flux in~\cite{DeMarchi:2025xag}).
 By comparing these fluxes with the BBDM flux from TXS 0506+056 shown in Fig.~\ref{fig:BBDM_Fluxes} for $m_\chi=10\,\text{MeV}$, we find that inelastic contributions play a significantly larger role in the blazar sample. This difference arises due to the minimal proton boost factor $\gamma'_{\min_p} = 10^2$ adopted in~\cite{Rodrigues:2023vbv} for the whole sample. As a result, elastic contributions, being more suppressed at higher proton energies, are relatively weaker 
 compared to inelastic ones. The fact that $\gamma'_{\min_p}=10^2$ for the blazar sample also exacerbates the different dependence on $T_\chi$ 
 of the elastic contributions between the spin-0 (scalar and pseudoscalar mediators) and spin-1 cases (vector and axial mediator).

\subsection{The impact of the Earth attenuation}
\label{subsec:attenuation}

The flux of BBDM will be attenuated from the Earth surface to the location of the detector due to the scatterings with nuclei in the Earth crust. As a result, a detector can develop a blind spot to the incoming BBDM flux if the DM–nucleon interaction is sufficiently strong, causing significant depletion before reaching the detector.
Various methods with varying levels of complexity have been developed to estimate the energy loss of DM as it travels inside the Earth. The simplest approach consists of approximating the energy loss by its averaged value and evolve it using a corresponding differential equation under the assumption of forward scattering, see, e.g., \cite{Bringmann:2018cvk, Ema:2018bih, Ema:2020ulo, Das:2024ghw}. We give more details on this approach applied to our toy models in App.~\ref{app:EarthAtt}. 
More advanced methods employ Monte Carlo simulations to track the stochastic nature of the scatterings and energy loss in detail, often incorporating a three-dimensional description of particle propagation (see, e.g., \cite{Xia:2021vbz}). In this work, we decide to adopt a simplistic but more conservative treatment by requiring that DM gets lost after a single scattering in the Earth crust.  

To estimate how many scatterings DM undergo in the Earth crust, we assume that the interactions take place only with protons $p$ and neutrons $n$ at rest. This treatment is valid as long as $Q^2\gtrsim \mathcal{O}(0.04\,\text{GeV}^2)$, whereas below this value the DM particles can probe the nuclear structure of the target nuclei. Then, we consider only the elastic contribution to the DM-nucleon scattering, but evaluate all the nucleon form factors in $\sigma_{Y\chi N}^{\EL}$ at zero momentum transfer, noting that this approach is more conservative than evaluating the cross section including the full $Q^2$-dependence of the form factors (see, e.g., \cite{Xia:2021vbz}) and the DIS contribution. Also, in our analysis, we consider $m_N\equiv m_p \simeq  m_n$ and concentrate on quark couplings to the mediator such that $g_{qY}\equiv g_{uY} = g_{dY}$, and thus $g_{NY} \equiv g_{pY} = g_{nY}$. Under these assumptions, and following the derivation presented in App.~\ref{app:elastic}, the elastic scattering cross sections for pseudoscalar, vector, and axial toy models are the same for proton and neutron. The same goes for the scalar mediator case if we neglect small corrections arising from nuclear isospin symmetry breaking. We thus adopt the approximation $\sigma_{Y\chi N}^\EL \equiv \sigma^{\EL}_{Y\chi p} \simeq \sigma^{\EL}_{Y\chi n}$ throughout our subsequent analysis. Then, the expressions for $d\sigma^{\EL}_{Y\chi p}/dT_p$ 
that are needed for our treatment can be readily obtained from those in Eqs.~(\ref{eq:scalarelastic}-\ref{eq:axialelastic}) by replacing $T_\chi^\max(T_p)\to T_p^\max(T_\chi^x)$, evaluating the form factors at $Q^2 = 0$, performing the trivial integration over $\mu_s$. These can then be integrated from 0 to $T_p^{\max}(T_\chi^x)$ to get the total cross section.
Furthermore, we consider the average mass density of the Earth crust to be $\rho_{p+n} \simeq 2.7 \,\text{g}/\text{cm}^3$ \cite{Emken:2018run}, and fix the proton and neutron number densities by assuming an equal ratio, $n_N\equiv n_p \simeq n_n \simeq \rho_{p+n}/(2m_N)$.

\begin{figure}[t!]
        \centering
        \includegraphics[width = 0.48\textwidth]{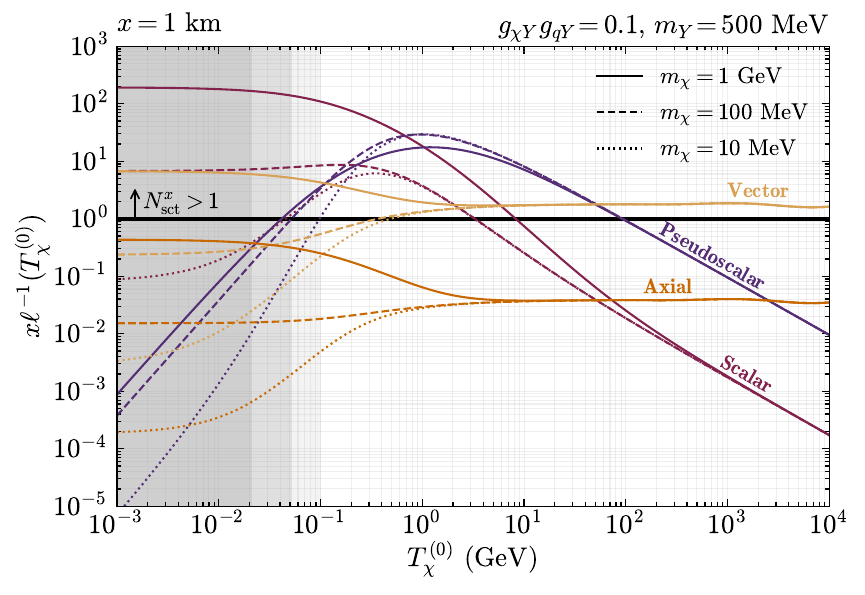}
                \includegraphics[width = 0.48\textwidth]{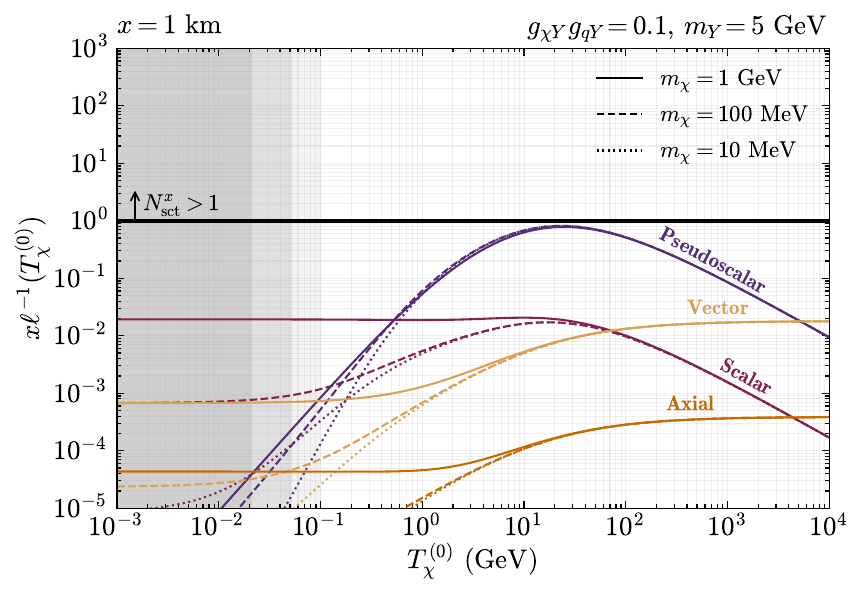}
        \caption{Distance $x$ travelled by DM in the Earth crust, in units of the mean path $\ell$ travelled by DM before it undergoes a scattering with nucleons, as a function of the DM kinetic energy at surface $T_\chi^{\zero}$. We consider the cases of scalar, pseudoscalar, vector and axial mediators and we choose as benchmark values $x=1\,\text{km}$, $m_Y=500\,\text{MeV}$ ($5\,\text{GeV}$), $m_\chi=10\,\text{MeV}$, 100\,\text{MeV}$, 1\,\text{GeV}$ and  $g_{\chi Y} g_{q Y} = 0.1$ for the left (right) panel.
        Attenuation effects become relevant when $ N_{\rm sct}\geq 1$ (thick black line). Inside the gray shaded areas, one for each value of DM mass, we have $Q^2<0.04\,\text{GeV}^2$ and the DM scattering with nuclei, which we have not included in our treatment, is dominant.  We stress that the differences among the various mediator cases, at large $T_\chi^{(0)}$ arise as an artifact of our conservative approximation of considering the elastic scattering cross sections with form factors evaluated at zero momentum transfer.}
            \label{fig:meanfreepath}
\end{figure}

After all these steps, we investigate the 
quantity $\ell^{-1}\equiv 2 n_N \sigma_{Y \chi N}^{\EL}$, 
which is the mean path travelled by the DM particle before undergoing a scattering. In particular, we show in Fig.~\ref{fig:meanfreepath} the product $x\ell^{-1}(T_\chi^{\zero})$ versus $T_\chi^\zero$, $x$ being the distance travelled by DM in the Earth crust, which we fix at $x = 1\,\text{km}$ for the figure. In the plot, the other parameters are fixed as $g_{\chi Y}g_{qY} = 0.1$, $m_Y = 500 \,\text{MeV}$ (left panel), $5 \, \text{GeV}$ (right panel), and $m_\chi = 1\,\text{GeV}$ (solid), $100\,\text{MeV}$ (dashed), $10\,\text{MeV}$ (dotted). The different colours correspond to the four considered toy models for the DM-nucleon interaction, as specified in the figure. It is immediate to rescale the plots for different values of $x$ and couplings $g_{\chi Y} g_{qY}$ since $x\ell^{-1} \propto x \,g_{\chi Y}^2 g_{qY}^2$. For instance, the same curves can be obtained for $x = 10^4\,\text{km}$ and $g_{\chi Y} g_{qY} = 10^{-3}$. The dependence on $m_Y$ is less straightforward, this being the reason why we show two benchmark cases. We also point out that, if $Q^2<0.04\,\text{GeV}^2$ (shaded gray areas in the figure), the dominant contribution to the Earth attenuation originates from DM scattering with nuclei, hence we are underestimating the attenuation in this kinematic region. When $x\ell^{-1}\geq 1$, the average number of scattering over a distance $x$, $N_{\rm sct}^x$, exceeds one. In the figure, the threshold $N_{\rm sct}^x=1$ is marked by a thick black horizontal line.
The Earth attenuation becomes relevant when $N^x_{\rm sct} \gtrsim \,\text{few}$, with the DM undergoing multiple scatterings and its energy dropping exponentially (see App.~\ref{app:EarthAtt}, and the left panel of Fig.~\ref{fig:Flux_Attenuation} there, where we prove this in a more refined approach). According to the right panel of Fig.~\ref{fig:meanfreepath}, this never happens for $m_Y = 5\,\text{GeV}$, and thus, for the parameters of the figure, the effects of the Earth attenuation can be neglected.  For the case $m_Y = 500\,\text{MeV}$ the situation is more subtle: the Earth attenuation remains negligible for the axial toy model; it becomes more important for the vector case for relatively large masses; it affects the pseudoscalar and scalar mediator cases for all the depicted DM mass benchmarks. We note, however, that attenuation in this case is significant only for relatively large couplings. A simple rescaling of $g_{\chi Y}g_{qY}$ to smaller values would largely reduce its impact.

Based on this qualitative description and, \textit{a posteriori}, on our final results, we note that Earth attenuation on the kilometer distance affect our conclusions only for relatively large couplings. In the scalar and vector scenarios, the region where attenuation starts playing a role is already excluded by other DM searches, while, in the axial case, it matters when couplings approach, if not exceed, the non-perturbative regime.  \textit{De facto}, attenuation is crucial to determine an upper limit on DM-nucleon interaction, in a region that is still allowed by current constraints, only in the pseudoscalar scenario for $m_\chi \gtrsim 300$ MeV. Therefore, for our subsequent analysis, we account for the Earth attenuation on the kilometer distance in the pseudoscalar scenario only, and we do it as follows. We set the attenuated BBDM flux to be equal to that at the Earth surface for $T_\chi^{\zero}$ as long as $N_{\rm sct}^{x_\det} < 1$, i.e.~$x_{\det}\ell^{-1}(T_\chi^{\zero})<1$, being $x_{\det}$ the depth of a given detector. For the energies such that $x_{\det}\ell^{-1}(T_\chi^{\zero})\geq 1$, a DM particle undergoes on average at least one scattering. When such condition is satisfied, we artificially set the BBDM flux to zero. This method is clearly conservative, as it does not account for the fact that the BBDM flux at these energies is not necessarily stopped, but rather redistributed to lower energies (see App.~\ref{app:EarthAtt}). However, a more accurate analysis would require substantial computational effort, and we prefer to proceed with our conservative approach to ensure the numerical evaluation remains tractable.
 We further highlight that the differences in $x_{\det}\ell^{-1}$ at large $T_\chi^{\zero}$, between the considered mediator cases, arise from the adopted conservative approximation, where the cross section is computed including only the elastic contribution and with form factors evaluated at $Q^2 = 0$. Accounting for the DIS contribution would make the cross sections to have the same behaviour at high energies (albeit with different normalisation), thereby reducing the apparent differences among the mediator cases, while suppressing the overall normalisation. We could in principle include the DIS, but since Earth attenuation effects would mostly affect the results in regions of the parameter space that are already excluded by other laboratory experiments, we do not pursue such refinements here.

\begin{figure}
    \centering
    \includegraphics[width=0.7\linewidth]{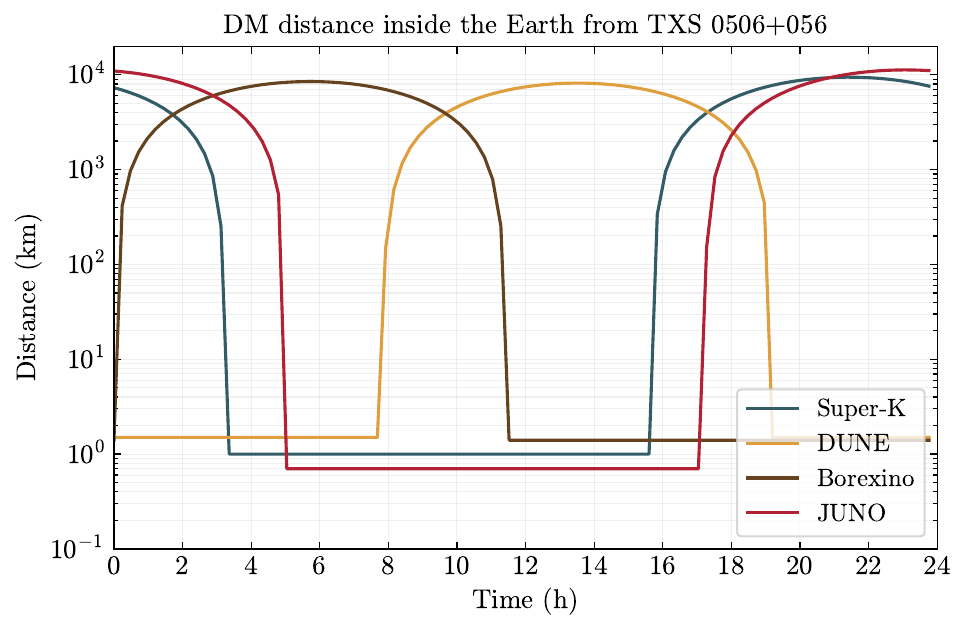}
    \caption{Daily variation of the DM path length from TXS 0506+056 to various underground detectors, measured from the Earth's surface. The constant $x$ intervals correspond to times where the source is above the horizon of the detector, hence $x\simeq x_{\text{det}}$, where $x_{\text{SK}}=1$ km, $x_{\text{DUNE}}=1.5$ km, $x_{\text{JUNO}}=0.7$ km and $x_{\text{Borexino}}=1.4$ km. We refrain to show KamLAND and Hyper-Kamiokande as their associated daily variation of the DM path inside the Earth differ from the one of Super-K only due to the different depths of the detectors, which we take as $x_{\text{KamLAND}}=1$ km and $x_{\text{HK}}=0.65$ km. }
    \label{fig:DMpath}
\end{figure}

Finally, we note that the effect of Earth attenuation depends on the path length travelled by DM particles through the Earth from the surface to the detector, which varies over time as the position of each source in the sky changes with the Earth's rotation. When the blazar goes below the horizon, the distance DM must travel through the Earth increases significantly, making attenuation rapidly more important. To remain conservative, we set the BBDM flux to zero whenever the originating blazar goes below the horizon, for each mediator case,. In practice, this implies multiplying the BBDM flux by an overall factor $f_{\det}$ that accounts for the fraction of the day during which the source remains above the horizon, as seen from the detector's location. We depict in Fig~\ref{fig:DMpath} the depth in the direction of TXS 056+056 for the neutrino detectors Super-Kamiokande (SK, or Super-K), Borexino, JUNO and DUNE. According to the figure, for TXS 0506+056 the fraction of the day for which it stays above the horizon for each detector is $f_{\det} \simeq 1/2$. We compute analogously $f_{\det}$ for each blazar of the sample in \cite{Rodrigues:2023vbv}.

We summarise our methodology for incorporating the effects of the Earth attenuation by writing the BBDM flux from a single blazar at a given detector as:
\begin{equation}\label{eq:attenuationconservative}
    \frac{d\Phi^{\det}_\chi}{dT_\chi^{\zero}} = \frac{d\Phi_\chi}{dT_\chi^{\zero}}f_{\det}\Theta[1-x_{\det} \ell^{-1}(T_\chi^{\zero})],
\end{equation}
where the Heaviside $\Theta$-function ensures that $x_{\det} \ell^{-1}(T_\chi^{(0)})\leq 1$.

\subsection{The recoil spectrum and number of events at neutrino detectors} 

Once a DM particle reaches a detector, it may transfer part of its kinetic energy to a target nucleus. In our study, we focus on the signal produced by elastic DM scatterings at neutrino detectors. Given their typical energy sensitivity, DM dominantly scatters off individual nucleons of the detector material, thus we consider DM scatterings on $N=p,n$, using $\sigma^{\EL}_{Y \chi N} \equiv \sigma_{Y \chi p}^{\EL} \simeq \sigma^{\EL}_{Y \chi n}$ and $m_N\equiv m_p \simeq m_n$. We denote as
$T_N$ the nucleon kinetic energy upon DM scattering. We compute the differential recoil rate $dR^{\det}_N/dT_N$ from a single blazar at a given detector as
\begin{equation}\label{eq:DM-N rate}
    \frac{dR^{\det}_N}{dT_N} \simeq N^{\det}_{\text{target}}\int_{T_\chi^{\min}(T_N)}^{+\infty}dT_\chi^{\zero} \frac{d\sigma^{\EL}_{Y \chi N}}{dT_N} \frac{d\Phi_\chi^{\det}}{dT_\chi^{\zero}}\, \,, 
\end{equation}
where $N^{\det}_\text{target}$ is the total number of target nuclei in the detector, $T_\chi^{\min}$ the minimal DM kinetic energy needed to produce a recoil energy $T_N$ and the DM flux is evaluated according to Eq.~\eqref{eq:attenuationconservative}.

We can now  extract the sensitivities on DM-nucleon interactions by computing the expected number of BBDM events at the selected detectors. This is given by 
\begin{equation}\label{eq:events}
    N^{\det}_{\text{BBDM}}=t_\text{exp}^{\det}N^{\det}_{\text{target}}\int_{T_{\min}^{\det}}^{T_{\max}^{\det}}
dT_N\,\epsilon^{\det}(T_N)\frac{dR^{\det}_N}{dT_N} < N^{\det}_\text{events}\,,
\end{equation}
where $t^{\det}_\text{exp}$ denotes the exposure time, $\left[ T_{\min}^{\det},\, T_{\max}^{\det} \right]$
 the energy range sensitivity, $\epsilon^{\det}$ the efficiency and $N^{\det}_\text{events}$ the experimental upper limit on the number of events.
We calculate limits using Super-K, KamLAND and Borexino data and we compute the projected sensitivities at Hyper-Kamiokande (HK, or Hyper-K), JUNO and DUNE. We discuss here how we extract each specific limit and future sensitivity, and postpone the comparison with current constraints and the neutrino predictions of \cite{DeMarchi:2024riu, DeMarchi:2025xag} to the next section.

We extract current bounds on DM-proton interaction from Super-K based on the search~\cite{Super-Kamiokande:2022ncz} performed by the collaboration investigating interactions with protons. The efficiency factor $\epsilon(|\vec{p}_p|)$ is given in~\cite{Super-Kamiokande:2022ncz} in the proton momentum range $1.2\leq |\vec{p}_p|/\text{GeV} \leq 2.3$, with the corresponding kinetic energy range $0.58\lesssim T_p/\text{GeV}\lesssim 1.55$.
The total exposure time in Super-K is $t^\text{SK}_\text{exp} = 6050.3\,\text{days}$, the number of proton targets (including those in oxygen, as these should also be relevant when $Q^2 =  2m_NT_N > \text{GeV}^2$) in its 22.5 kton fiducial mass of water is $N_{\rm target}^\text{SK} = 7.5\times 10^{33}$ and by selecting only the 60 events in~\cite{Super-Kamiokande:2022ncz} where the signal comes from above the horizon we extract the limit on the BBDM events $N_{\text{events}}^{\rm SK}(0.58 < T_p /\text{GeV}< 1.55) = 18$.

Hyper-K will be a 187-kton fiducial volume detector, $N_{p}^{\rm HK} = 6.2\times 10^{34}$, and is expected to start taking data in 2027~\cite{Hyper-Kamiokande:2018ofw}. We obtain its projected limits by setting the exposure time $t_\text{exp}^\text{HK} = 10\,\text{yr}$ and  assuming the same background event rate, energy range  and efficiency as Super-K. This procedure gives us the upper limit $N_{\text{events}}^{\rm HK}(0.58 < T_p /\text{GeV}< 1.55) = 37$.

The DUNE detector will consist of four modules, with a combined fiducial volume of 40 ktons~\cite{DUNE:2018tke}. As DUNE relies on the scintillation light and not on the Cherenkov one, its energy threshold will be lower than Super-K. We take it as $T_p > 40\,\text{MeV}$ as given by~\cite{DUNE:2024qgl}. We set the upper energy threshold as 10 GeV, although our results are insensitive to the precise choice as long as it is much larger than $\sim 50\,\text{MeV}$. In this energy range, DM dominantly interacts with individual nucleons and so we take the number of the target particles as $N_{\rm target}^{\rm DUNE} = 2.4\times 10^{34}$ where we summed over protons and neutrons. We set our bounds by choosing an exposure time of $t_\text{exp}^\text{DUNE} = 5\,\text{yr}$, $\epsilon^{\rm DUNE}=1$ and by adopting the atmospheric neutrino background modelling of \cite{Berger:2019ttc}, which gives the limit on the signal $N_{\text{events}}^{\rm DUNE}(0.04 < T_p /\text{GeV}< 10) = 21$.  The expected angular resolution of DUNE \cite{Berger:2019ttc} could in principle allow to better estimate the background from specific directions, thus improving considerably the sensitivity to boosted DM from single blazars. Here, we remain conservative by considering the background across the full sky.

KamLAND observed one event in the energy bin  $\left[13.5, 20 \right]\,\text{MeV}_{\rm ee}$ \cite{KamLAND:2011fld}, where $\text{MeV}_{\rm ee}$ stands for MeV electron equivalent. From the detector information and following the approach performed in \cite{Cappiello:2019qsw} (see Eq.~(11) therein), we infer the equivalent proton recoil energy $22.5\lesssim T_p/\text{MeV} \lesssim 30.7$ and impose our limit on the BBDM signal in this energy bin as $N_{\text{events}}^{\text{KamLAND}}(22.5 \lesssim T_p /\text{MeV} \lesssim 30.7) = 3$. The exposure time is $t_\text{exp}^\text{KamLAND} = 123\,\text{days}$ and the number of proton targets in the detector $N_{\rm target}^{\text{KamLAND}}=3.4 \times 10^{32}$, including those in carbon nuclei.

JUNO is a reactor neutrino experiment with a 20 kton liquid scintillator detector volume~\cite{JUNO:2021vlw} expected to be operational within 2025. The JUNO projection~\cite{Chauhan:2021fzu} estimates the SM background from elastic scatterings on hydrogen and quasi-elastic scatterings on carbon, in the visible scintillation energy ($E_\text{vis}$) range $15\lesssim E_\text{vis}/\text{MeV} \lesssim 100$, corresponding to a kinetic proton recoil energy within $21.2\lesssim T_p/\text{MeV} \lesssim 112$, where the relation between $E_\text{vis}$ and $T_p$ is given in \cite{Chauhan:2021fzu}. We consider for our signal elastic scatterings on all the protons in the detector, $N_{\rm target}^{\text{JUNO}}=6.8 \times 10^{33}$.  Setting an exposure time $t_\text{exp}^\text{JUNO} = 10\,\text{yr}$, $\epsilon^{\text{JUNO}}=1$ and adopting the expected background rate from atmospheric neutrinos as~\cite{Chauhan:2021fzu} yields the limit on the signal $N_{\text{events}}^\text{JUNO}(21.2 \lesssim T_p /\text{MeV} \lesssim 112) = 79$.

We finally compute limits from Borexino~\cite{Borexino:2013bot}. We use the number of proton targets $N_{\rm target}^{\text{Borexino}}=3.2 \times 10^{31}$, with an exposure time $t_\text{exp}^\text{Borexino} = 446.1\text{ days}$ and then impose our signal $N_{\text{events}}^{\text{Borexino}}<3$ in the energy range $21.2\lesssim T_p/\text{MeV} \lesssim 24.5$.
This limit follows from the absence of events in the interval $[12.5, 15]\,\text{MeV}_\text{ee}$, using the same procedure as for KamLAND to convert $\text{MeV}_\text{ee}$ into $T_p$.\footnote{In \cite{Wang:2025ztb}, limits on BBDM from TXS 0506+056 for the vector mediator case have been computed for Borexino and LUX-ZEPELIN experiments. The analysis performed for Borexino, however, differs from our in terms of energy range and number of events.}

We summarise the relevant parameters adopted in our analysis for each detector in Table \ref{tab:Experimentparameters}. In addition, to facilitate a direct comparison of the recoil spectrum with the energy range within which each specific detectors operate, we show in Fig.~\ref{fig:Recoil_Spectrum} the proton recoil spectrum for both the individual source TXS 0506+056 and the sample of blazars (excluding 3C 371). Note that the plots in the figure are obtained neglecting attenuation, as this would depend on the specific detector's depth and position.

In the computation of the recoil rate we have neglected the effects of the DIS. It has been shown in~\cite{Cappiello:2024acu}  for IceCube with a scalar or vector mediator, and in~\cite{Diurba:2025lky} for DUNE with a vector or axial mediator, that DIS at detection leads to a (relatively mild for DUNE) increase in the detector sensitivity to galactic cosmic ray-upscattered DM. The DIS contribution at detection could also increase the sensitivity to BBDM fluxes in the scenarios considered by us, especially at $m_\chi \sim \,\text{GeV}$, but a proper evaluation would deserve a dedicated study, which we postpone to future work.
Still considering IceCube, leveraging its low-energy capabilities could also provide sensitivities to BBDM fluxes, as it does for cosmic ray-upscattered DM~\cite{Cavicchi:2025kxi}.

\setlength{\arrayrulewidth}{.4pt}

\setlength{\tabcolsep}{10pt}
\newcolumntype{C}{@{}>{\centering\arraybackslash}X}
\begin{table}
\centering
\begin{tabularx}{\linewidth}{@{}C|CCCCCC@{}}
    \hline
    \rule{0pt}{2.5ex}
Experiment & Mass &$N^{\det}_{\text{target}} (N)$ & $t_\text{exp}^{\det}$& $\left[ T_{\min}^{\det},\, T_{\max}^{\det} \right]$& $N^{\det}_{\text{BBDM}}$ & $\epsilon^{\det}$\\ 
 & (kton) & &(years) & (GeV) &  & \\ 
\hline
    \rule{0pt}{2.5ex}
SK & 22.5 & $7.5 \times 10^{33} (p)$  & 16.56 & $\left[ 0.58, 1.55 \right]$ & 18 & \cite{Super-Kamiokande:2022ncz}\\
~HK  & 187 & $6.2 \times 10^{34} (p)$  & 10 & $\left[ 0.58, 1.55 \right]$ & 37 & \cite{Super-Kamiokande:2022ncz}\\
~DUNE & 40 & $2.4 \hspace{-0.015em}\times \hspace{-0.015em} 10^{34} (p,n)$  & 5 & $\left[0.04 , 10 \right]$ & 21 &   ~1\\
~JUNO & 20 & $6.8 \times 10^{33} (p)$  & 10 & $\left[0.0212, 0.1120 \right]$ & 79 & ~1\\
~KamLAND & 1 & $3.4 \times 10^{32} (p)$  & 0.337 & $\left[0.0225, 0.0307 \right]$ & 3 & ~1\\
~Borexino & 0.278 & $3.2 \times 10^{31} (p)$  & 1.22 & $\left[ 0.0212, 0.0245 \right]$ & 3 & ~1\\
\hline
\end{tabularx}
\caption{Specification of the detector parameters that we adopted in our analysis, see Eq.~\eqref{eq:events}. We consider the free nucleons $N=p,n$ as targets as specified for each experiment.}\label{tab:Experimentparameters}
\end{table}

\begin{figure}
    \centering
        \includegraphics[width=.43\textwidth]{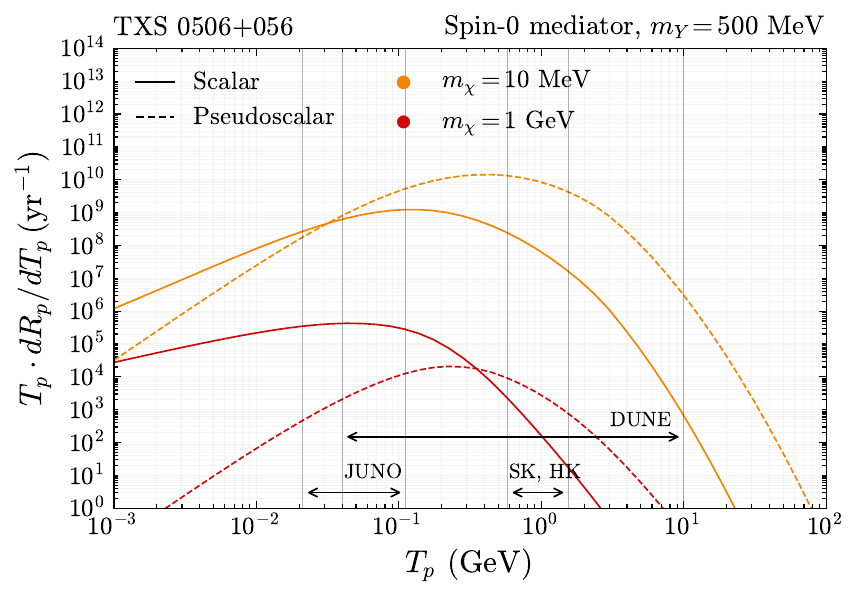}      \includegraphics[width=.43\textwidth]{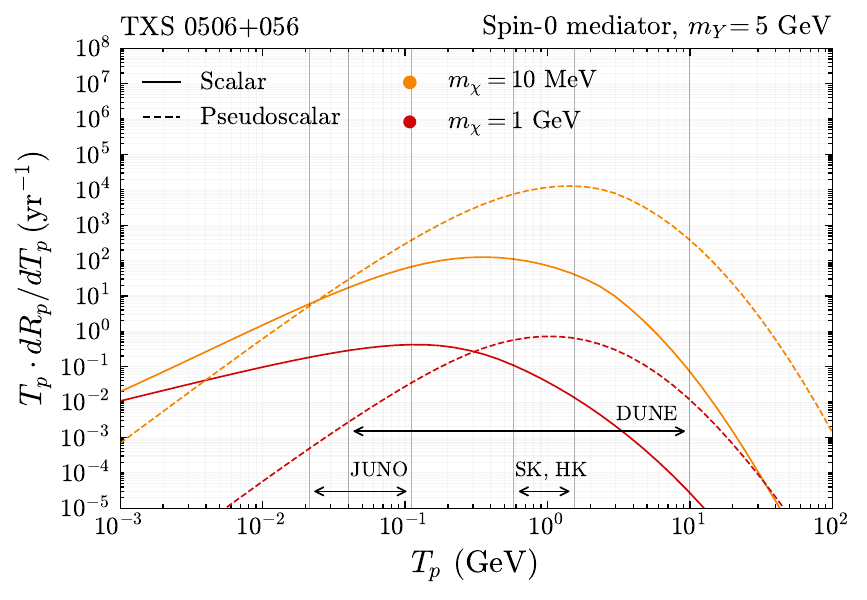}      
        \includegraphics[width=.43\textwidth]{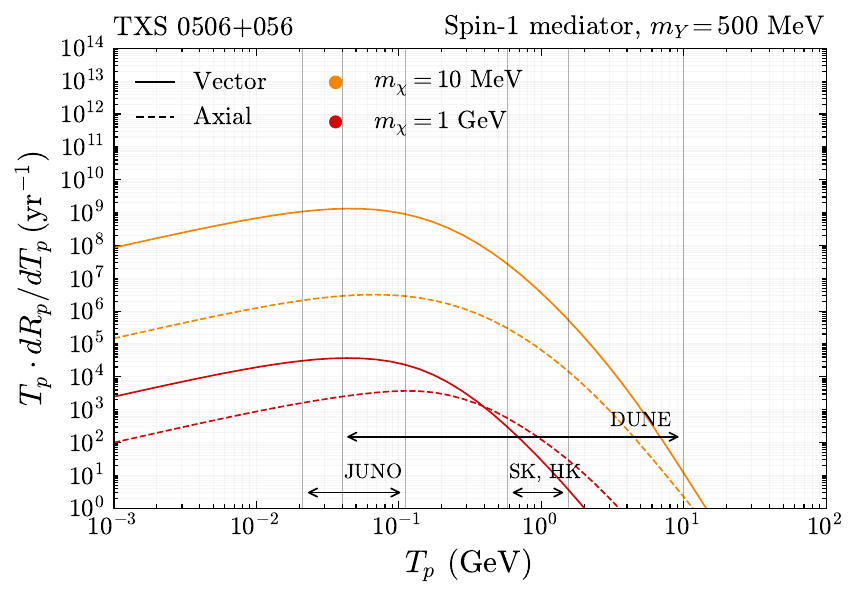} 
       \includegraphics[width=.43\textwidth]{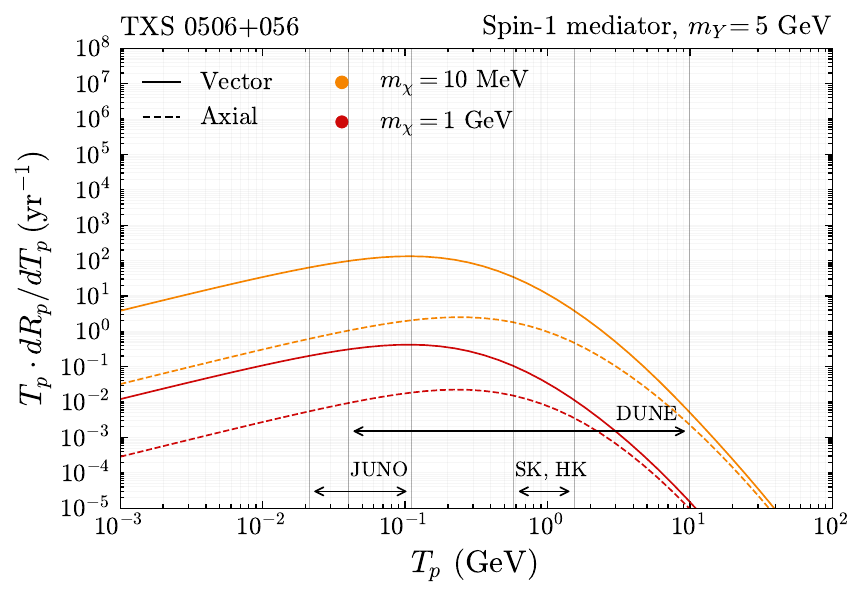}
       \includegraphics[width=.43\textwidth]{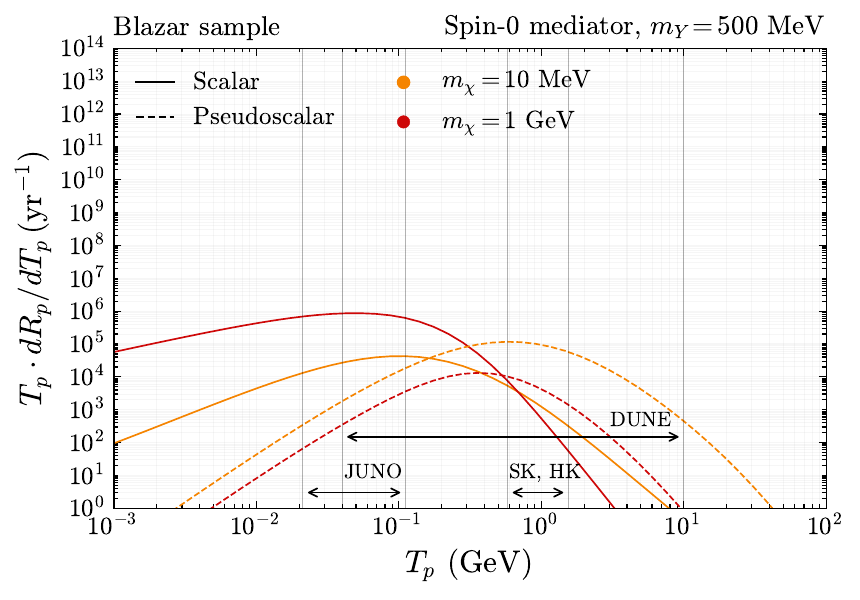}             
        \includegraphics[width=.43\textwidth]{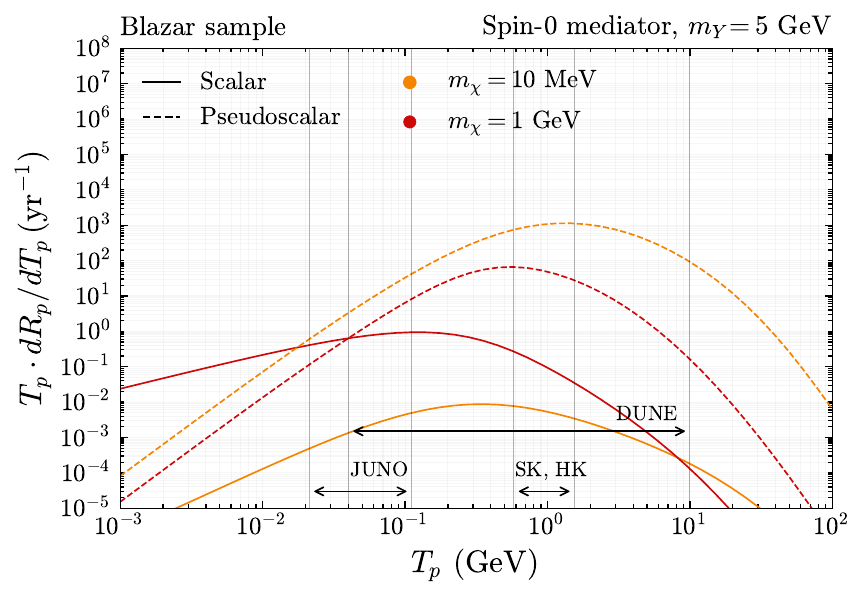}      
        \includegraphics[width=.43\textwidth]{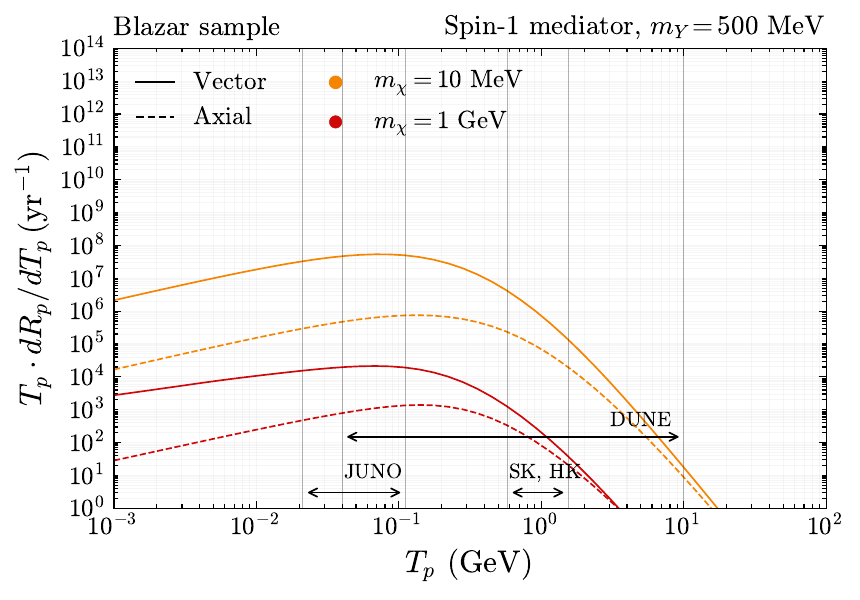} 
       \includegraphics[width=.43\textwidth]{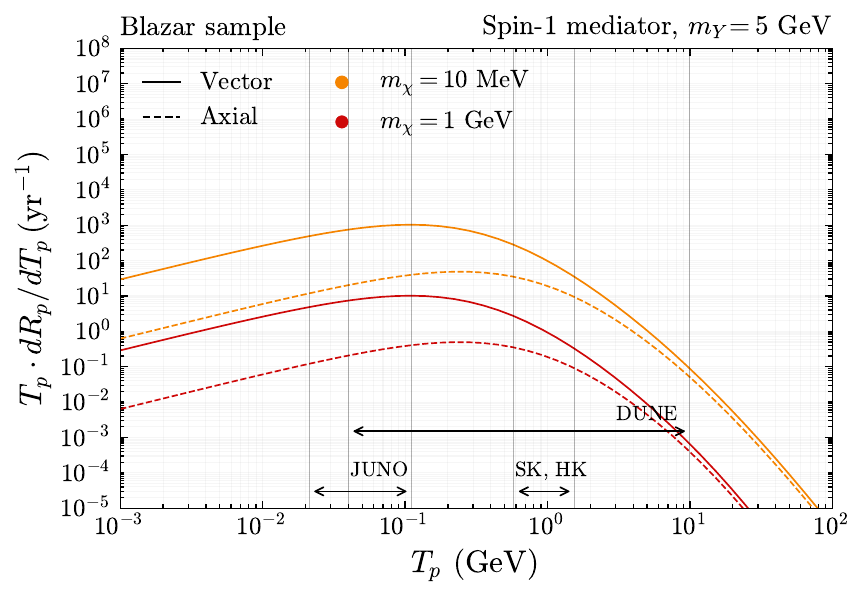}
        \caption{
        The proton recoil spectrum from BBDM from TXS 0506+056 (first and second rows) and the blazar sample excluding 3C 371 (third and fourth rows), for $m_\chi = 10\,\text{MeV}$ (orange), $1\,\text{GeV}$ (red), before attenuation. The first and third rows are for spin-0 mediators, while the second and fourth rows for spin-1 mediators, with $m_Y=500\,\text{MeV}$ (left panels) and $5\,\text{GeV}$ (right panels), $g_{\chi Y} g_{u Y} =  g_{\chi Y} g_{dY} = 0.1$, $\Sigma_\DM$ as in BMC I and $N_{\rm target} = 7.5\times 10^{33}$. The arrows mark the energy range within which Super-K, Hyper-K, DUNE and JUNO operate (not shown those for KamLAND and Borexino, similar to that of JUNO).
        Note: the range of the vertical axes differ between left and right panels.
        }
                \label{fig:Recoil_Spectrum}

\end{figure}

\section{Searches for sub-GeV dark matter test blazar-dark matter signals}\label{sec:limits}

In this section, we project the limits and sensitivities to BBDM computed earlier 
in the $\sigma^{\NR}_{Y \chi p} - m_\chi$ plane for the scalar, vector and axial mediator cases ($g_{\chi Y} g_{q Y} - m_\chi$ plane for the pseudoscalar mediator case), and compare them against current constraints from various DM searches, as well as to the neutrino prediction via the same DM-proton scatterings in blazar jets estimated in \cite{DeMarchi:2024riu, DeMarchi:2025xag}.

\begin{figure}
        \centering
                \includegraphics[width = 0.4\textwidth]{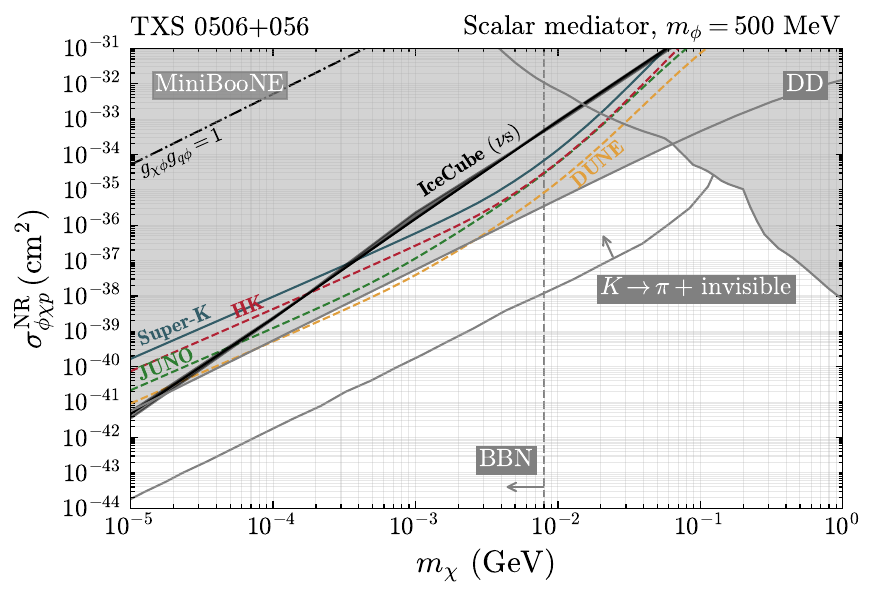}
                 \includegraphics[width = 0.4\textwidth]{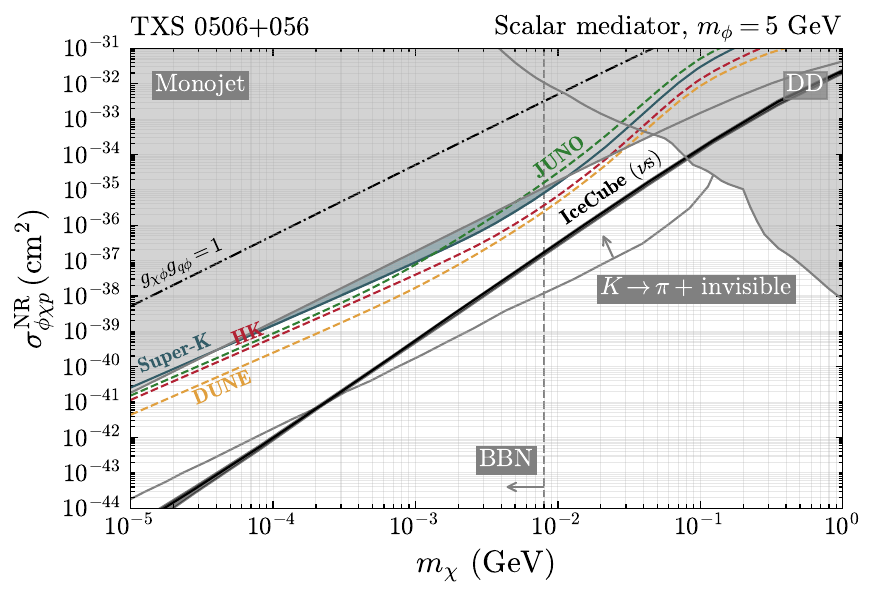}
                  \includegraphics[width = 0.4\textwidth]{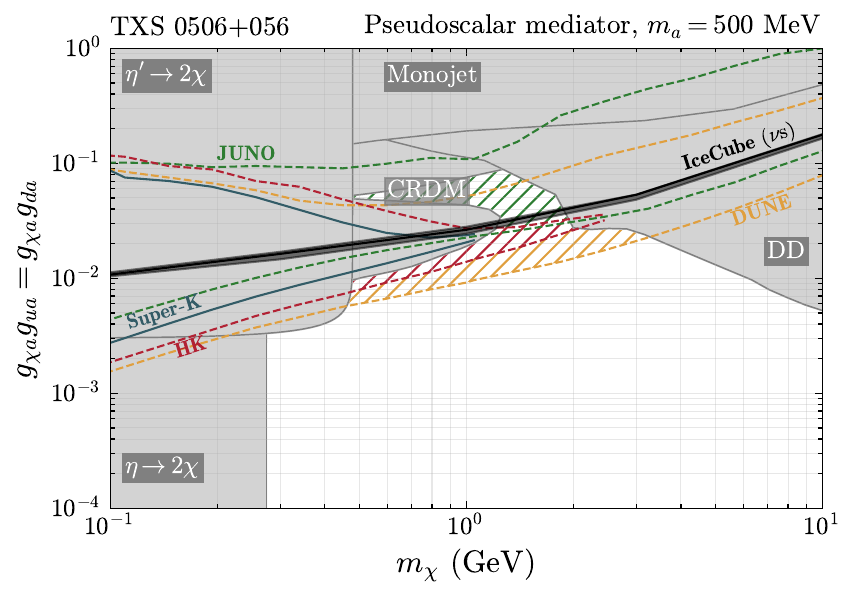}
                \includegraphics[width = 0.4\textwidth]{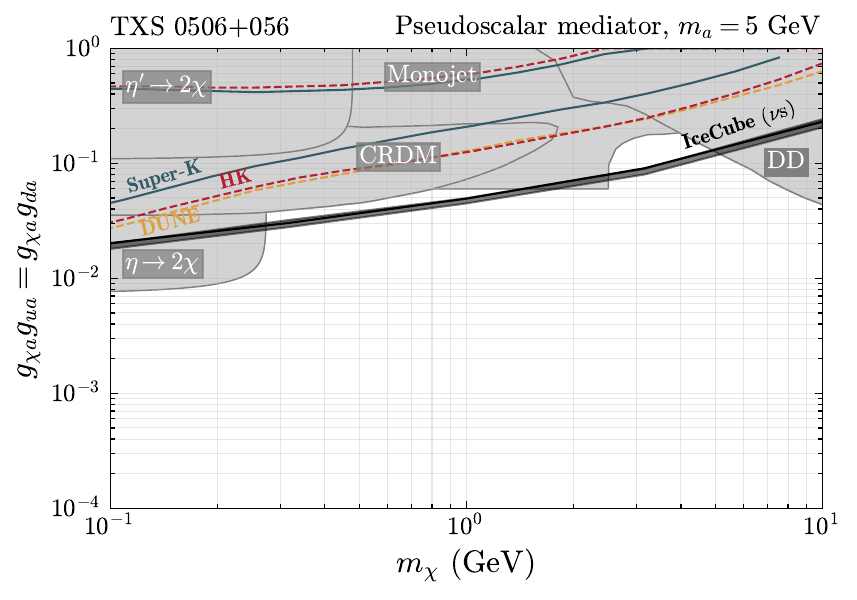}
                \includegraphics[width = 0.4\textwidth]{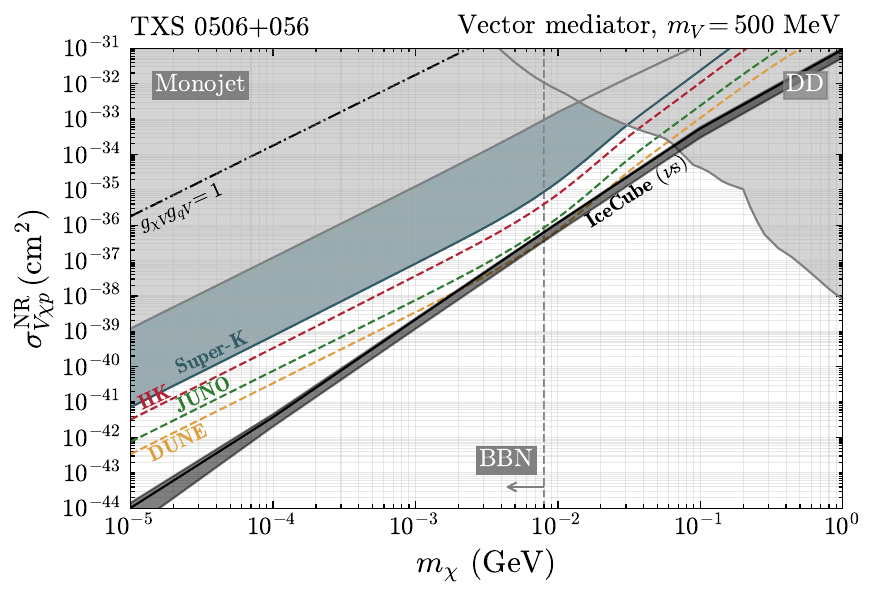} 
                \includegraphics[width = 0.4\textwidth]{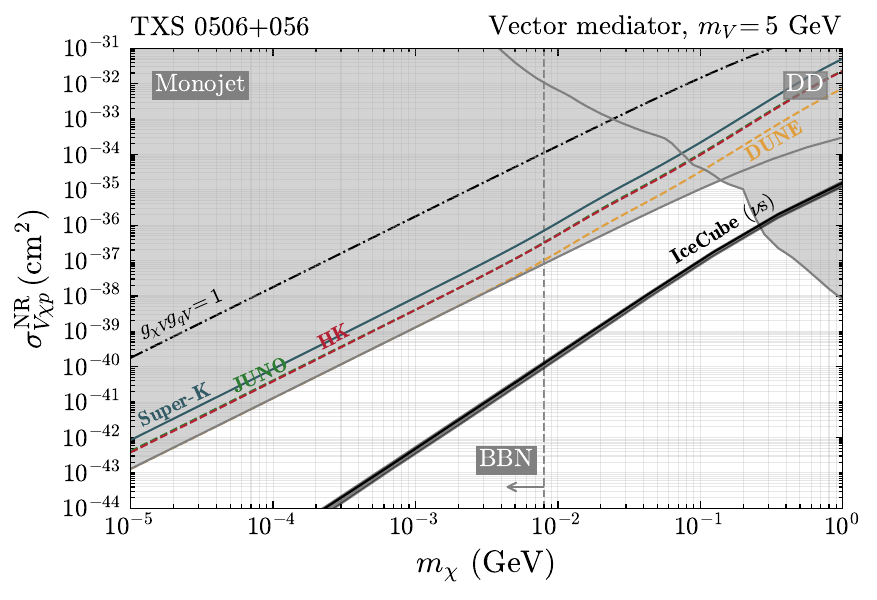}    
                \includegraphics[width = 0.4\textwidth]{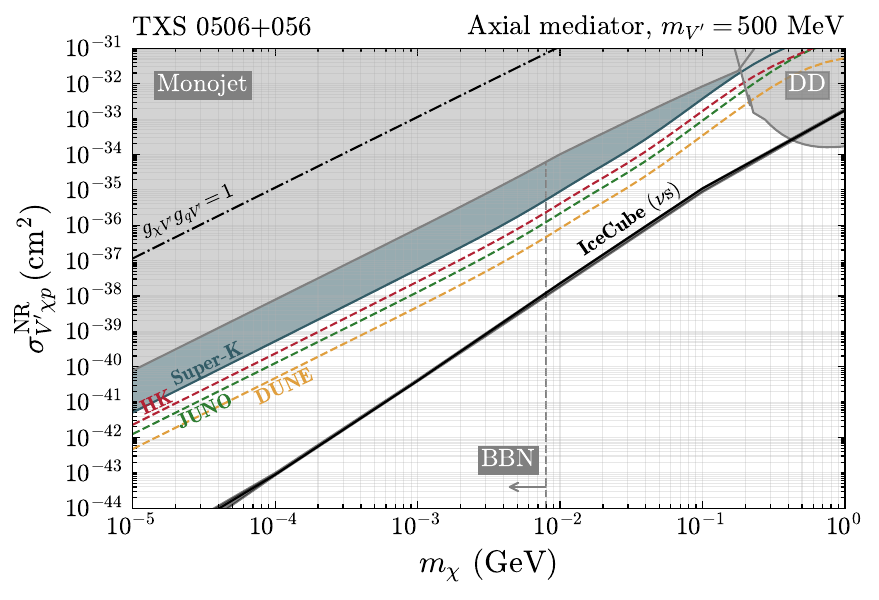}
                \includegraphics[width = 0.4\textwidth]{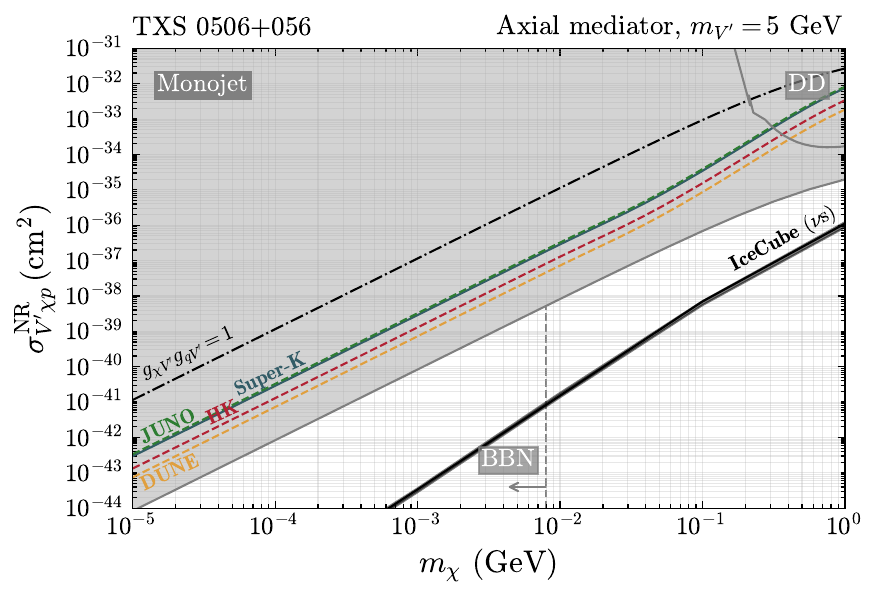}
    \caption{{\bf TXS 0506+056.} Limits and sensitivities from TXS 0506+056 in the plane of $\sigma_{Y\chi p}^{\NR}$ against $m_\chi$ for a scalar (top), vector (third row), and axial (bottom) mediator.  For the pseudoscalar case (second row), the limits are shown in the $g_{\chi a} g_{q a}$ -- $m _\chi$ plane. We set $m_Y=500\,\text{MeV}$ (left panels) and $m_Y = 5\,\text{GeV}$ (right panels) for the mass of the mediator and BMCI for the DM LOS integral. 
 Blue shaded areas denote the limits set by SK. Future sensitivities are shown in dashed red (HK), orange (DUNE) and green (JUNO). The 2017 neutrino detected from TXS 0506+056 is explained by the same DM interactions along the black lines~\cite{DeMarchi:2025xag}. The black dot-dashed line ($g_{\chi Y}g_{qY} = 1)$ marks the non-perturbative regime.  
 The shaded grey areas are excluded by direct detection and various model-dependent searches, while the vertical dashed gray line marks the Big Bang nucleosynthesis limit, evadable if the dark sector thermalises with the SM below photon-neutrino decoupling. See Sec.~\ref{sec:constraints_modeldep} for extensive details on these constraints.}
\label{fig:TXS_Sigma_vs_mx}
\end{figure}

\begin{figure}[t!]
        \centering
                \includegraphics[width = 0.4\textwidth]{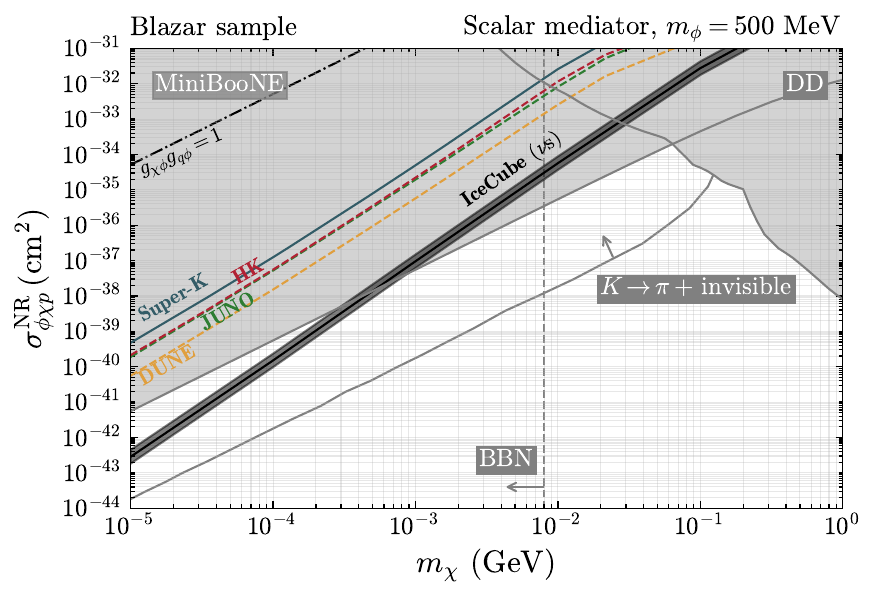}
        \centering
                \includegraphics[width = 0.4\textwidth]{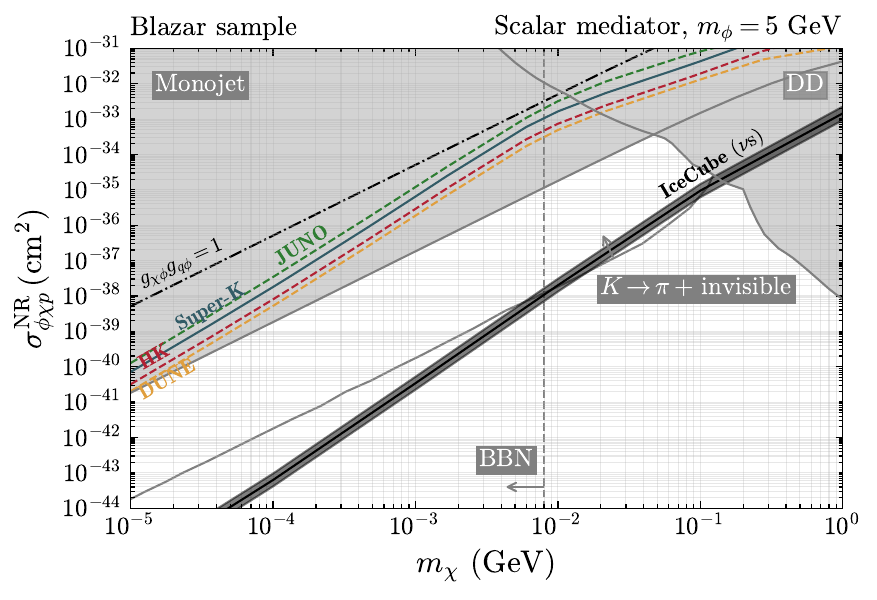}
                \includegraphics[width = 0.4\textwidth]{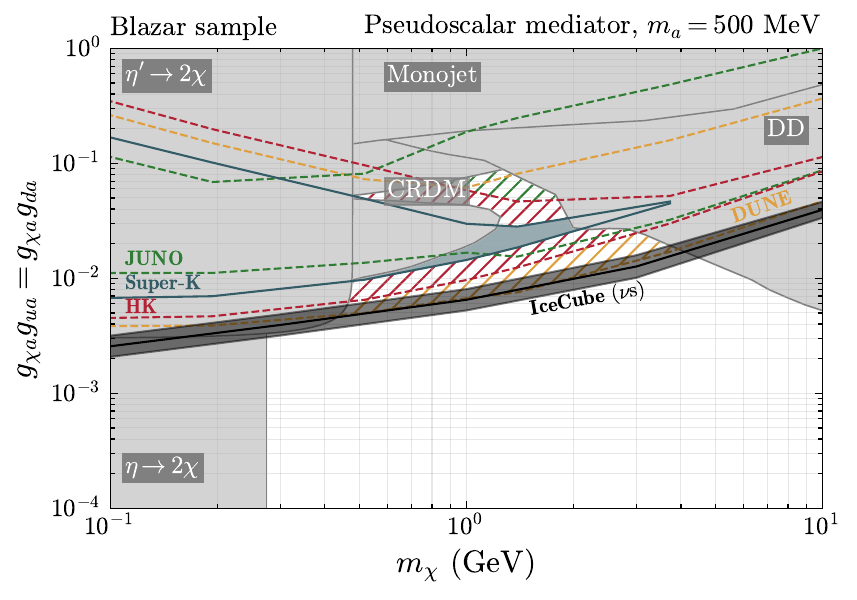}
                \includegraphics[width = 0.4\textwidth]{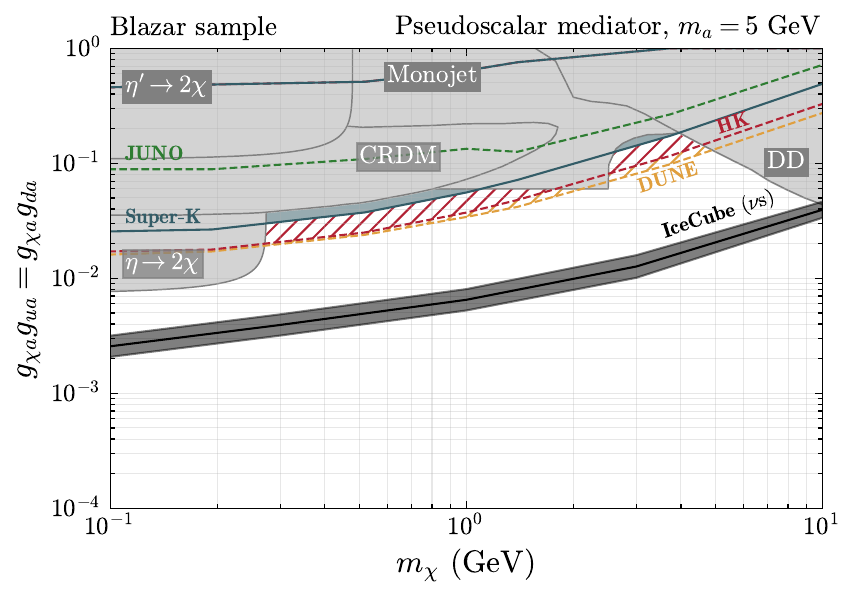}
                \includegraphics[width = 0.4\textwidth]{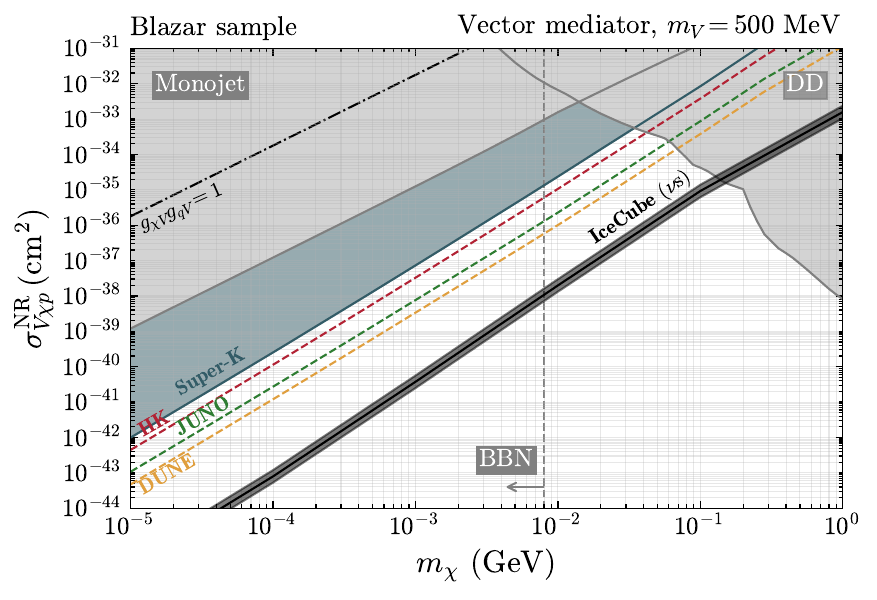}  
                \includegraphics[width = 0.4\textwidth]{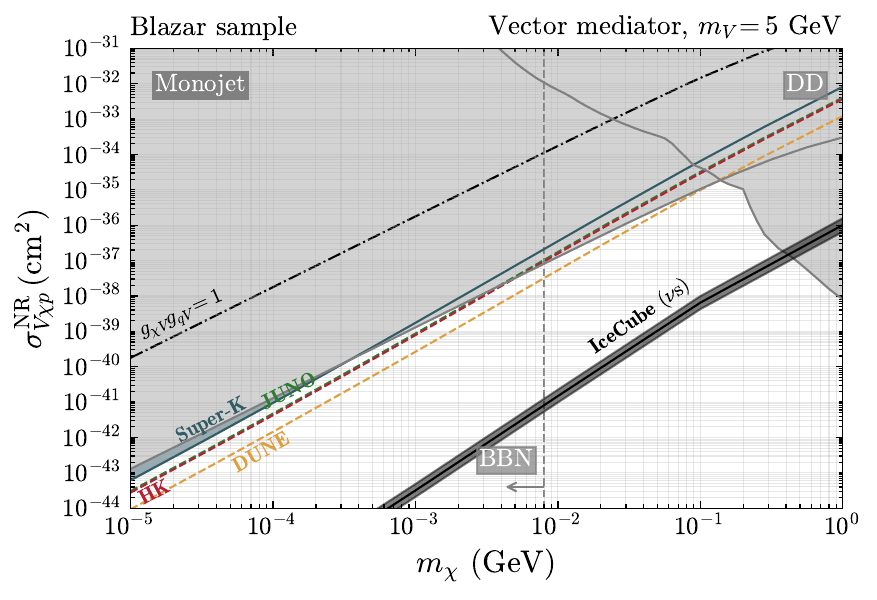}  
            \includegraphics[width = 0.4\textwidth]{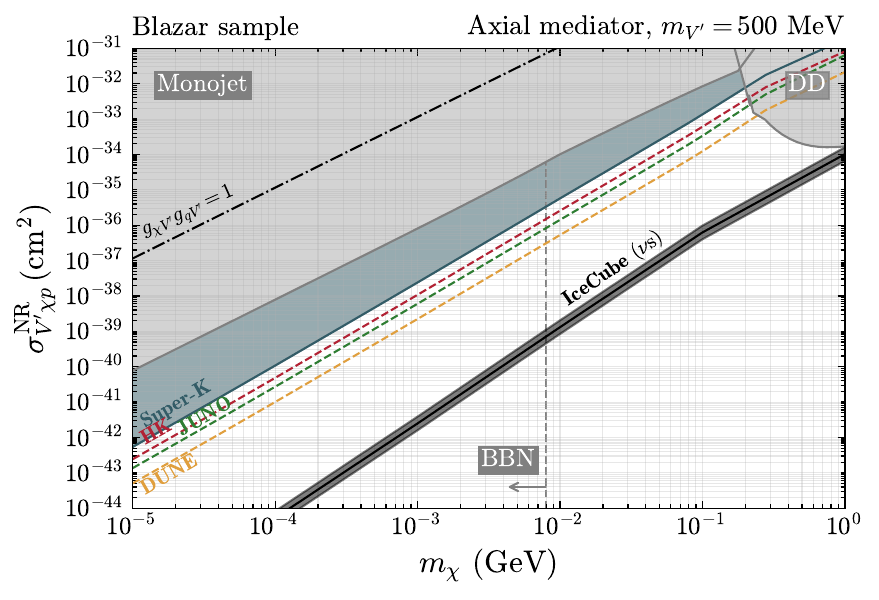}
            \includegraphics[width = 0.4\textwidth]{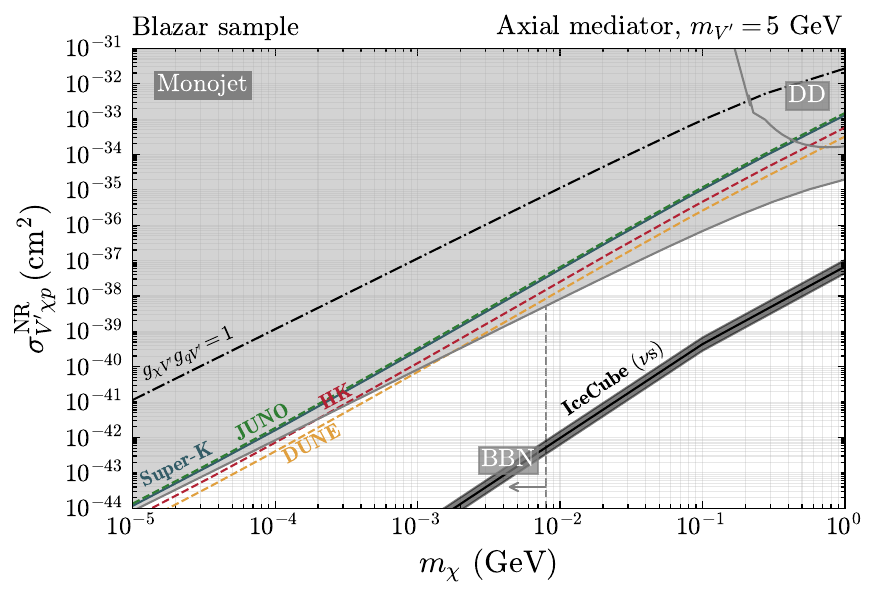}
    \caption{{\bf Blazar sample.} Same as in Fig.~\ref{fig:TXS_Sigma_vs_mx} but for the sample of blazars taken from~\cite{Rodrigues:2023vbv} (excluding 3C 371).
    Along the black continuous lines, the diffuse neutrino flux detected by IceCube~\cite{Abbasi:2021qfz} 
    is explained by DM-$p$ DIS around blazars. 
    See the description of Fig.~\ref{fig:TXS_Sigma_vs_mx} and the main text for more details.}
                \label{fig:Sample_Sigma_vs_mx}
\end{figure}

We show our results in Figs.~\ref{fig:TXS_Sigma_vs_mx} and \ref{fig:Sample_Sigma_vs_mx} 
for TXS 0506+056 and the blazar sample~\cite{Rodrigues:2023vbv}, respectively, focusing on two benchmark values of the mediator mass $m_Y = 500\,\text{MeV}$ and $5\,\text{GeV}$.
For these plots, to keep the discussion minimal, we have fixed $R_\min = 10^2\, R_S$ as in BMCI for the DM spike.
However, we recall that the BBDM and neutrino signals have distinct dependences on the DM column density. Since the former scales quadratically with the cross section, involving both DM upscattering at the source and elastic scattering at detection, and the neutrino signal depends linearly on the cross section, as detection proceeds via SM interactions, the corresponding bounds scale with the LOS integral as $\Sigma_{\DM}^{-1/2}$ for BBDM and as $\Sigma_{\DM}^{-1}$ for the neutrino signal. Knowing such scaling, the computed limits and sensitivities to BBDM, and to the blazar neutrinos from DM-proton interaction, can be straightforwardly rescaled for different values of $\Sigma_\DM$. We show the results for BMCII ($R_\min = 10^4\,R_S$) in App.~\ref{app:BMCII}. 

In the remainder of this section, we provide a detailed discussion of the model-dependent and model-independent constraints relevant to our DM benchmark scenarios, and the comparison with the predicted neutrino emission from the same DM-proton interactions in blazars. We also discuss how the same DM-proton interactions responsible for the signals discussed in this work can deplete the DM spike and potentially compromise our final results for TXS 0506+056 (but not for the blazar sample). In addition, we complement this section with App.~\ref{app:Spikedepletion}, where we estimate to what extent $4\chi \to 2\chi$ annihilations and $2\chi \to 2\chi$ scattering processes can also alter the DM spike.

\subsection{Model-independent constraints}
The effect of a new particle species $\psi$ with thermal abundance and energy density $ \rho_\psi$ is to increase the Hubble parameter $H$ at a given epoch as $H^2 \propto \rho_{\text{tot}}$. Assuming that $\psi$ does not have a large chemical potential, $\rho_\psi$ is controlled only by its mass $m_\psi$ and temperature T. As in our benchmark points we assume for the mediators  $m_Y\sim \mathcal{O} (1)$ GeV, for the range of couplings that we probe, $Y$ decays well before Big Bang nucleosynthesis (BBN), thus not altering its predictions. However, the presence of thermal DM affects the number of effective relativistic neutrino species, $N_\text{eff}$, and hence the values of helium and deuterium primordial abundances produced at BBN,  requiring $m_\chi \geq\mathcal O (10)$ MeV.
This limit can be relaxed to $m_\chi \geq O(1)$~MeV by switching on a coupling of the mediators to neutrinos, with a specific size.
Otherwise, if the dark sector thermalizes with the SM after neutrino-photon decoupling, then the BBN limit can be evaded~\cite{Berlin:2017ftj} while large DM-nucleon cross sections can be maintained~\cite{Berlin:2018ztp}.

Constraints from telescope searches of DM annihilation products, including those coming from cosmic microwave background observations, are more dependent on the cosmological DM history. For example, they (almost~\cite{Graesser:2011wi,Baldes:2017gzw,Baldes:2017gzu}) disappear in the case of asymmetric DM, where instead our mechanism works the same, both in explaining IceCube's neutrinos and in BBDM. Therefore, we do not display any indirect detection limit in Figs.~\ref{fig:TXS_Sigma_vs_mx} and~\ref{fig:Sample_Sigma_vs_mx}, keeping in mind that possible signals of sub-GeV DM at telescopes could test our mechanism in some of its specific realisations. 
Analogously, the DM abundance can be achieved orthogonally to the DM parameters on which our signals rely (e.g. from an asymmetry), hence we refrain from showing such lines, or exclusions by overclosure, on our plots.

The parameter space is constrained also by conventional direct detection (DD) experiments. 
For the spin-independent DM-nucleon interaction cross section (vector, scalar, and pseudoscalar induced at one-loop, see Section \ref{sec:ToyModels}),
we combine the constraints from TESSERACT~\cite{TESSERACT:2025tfw}, SuperCDMS~\cite{SuperCDMS:2023sql}, SENSEI~\cite{SENSEI:2023zdf}, CRESST-III~\cite{CRESST:2019jnq}, DarkSide-50~\cite{DarkSide:2018bpj}, LUX-ZEPELIN (LZ)~\cite{LZ:2024zvo}, XENONnT~\cite{XENON:2023cxc,XENON:2025vwd}, and we show the resulting DD bounds in Figs.~\ref{fig:TXS_Sigma_vs_mx} and \ref{fig:Sample_Sigma_vs_mx}.
For spin-dependent interactions (axial mediator), we combine bounds from Borexino~\cite{Bringmann:2018cvk}, delayed
coincidence searches in near-surface detector~\cite{Collar:2018ydf} and NEWS-G~\cite{NEWS-G:2024jms}.
We do not include in our plots bounds based on the Migdal effect in liquid Xenon induced by DM (e.g.~\cite{PandaX:2023xgl, XENON:2019zpr}), because the theory prediction of this effect in liquid Xenon is in conflict with SM experiments~\cite{Xu:2023wev} and thus requires further study.\footnote{A number of neutron-induced Migdal events in liquid Xenon compatible with zero was reported in~\cite{Xu:2023wev}, whereas the theory prediction is one order of magnitude larger than the experimental uncertainty.}

\subsection{Model-dependent constraints}
\label{sec:constraints_modeldep}

While \textsc{BBN} and \textsc{DD} constraints are largely model-independent, laboratory bounds crucially depend on the nature of the DM interactions with SM particles. We summarise below the most relevant limits for each of the mediator cases studied in this work.\footnote{DM-proton interactions could also induce AGN cooling, as studied for a scalar mediator in~\cite{Herrera:2023nww} and for a vector one in~\cite{Gustafson:2024aom, Mishra:2025juk}. The benchmark choices for the DM spike considered in those references, however, are more aggressive than ours. This makes them challenging to reconcile with the cross sections tested by AGN cooling if one takes into account the spike-softening induced by DM self-interactions, see App.~\ref{app:Spikedepletion}. This is unavoidable even for asymmetric DM, but could be avoided for large enough inelasticities as those considered in~\cite{Gustafson:2024aom}. This limitation is on top of the inherent degeneracy between astrophysical uncertainties and DM-$p$ interactions in inducing AGN cooling.}

\vspace{-.5em}
\begin{itemize}
\item {\bf Scalar mediator}

Scalar mediators can be constrained by both meson decays and collider searches.
 Rare kaon decays~\cite{NA62:2021zjw} impose strong bounds on decays to invisible final states, including DM. We report as solid grey line in our Figs.~\ref{fig:TXS_Sigma_vs_mx} and \ref{fig:Sample_Sigma_vs_mx} the bound for $g_{t \phi}=0$, corresponding to the gluon coupled case in Fig.~2 of~\cite{Cox:2024rew}; we do not shade it as it can be evaded by a specific combination of $g_{u\phi}$ and $g_{d\phi}$ \cite{Pascoli:atmospheric}.  DM pairs can be produced at colliders together with SM radiation via processes like $q\bar{q} \to g\,\phi \to g\,\chi\bar{\chi}$. Based on the recasting performed in~\cite{Ema:2020ulo}, we show constraints from monojet plus missing energy searches at LHC~\cite{ATLAS:2021hza}.
 These are stronger than the ATLAS~\cite{ATLAS:2024kpy} and CMS~\cite{CMS:2021far} limits, which moreover rely on the mediator coupling to top quarks.

 Stringent constraints also arise from searches for cosmic ray accelerated DM (CRDM) \cite{Bringmann:2018cvk, Ema:2018bih}. CRDM limits imposed by SK and KamLAND searches in the same scalar scenario considered by us has been computed in \cite{Ema:2020ulo} for $m_\phi = 1\,\text{GeV}$ and $3 \,\text{GeV}$ (which we consider in place of $m_\phi = 500\,\text{MeV}$ and $5\,\text{GeV}$, respectively), and these are stronger than the constraints from monojet searches at colliders when $m_\phi = 500\,\text{MeV}$.

The most stringent limits when $m_\phi = 500\,\text{MeV}$, however,  come from DM searches from proton beam dumps in MiniBooNE \cite{MiniBooNE:2017nqe}. We employ the recast of those limits, to the case of a scalar mediator decaying invisibly, as performed in~\cite{Batell:2018fqo}.

\item {\bf Pseudoscalar mediator}

Processes such as $\pi^0 \to 2\chi$, $\eta \to 2\chi$, and $\eta' \to 2\chi$ constrain this scenario through the mixing between the pseudoscalar mediator and  light mesons. Bounds are derived from the upper limits on the invisible decay width of these states~\cite{ParticleDataGroup:2024cfk}. As shown in~\cite{Ema:2020ulo}, introducing a coupling $g_{s a}$ between the strange quark and the pseudoscalar mediator can relax the constraints. However, we conservatively shade the excluded regions assuming $g_{s a} = 0$ and $g_{u a}= g_{d a}$.
Bounds from monojet searches at the LHC apply similarly to the scalar case.  We also consider CRDM limits from SK and KamLAND for this scenario, recasting them from  \cite{Ema:2020ulo}.

\item {\bf Vector and axial mediators}

Limits from monojet searches at TeVatron and the early LHC, as recasted~\cite{Shoemaker:2011vi}, read $g_q \leq 0.02$ and are independent of the mediator mass, as long as it is significantly lighter than the cuts used in those searches, which is true for all $m_Y$ benchmarks that we use. These limits are stronger than monojet CMS ones on light spin-1 mediators~\cite{CMS:2021far}.

It would be interesting to quantify the impact of the MiniBooNE limits of~\cite{MiniBooNEDM:2018cxm} on the vector mediator case for $m_V = 500\,\text{MeV}$, but their recast goes beyond the purposes of this study.

\end{itemize}

Among the limits discussed so far, the only ones that do not scale with powers of $g_{\chi Y} g_{q Y}$ are those coming from monojet searches for $m_Y > 2 m_\chi$, which are proportional to $g_{q Y}$ only.
In drawing them in Figs.~\ref{fig:TXS_Sigma_vs_mx} and~\ref{fig:Sample_Sigma_vs_mx} we then need to make an assumption for one of the couplings, and we choose $g_{\chi Y} = 1$.
Had we chosen smaller values of $g_{\chi Y}$, our limits would have been unchanged but this would have implied larger values of $g_{qY}$ to yield the same cross section. Monojet limits on our scenario (i.e. on $g_{\chi Y} g_{q Y}$) would have then become relatively stronger, by the same factor that rescaled $g_{\chi Y}$.
In particular, in the case $g_{\chi Y} \simeq g_{q Y}$, the monojet limits on the couplings $g_{\chi Y} g_{q Y}$ become stronger by a factor $(g_{\chi Y} g_{q Y})_\text{lim}^{1/2} \simeq (g_{q Y})_\text{lim}$, where $(g_{\chi Y} g_{q Y})_\text{lim}$ is the monojet limit that can be directly read-off the plots.
On the $\sigma_{Y\chi p}^{\text{NR}}-m_\chi $ plane the limit is instead rescaled by a factor $(g_{\chi Y} g_{q Y})_\text{lim} \simeq (g^2_{q Y})_\text{lim}$, since $\sigma_{Y\chi p}^\NR \propto (g_{\chi Y} g_{qY})^2$.
Compared to our $g_{\chi Y} = 1$ benchmark, the choice $g_{\chi Y} \simeq g_{q Y}$ would then restrict the parameter space where blazar signals (either neutrinos produced by DM-$p$ scatterings or BBDM) are the dominant probe of DM, but would not change the qualitative message of our study.

While we have assumed the mediators to couple to $u$ and $d$ quarks only, let us briefly comment that new strong limits would arise if they also coupled to heavier quarks.
A coupling to top quarks would for example induce, at one loop, sizeable decays of $B$ mesons into lighter mesons plus DM, see e.g.~\cite{Knapen:2017xzo}.
Couplings to the charm or bottom quarks would induce invisible (plus possibly other particles, depending on the mediator) decays of the QCD vector resonances $J/\psi$ or $\Upsilon$. Searches for their invisible decays at BES and BABAR would translate into limits on those couplings, see e.g.~\cite{Kim:1986ax} for light scalars and~\cite{Batell:2014yra} for light vectors.

We finally comment on possible EW-invariant realization of our toy-models of Eqs~\eqref{eq:L_scalar}, \eqref{eq:L_vector}, \eqref{eq:L_axial} and~\eqref{eq:L_pseudoscalar} and on their implications for the DM couplings.
While the couplings $g_{\chi Y}$ do not involve SM fields and therefore pose no problems in this respect, those to quarks $g_{q Y}$ require some specification.
They can for instance be obtained via mixing of SM fermions with vector-like heavy quarks, see e.g.~\cite{Ema:2020ulo} for scalar and pseudoscalar mediators and~\cite{Kahn:2016vjr} for vector and axial vector ones.
Such UV-completions also allow to obtain the hierarchy $g_{\chi Y}/g_{q Y} \gg 1$, that we assumed for monojet limits in Figs~\ref{fig:TXS_Sigma_vs_mx} and~\ref{fig:Sample_Sigma_vs_mx}.\footnote{In such UV-completions, both axial and vector couplings are generated unless one is willing to tune the UV parameters. This does not pose a problem to our mechanism, it just makes DD limits slightly stronger for the axial case.} However, it can be challenging to obtain $g_{\chi Y}/g_{q Y} \gg 1$ in other UV completions, see~\cite{Gori:2025jzu}.

\subsection{Limits and sensitivities on blazar-boosted dark matter at neutrino detectors}

In general, we find that the most constraining limits from the BBDM signals are currently set by Super-K (we do not show in the plots the limits from KamLAND and Borexino as the ones from SK are more stringent across all the scenarios considered in this work) and are expected to be tightened in the near future by Hyper-K, JUNO, and DUNE: the former due to its larger volume and the latter thanks to their lower energy thresholds, which encompass the peak of the proton recoil spectrum (see Fig.~\ref{fig:Recoil_Spectrum}).

For $m_Y = 500\,\mathrm{MeV}$, we find that BBDM can probe regions of the parameter space that are currently allowed by existing experiments, for all the four DM-quark interactions that we studied.  
On the other hand, increasing the mediator mass to $m_Y = 5\,\mathrm{GeV}$ leads to a degradation of the BBDM sensitivity, as collider constraints are only weakly dependent on the value of $m_Y$, while our BBDM signal gets suppressed. We find indeed that for $m_Y = 5\,\mathrm{GeV}$ BBDM limits are no longer competitive with existing constraints, even in the optimistic BMCI scenario, as shown in Figs.~\ref{fig:TXS_Sigma_vs_mx} and~\ref{fig:Sample_Sigma_vs_mx}.

It is interesting to compare the projected sensitivities of detectors across different interactions and DM masses. In particular, we obtain that JUNO performs relatively worse than the other detectors in the pseudoscalar case for each of the considered benchmark mediator masses, and in all the cases for $m_Y = 5\,\mathrm{GeV}$. The loss of sensitivity of JUNO detector in the pseudoscalar case is due to the suppression at low momentum transfer $Q^2$ of the pseudoscalar interaction cross section. The dependence on the mediator mass instead arises from the shift of the proton recoil spectrum peak toward higher energies as the mediator mass increases, moving outside JUNO's sensitivity range.
For the same reason, we find that DUNE, which has a lower energy threshold than Super-K, leads only to a marginal improvement in the pseudoscalar mediator case, as shown in Figs.~\ref{fig:TXS_Sigma_vs_mx} and \ref{fig:Sample_Sigma_vs_mx}.

Overall, we find that the hierarchy among detectors is non-trivial and strongly depends on the choice of the DM mass, mediator mass, and interaction type, highlighting the importance of using explicit DM models when comparing the reach of different experiments.

\subsection{Dark matter-induced neutrino signals survive blazar-boosted dark matter searches}

Inelastic DM-proton interactions can induce proton disintegration, leading to the emission of high-energy neutrinos which can be detected at large-volume neutrino telescopes, such as IceCube. Clearly, the parameters governing the production of neutrino from DM-proton inelastic scatterings are the same as those that control the BBDM signal discussed in this work. This naturally raises the question of whether the two types of signals are mutually compatible, if one should be observed before the other, or if the absence of one of the two could constrain the viability of the other.

In \cite{DeMarchi:2024riu}, we computed the neutrino flux from DM-proton inelastic interactions around the blazar TXS 0506+056, and identified the regions in the $m_\chi-\sigma_{Y\chi p}^{\NR}$ plane allowing to explain the 2017 IceCube neutrino event via this mechanism, specifically for vector and scalar mediator models with $m_Y = 5\,\text{GeV}$. We have extended such analysis for TXS 0506+056 in \cite{DeMarchi:2025xag} covering the same DM models considered here, both for $m_Y = 500\,\text{MeV}$ and $5\,\text{GeV}$. Furthermore, in \cite{DeMarchi:2025xag}, we have also evaluated the diffuse neutrino flux arising from DM-proton interactions across the same blazar sample analysed in this study and extracted the corresponding combinations of DM parameters that leads to saturation of the diffuse neutrino flux observed by IceCube \cite{Abbasi:2021qfz}. The results of the analyses performed in \cite{DeMarchi:2024riu, DeMarchi:2025xag} for BMCI are represented in Figs.~\ref{fig:TXS_Sigma_vs_mx} and \ref{fig:Sample_Sigma_vs_mx} in black, in order to compare them against BBDM limits and sensitivities. The widths of the black lines span the 90\% C.L. of the energy $E_\nu$ of the neutrino observed from TXS 0506+056 and, for the blazar sample, the analogous interval on the observed diffuse neutrino flux.

A non-trivial hierarchy immediately emerges between the BBDM and neutrino signals: the former is primarily controlled by elastic interactions, while the latter relies on inelastic processes. As a result, increasing the mediator mass tends to enhance the relative strength of the latter.

We highlight that when the region that corresponds to the neutrino prediction lies inside the BBDM sensitivity regions, the observation (non-observation) of a BBDM signal at future neutrino detectors would support (reject) the DM interpretation of the IceCube neutrinos. According to Figs.~\ref{fig:TXS_Sigma_vs_mx} and \ref{fig:Sample_Sigma_vs_mx}, this situation arises in the case of TXS 0506+056 when $m_Y=500\,\text{MeV}$: for the scalar mediator scenario if $m_\chi \lesssim 0.2\,\text{MeV}$, for the pseudoscalar case at $m_\chi \sim \mathcal{O}(\text{GeV})$, and for the vector mediator at $m_\chi \sim \mathcal{O}(10 \, \text{MeV})$. For the blazar sample, a similar overlap occurs only mildly, and only for DUNE. 

On the other hand, in the situation for which the neutrino prediction region lies below the BBDM ones, the DM interpretation of the neutrino signal would survive BBDM null searches.
In this case, a future observation of a BBDM signal would imply that a corresponding neutrino flux should have already exceeded observations, thereby disfavouring the DM origin of the BBDM signal under the same interaction model.
We find that this situation occurs in all the other cases we considered, except for the scalar mediator scenario for TXS 0506+056 with $m_\phi = 500\,\text{MeV}$ and $m_\chi \gtrsim 0.2\,\text{MeV}$, where the DM hypothesis for the IceCube neutrinos is incompatible with the null observations of BBDM signals at Super-K.

Finally, we comment on the differences between the signals from TXS 056+056 and the blazar sample. These are mainly due to the astrophysical parameters of the lepto-hadronic fits. In the case of the blazar sample, the proton spectral index and the minimum proton Lorentz factor in the blob are taken to be $\alpha_p = 1$ and $\gamma'_{\min_p} = 100$, respectively, thus enhancing the contribution of higher-energy protons. In contrast, for TXS 0506+056, the adopted fit gives $\alpha_p = 2$ and $\gamma'_{\min_p} = 1$, thereby giving more weight to lower-energy protons. This, overall, leads to differences in the resulting BBDM flux and nuclear recoil spectra (see Figs.~\ref{fig:BBDM_Fluxes}, \ref{fig:BBDM_Fluxes_SAMPLE} and \ref{fig:Recoil_Spectrum}), which ultimately translate into distinct projected limits and sensitivities for the two cases.  
Regarding neutrino production from DM-proton scattering, this is more prominent in scenarios where high-energy protons dominate, as they are more efficient at producing neutrinos through inelastic collisions. It is worth emphasising that this choice of $\alpha_p$ is not strongly motivated for the considered sample, and was effectively fixed in the fit by hand \cite{Rodrigues:2023vbv}. Indeed, variations in $\alpha_p$ would significantly affect the relative strength of the neutrino and BBDM signals and could make the latter a test of the former signal over a wider range of parameters.

The above discussion regarded BMCI, but the testability of both the BBDM and neutrino signals depends on where the jet starts, see App.~\ref{app:BMCII} for BMCII.

\subsection{Depletion of the spike by the jets}\label{subsec:spikedepletion}

We now investigate processes which can deplete the assumed DM density profile around the BHs and 
estimate their characteristic timescales. We compare these to the typical timescale $t_{\mathsmaller{\rm Accr.}}$ for the BH accretion, as this is tied to the formation of the DM spike and its replenishment.

A rough estimate of $t_{\mathsmaller{\rm Accr.}}$ can be obtained by assuming that the BH's accretion saturates the Eddington limit \cite{Shakura:1972te}
\begin{equation}
\frac{M_\text{BH}}{\dot{M}_{\text{BH}}} \sim 3\times 10^7 \text{yr}\,\left(\frac{\varepsilon}{0.06} \right),
\end{equation}
where $\varepsilon$ is the radiative efficiency, i.e.~the fraction of energy that gets converted into radiation during the accretion process. The efficiency $\varepsilon$ reads 0.06 for a Schwarzschild BHs, but it can reach 0.4 for Kerr BHs. For definiteness, in the computations that follows, we consider $t_{\mathsmaller{\rm Accr.}}=10^8$~years.

The interaction of the blazar jet protons with the surrounding DM ejects DM particles and can eventually deplete the DM spike over time.\footnote{We thank Paolo Salucci for asking us about the importance of such an effect in this context.}
In this section, we estimate to what degree this happens and whether that invalidates our results. As a first step we compute the total cross section for the $\chi - p$ scattering  setting $g_{qY} = g_{\chi Y} = 1$, namely
\begin{equation}
\sigma_{\text{tot}}(s) = \int_0^{T_\chi^{\text{max}}(T_p)} dT_\chi \frac{d\sigma_{Y\chi p}^{\EL}}{dT_\chi} + \int_{Q^2_{\text{min}}/s}^{1} dx\int_{Q^2_{\text{min}}}^{xs}dQ^2\;\frac{d\sigma^{\DIS}_{Y\chi p}}{dxdQ^2},
\end{equation}
where $s = m_p^2 + m^2_\chi + 2m_\chi (m_p + T_p)$, and convolute it with the proton spectrum in the blazar jet as
\begin{equation}
    \langle\sigma\rangle \equiv \int_{1}^{\Gamma_B \gamma_{\max_p}} d\gamma_p \; \sigma_{\text{tot}}(s) \int d\Omega\,\frac{1}{\kappa_p}\frac{d\Gamma_p}{d\Omega dT_p},
\end{equation}
where we point to Sec.~\ref{subsec:blazar_jet} for the definitions related to the blazar jet.
We then consider a spherical shell of thickness $dr$ at a distance $r$ from the central BH and count how many particles the shell contains, $N_\DM(r, t)$, as well as the rate of depletion, $\dot{N}_\DM(r, t) \equiv dN_\DM(r,t)/dt$, according to
\begin{eqnarray}
\begin{aligned}
     N_\DM(r, t) &= 4\pi r^2 \frac{\rho_{\DM}}{m_\chi}\; dr, \\
     \dot{N}_\DM(r, t) &= -(g_{qY}g_{\chi Y})^2\kappa_p \langle\sigma\rangle \frac{\rho_{\DM}}{m_\chi} dr.
\end{aligned}
\end{eqnarray}
\begin{figure}[t!]
        \centering
                        \includegraphics[width = 0.4\textwidth]{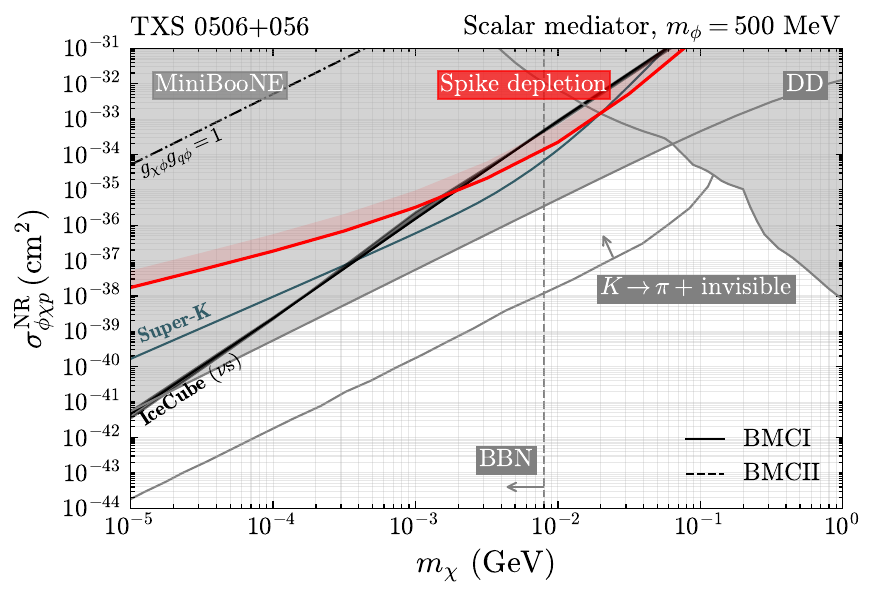}
                        \includegraphics[width = 0.4\textwidth]{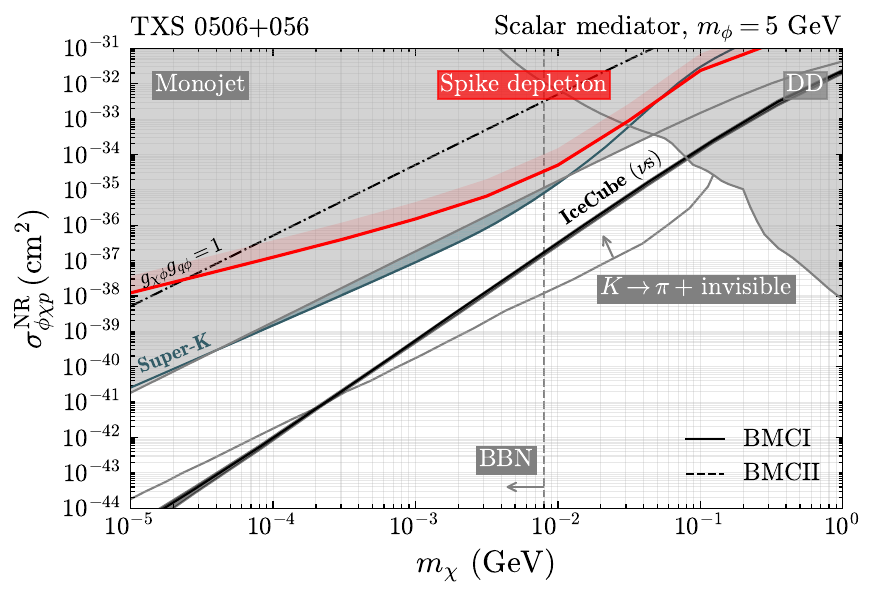}
                        \includegraphics[width = 0.4\textwidth]{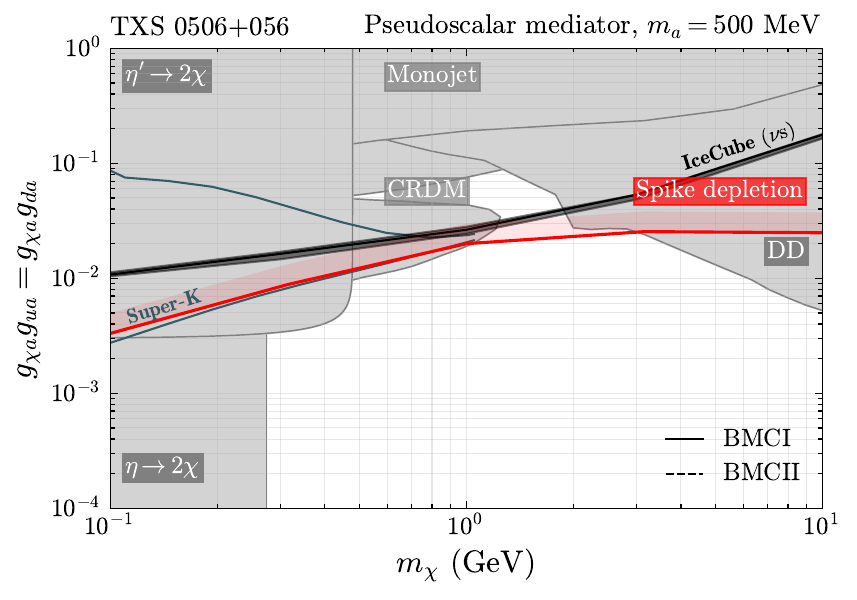}
                        \includegraphics[width = 0.4\textwidth]{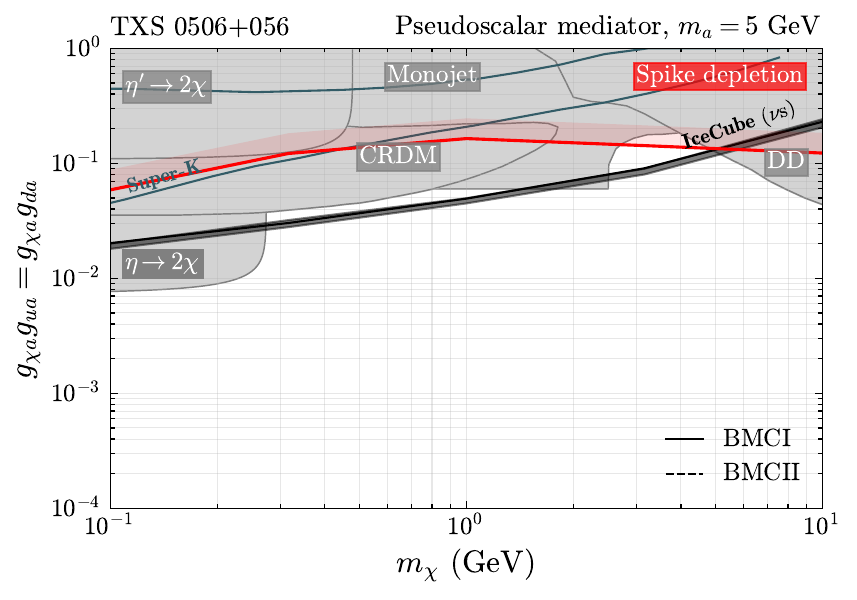}
                        \includegraphics[width = 0.4\textwidth]{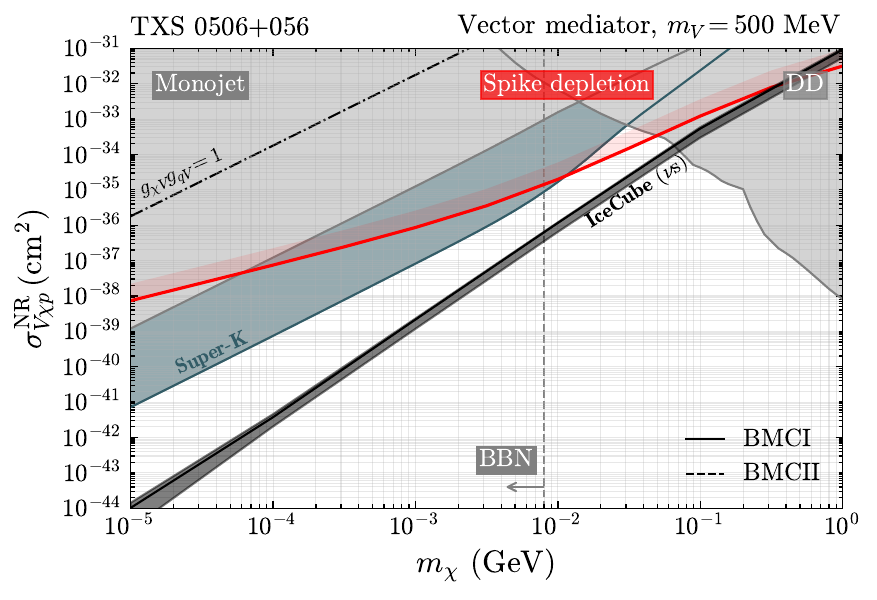}
                        \includegraphics[width = 0.4\textwidth]{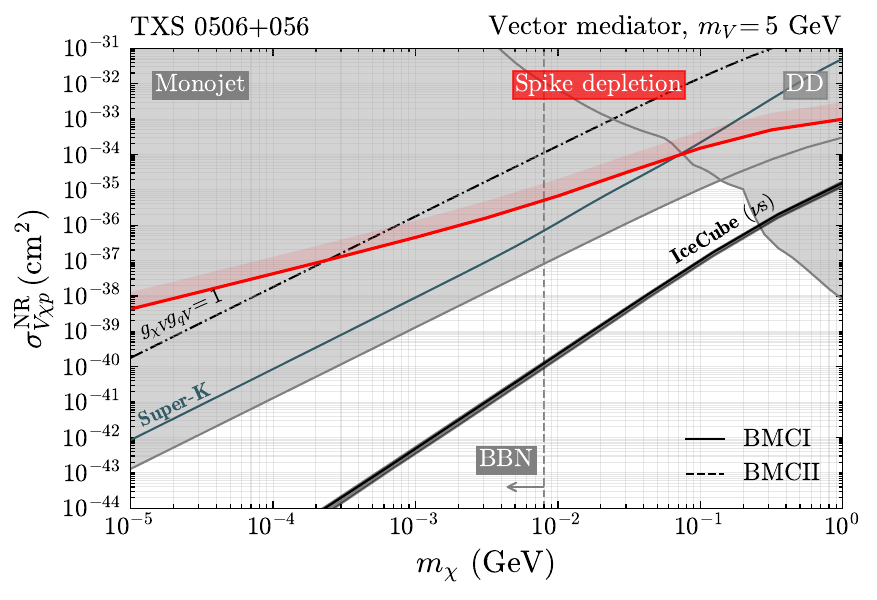}
                        \includegraphics[width = 0.4\textwidth]{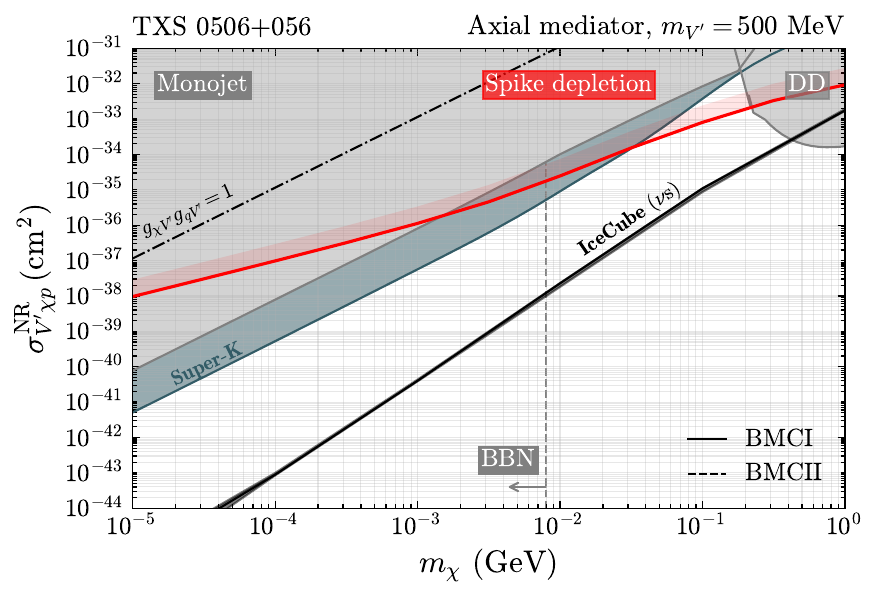}
                        \includegraphics[width = 0.4\textwidth]{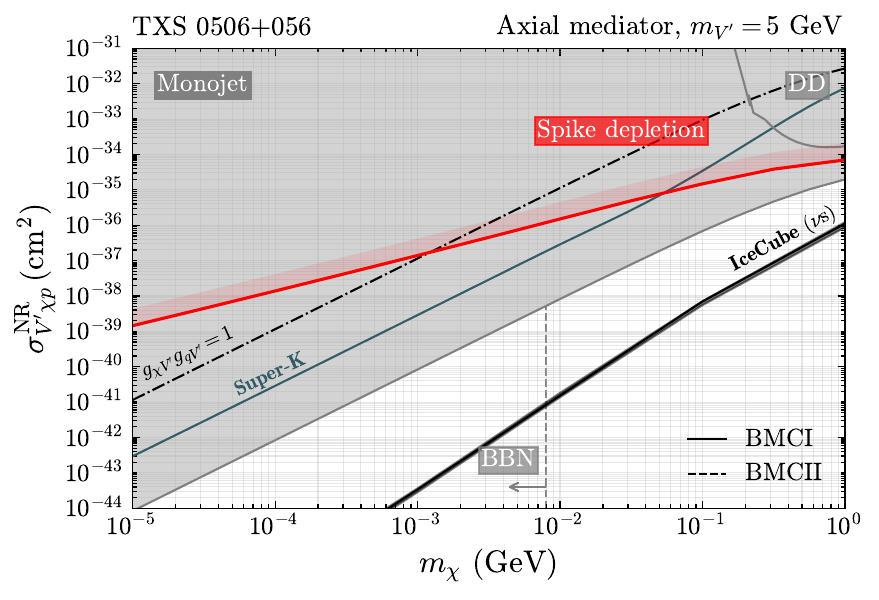}
    \caption{Same as Fig.~\ref{fig:TXS_Sigma_vs_mx}, but here we show only the projected limits on BBDM from TXS 0506+056 at SK (blue) and the region for which the DM hypothesis for the 2017 IceCube neutrino from the same blazar holds (black) for BMCI. We overlay in red the spike depletion limits that we estimate according to Eq.~\eqref{eq:bounddepletion}. See App.~\ref{app:BMCII} for BMCII.
    All other details are as in Fig.~\ref{fig:TXS_Sigma_vs_mx}.}
                \label{fig:Spike_depeletion}
\end{figure}
We thus obtain a differential equation for $\rho_{\DM}(r, t)$:
\begin{equation}
    \frac{d\rho_\DM(r, t)}{dt} = -(g_{qY}g_{\chi Y})^2\frac{\kappa_p \langle\sigma\rangle}{4\pi r^2}\rho_\DM(r, t),
\end{equation}
which can be easily solved to find
\begin{equation}
    \rho_\DM(r, \tau) = \rho_\DM(r, 0)\,\text{exp}\left[{-\frac{\kappa_p \langle\sigma\rangle\tau}{4\pi r^2}}\right],
\end{equation}
having defined, for convenience, the time variable  $\tau \equiv (g_{qY}g_{\chi Y})^2 t$.
The DM column density at $t = \tau$ then reads
\begin{equation}
    \Sigma_{\DM}^{\rm spike}(\tau) = \int_{R_\min}^{R_\text{sp}} dr\; \rho_\DM(r, 0)e^{-\kappa_p \langle\sigma\rangle\tau /(4\pi r^2)},
\end{equation}
We further define $\tau_{1/2}$ as the time at which the DM column density gets halved by the interactions with the jet, that is
$ \Sigma_{\DM}^{\rm spike}(\tau_{1/2}) \equiv \Sigma_{\DM}^{\rm spike}(0)/2$.
After computing $\tau_{1/2}$ as a function of the DM mass, we set a limit on the combination $g_{qY}g_{\chi Y}$ by imposing:
\begin{equation}\label{eq:bounddepletion}
  g_{qY}g_{\chi Y}\leq \sqrt{\frac{\tau_{1/2}}{t_{\mathsmaller{\rm Accr.}}}}. 
\end{equation}
When the above condition is satisfied, the DM column density is not depleted substantially over a timescale compatible with the accretion onto the BH. By that epoch, we assume that there is enough time for the DM spike to reform and thus that there is no dramatic loss of DM along the blazar jet.

We show in Fig.~\ref{fig:Spike_depeletion} the same SK limits on BBDM and the curve corresponding to the DM hypothesis for the 2017 IceCube neutrino from TXS 0506+056, as in Fig.~\ref{fig:TXS_Sigma_vs_mx}, but superimposing in red the constraints from the DM spike depletion given by Eq.~\eqref{eq:bounddepletion} for BMCI, see App.~\ref{app:BMCII} for BMCII. While this effect is clearly not relevant for BMCII in neither of the cases that we have considered, it is important for BMCI. Concerning the BBDM signal, such limit largely reduce the associated testable region in all the considered toy models. 

We emphasize, however, that throughout this work we have assumed that the blazar TXS 0506+056 is characterised for much of its history by parameters corresponding to its flaring state in 2017. In reality, this is an aggressive assumption: the source undergoes flaring episodes only intermittently, and its time-averaged luminosity is most likely lower than our reference value. As a consequence, the efficiency of DM depletion via DM-proton upscattering is reduced on average with respect to what we have estimated. Accounting for the fact that the source is in a flaring state only for a fraction of the time would naturally relax the corresponding constraints, while also modifying the expected BBDM signal from this source. A proper time-dependent treatment of the source activity is beyond the scope of this work, but we remark that this aspect strengthen the robustness of our computation for the blazar sample, which is based on steady activities rather than flaring states. In contrast, the DM interpretation of the high-energy IceCube neutrino from TXS 0506+056 does not rely on sustained flaring activity over the blazar lifetime, as neutrinos arrive almost simultaneously with photons from a flare. As a result, this interpretation is potentially unaffected by the depletion of the DM spike.

For the case of the blazar sample, the different power spectrum slope leads to much smaller values of $\kappa_p$, as listed in Table \ref{tab:AGNparameters}, while giving more weight to higher-energy protons. The mild increase of $\sigma^{\DIS}_{Y \chi p}$ at higher proton energies is, however, not enough to counterbalance the effect of the smaller $\kappa_p$. As a result, in this case we find that the spike depletion due to DM-proton interaction in blazar jets is not efficient in depleting the DM column densities along the blazar jets over billions of years, for DM parameters in the viable sub-GeV DM parameter space, neither for the considered BMCI nor BMCII, for both the benchmark values of the mediator mass and all the toy models that we have considered. Thus, the blazar DM signals that we have evaluated here and in \cite{DeMarchi:2025xag} for the blazar sample are safe from the spike depletion induced by the same DM-proton scatterings that generate such signals. 

The DM spike could in principle be depleted by other effects, such as DM-only $4 \to 2$ and $2 \to 2$ interactions.
We estimate their effects on the DM spike in App.~\ref{app:Spikedepletion}.
We anticipate here that the spike depletion that they induce is in most cases less severe than the one induced by DM-$p$ interactions, discussed in this subsection, and can anyway be more easily evaded with certain choices for the DM dynamics.

\section{Summary and outlook}
\label{sec:summary}

Sub-GeV dark matter (DM) is motivating a vibrant experimental program to look for its non-gravitational interactions. At the same time, observations of high-energy neutrinos and the quest for their origin are attracting attention to blazars. It has recently been noticed that the combination of these two distinct physics avenues could shed light on both. Indeed, the interactions of sub-GeV DM around blazars with protons in their jet could i) induce a blazar-boosted DM (BBDM) flux detectable on Earth~\cite{Wang:2021jic}, and ii) explain the neutrino observed from TXS 0506+056 in 2017~\cite{DeMarchi:2024riu}, as well as diffuse ones~\cite{DeMarchi:2025xag}. 
In this paper, we have performed several needed steps forward in the exploration of the interplay between DM and blazars. 
\begin{enumerate}
\item
We have calculated the full energy-dependent elastic and inelastic DM-proton cross sections, see Sec.~\ref{sec:intro_particle}, for four explicit mediators of DM-quark interactions: scalar, pseudoscalar, vector and axial.
This is a necessary improvement over previous literature on BBDM, which focused on constant DM-proton cross sections (see \cite{Wang:2025ztb} for a recent study on the vector mediator case only), because we are not aware of any DM model that results in a cross section that is constant for all energies of blazar protons and that can result in any signal.
Analytical formulae for these cross sections, including terms omitted in previous 
literature but important in this context, are reported in App.~\ref{app:elastic} and~\ref{app:DIS}.
\item
Using these cross sections, we have computed the BBDM fluxes for the individual blazar TXS 0506+056 and, for the first time, for a
sample of 
324 blazars.
We have then computed the proton recoil spectra induced on Earth by such BBDM, and used them to derive limits and sensitivities to sub-GeV DM at Super-Kamiokande, KamLAND, Borexino, JUNO, Hyper-Kamiokande and DUNE, see Sec.~\ref{sec:BBDM} and App.~\ref{app:EarthAtt} for more details on how we treated Earth attenuation.
\item
We have compared these limits and sensitivities with the corresponding sub-GeV DM interpretation for the IceCube neutrino observations, as computed in \cite{DeMarchi:2024riu} for TXS 0506+056 and in \cite{DeMarchi:2025xag} for the blazar sample, see Sec.~\ref{sec:limits} and in particular Figs.~\ref{fig:TXS_Sigma_vs_mx} and \ref{fig:Sample_Sigma_vs_mx}.
We have repeated the comparison for two benchmarks for the DM spike (see Sec.~\ref{sec:intro_astro}), finding overall that the sub-GeV DM hypothesis for the origin of IceCube neutrinos remains consistent with BBDM searches at neutrino detectors.
More specifically, we find that one expects to observe neutrinos before BBDM in the aggressive spike benchmark, while in the conservative one there can be hope, at least in specific parameter space, for a BBDM test of the DM origin for high-energy neutrinos, see App.~\ref{app:BMCII}. This dependence on the spike benchmark is not a surprise, because the neutrino signal scales linearly in the DM density around the blazar, while the BBDM scales with the square root of it.

\item 
Figs.~\ref{fig:TXS_Sigma_vs_mx} and~\ref{fig:Sample_Sigma_vs_mx} also display our comparison of the BBDM and neutrino signals with other limits on the same sub-GeV DM models, coming from accelerators, direct detection and cosmology. 
While neutrino signals can survive such limits, notably for both spike benchmarks in the vector and axial mediator case, the possibility of BBDM signals is either excluded or expected to be accompanied by other sub-GeV DM signals. Our analysis allows to identify a strategy to test blazar-DM signals with DM searches that are completely independent of blazar physics.

\item
We have finally checked that, in the blazar sample case, both our spike benchmarks survive the depletion of the spike induced by DM-proton and DM-DM interactions, while this conclusion depends on the specific mediator for TXS 0506+056, see Sec.~\ref{subsec:spikedepletion} and App.~\ref{app:Spikedepletion}. We warn the reader that larger DM-proton cross sections and/or more aggressive spikes, as those used in other related literature, would have to be reconsidered in light of spike depletion.
\end{enumerate}

A limitation of our study, as well as of other works like~\cite{Wang:2021jic, Granelli:2022ysi, Bhowmick:2022zkj, CDEX:2024qzq, Jeesun:2025gzt, Wang:2025ztb}, regards that the computation of the BBDM flux from TXS 0506+056 relies on assuming a steady flux of BBDM based on jet parameters during a flare.
Indeed, the BBDM is delayed with respect to neutrinos and photons from the same source, see Fig.~\ref{fig:DMdelay}. Our computation of BBDM signals on Earth should then be regarded as optimistic, because a similar flaring activity should have persisted for a significant fraction of the blazar’s lifetime.
Additionally, given the high proton luminosities involved, it is difficult to avoid potential depletion of the DM spike due to interactions within the jet itself. These considerations further support the expectation that neutrinos should be observed prior to any detectable signal of DM accelerated by blazars in flare. In contrast, the results we have obtained for the BBDM diffuse flux from a sample of blazars are more robust, as they rely on a broader set of sources assumed to be in steady-state activity rather than transient flaring episodes,
and the spike depletion is less severe 
because of 
the different proton spectral index, as explained in Sec.~\ref{subsec:spikedepletion}.

To conclude, we note that our analysis relies on specific choices of lepto-hadronic parameters for the blazar jets. Varying these parameters can significantly affect the interplay between the BBDM signal and the associated neutrino emission. Moreover, DM itself can have an impact on the physics of blazar jets: the same DM-proton inelastic collisions that generate neutrinos also inject $\gamma$-rays into the jet environment, potentially altering the observed electromagnetic spectrum of blazars.\footnote{The DM-induced $\gamma$-ray emission in a vector mediator model has been performed in \cite{Wang:2025ztb} for TXS 0506+056, but without taking into account the effects of propagation of the photons in the blazar environment.} Ideally, one should incorporate DM-proton interactions directly into the modelling of blazars, allowing for a unified treatment of all the relevant signals (neutrinos, $\gamma$-rays and BBDM) and to properly assess how DM itself can influence the blazar fit parameters. Intriguingly, the presence of DM could even help alleviate existing tensions in lepto-hadronic models, such as the long-standing issue of super-Eddington proton luminosities.
We leave a comprehensive study of these aspects for future work.

\section*{Acknowledgements}
We thank Simon Knapen and Paolo Salucci for useful exchanges. FS and AGDM thank respectively the Galileo Galilei Institute for theoretical physics (GGI) and the Lawrence Berkeley National Laboratory (LBNL) for the kind hospitality offered during the completion of this work. We acknowledge the use of computational resources from the parallel computing cluster of the Open Physics Hub (\href{https://site.unibo.it/openphysicshub/en}{https://site.unibo.it/openphysicshub/en}) at the Department of Physics and Astronomy of the University of Bologna.
This work was supported in part by the European Union's Horizon research and innovation programme under the Marie Skłodowska-Curie grant agreements No.~860881-HIDDeN and No.~101086085-ASYMMETRY, by COST (European Cooperation in Science and Technology) via the COST Action COSMIC WISPers CA21106, and by the Italian INFN program on Theoretical Astroparticle Physics.


\appendix

\section{Details on the elastic dark matter-nucleon scattering cross section}
\label{app:elastic}

We detail in this section the derivation of the elastic scattering cross section in the cases of scalar, pseudoscalar, vector and axial mediators for the toy models described in Sec.~\ref{sec:ToyModels}. We derive the expressions for the scattering with a generic nucleon $N$, either the proton $p$ or the neutron $n$. However, we consider the proton and neutron to have equal masses $m_N \equiv m_p \simeq m_n$ when necessary.
\paragraph{\rule{1ex}{1ex} Scalar mediator.} In the case of a scalar mediator, the Feynman amplitude reads
\begin{equation}
i\mathcal{M}_{\phi \chi N} = g_{\chi \phi}\bar{u}(k_\chi)u(p_\chi)\frac{1}{m_\phi^2+Q^2}\sum_q g_{q\phi}\frac{m_N}{m_q}f_q^NG_S(Q^2)\bar{u}(k_N)u(p_N).
\end{equation}
The function $G_S$ is the scalar form factor accounting for the internal structure of the nucleon. We approximate it with a dipole-like function $G_S(Q^2)= 1/(1+Q^2/\Lambda_p^2)^2$ \cite{Dong:1995ec} and use the value of
$\Lambda_p$, for both protons and neutrons, that stems from measurements of the averaged electric charge radius $\left\langle r_\text{E}^2\right\rangle^{1/2} = \sqrt{12}/\Lambda_p \simeq 0.85\,\text{fm}$ \cite{Gao:2021sml}, giving $\Lambda_p \simeq 0.8\,\text{GeV}$.
The quantities $f_q^N$ are defined as $f_q^N = m_q\bra{N}\bar{q}q\ket{N}/(2m_N^2)$
computed at $Q^2=0$, and can be extracted from data and lattice computation: $ f_u^p\simeq 1.97 \times 10^{-2}$, $f_u^n \simeq 1.78\times 10^{-2}$, $f_d^p\simeq 3.83 \times 10^{-2}$ and  $f_d^n\simeq 4.23 \times 10^{-2}$ \cite{Hoferichter:2023ptl}.

After some manipulations, we arrive to the following expression for the differential elastic scattering cross section
in the case of a scalar mediator, in agreement with the results of, e.g., \cite{Ema:2020ulo}:
\begin{equation}\label{eq:scalarelastic_app}
    \boxed{\frac{d\sigma^{\EL}_{\phi \chi N}}{dT_\chi d\mu_s} =\, \frac{\sigma^{\NR}_{\phi \chi N}}{s}\frac{m_\phi^4}{16 \mu_{\chi N}^2}\frac{(4m_\chi^2+Q^2)(4m_N^2+Q^2)}{(m_\phi^2+Q^2)^2}G_S^2(Q^2)\frac{\delta(\mu_s-\mu_s^{\EL}(T_N, T_\chi))}{T_\chi^{\max}(T_N)},}
\end{equation}
where $\mu_{\chi N} = m_Nm_\chi/(m_N+m_\chi)$ is the reduced mass of the system, $T_N$ is the kinetic energy of the nucleon, and $\sigma^{\NR}_{\phi \chi N}$ is the cross section computed in the non-relativistic limit at zero momentum transfer $Q^2=0$ and for $s = (m_N+ m_\chi)^2$, namely
\begin{equation}
    \sigma^{\NR}_{\phi \chi N} \equiv \frac{g^2_{\chi \phi}g^2_{N\phi}}{\pi} \frac{\mu_{\chi N}^2}{m_\phi^4} \simeq   g_{\chi \phi}^2 g_{N \phi}^2 \left(\frac{\mu_{\chi N}}{1\,\text{GeV}}\right)^2\left(\frac{1\,\text{GeV}}{m_\phi}\right)^4 1.08\times 10^{-30}\text{cm}^2,
\end{equation}
with $g_{N\phi}\equiv m_N(g_{d \phi} f_d^N/m_d + g_{u \phi} f_u^N/m_u)$ and $m_u = 2.2$~MeV, $m_d = 4.7$~MeV~\cite{ParticleDataGroup:2024cfk}.

\paragraph{\rule{1ex}{1ex} Pseudoscalar mediator.} In the pseudoscalar case, the Feynman amplitude of the elastic process reads:
\begin{equation}
    i\mathcal{M}_{a \chi N} = g_{\chi a}\bar{u}(k_\chi)\gamma_5u(p_\chi)\frac{i}{m_a^2 +Q^2}\frac{m_N}{\hat{m}}\sum_q g_{qa}G^q_{5, N}(Q^2)\bar{u}(k_N)\gamma_5 u(p_N).
\end{equation}
where $G^q_{5,\,N}(Q^2)$ is the pseudoscalar form factor associated to the quark $q=u,\,d$ and nucleon $N$. 
In the above expression we considered the masses of the up and down quarks to be equal,
and approximated their values with their averaged mass, that is $m_u\simeq m_d \simeq \hat{m} = (m_u+m_d)/2$.
The combination $\sum_q g_{qa}G_{5,\,N}^q(Q^2)$ can be computed as a linear combination of the 
form factors $G^{u-d}_{5,\,N}(Q^2)$ and $G^{u+d}_{5,\,N}(Q^2)$ associated respectively to the isovector current $j_{u-d}^\mu = \bar{u} \gamma^\mu \gamma^5 u - \bar{d} \gamma^\mu \gamma^5 d$
and the isoscalar current $j_{u+d}^\mu = \bar{u} \gamma^\mu \gamma^5 u + \bar{d} \gamma^\mu \gamma^5 d$. the nuclear isospin symmetry, which heuristically means to exchange $u\leftrightarrow d$ to get $p\leftrightarrow n$, implies that $G^{u+d}_5 \equiv G^{u+d}_{5,\,p} \simeq G^{u+d}_{5,\,n}$, $G^{u-d}_5 \equiv G^{u-d}_{5,\,p} \simeq -G^{u-d}_{5,\,n}$. The partially conserved axial current (PCAC) relation
imposes that \cite{Alexandrou_2021, Alexandrou:2023qbg}
\begin{eqnarray}
G^{u-d}_5(Q^2) &=& G^{u-d}_A(Q^2) - \frac{Q^2}{4m_p^2}G^{u-d}_P(Q^2),\\
G^{u+d}_5(Q^2) &\simeq& G^{u+d}_A(Q^2) - \frac{Q^2}{4m_p^2}G^{u+d}_P(Q^2),
\end{eqnarray}
where $G_A^{u-d}(Q^2)$ and $G_A^{u+d}(Q^2)$ are the axial form factors, while $G_P^{u-d}(Q^2)$ and $G_P^{u+d}(Q^2)$ are the induced pseudoscalar form factors, respectively associated
to the isovector and isoscalar currents. In the isoscalar case we have neglected the contribution of the sub-leading anomalous gluon form factor.
The form factors $G_A^{u-d}(Q^2)$ and $G_A^{u+d}(Q^2)$ can be written under the dipole ansatz:
\begin{equation}\label{eq:GAu_pm_d}
G_A^{u\pm d}(Q^2) = \frac{g_A^{u\pm d}}{(1+Q^2/M_A^2)^2},
\end{equation}
with the axial charges $g_A^{u-d}$ and $g_A^{u+d}$ and axial mass $M_A$ 
extracted from lattice computations as
$g_A^{u-d} = 1.25$, $g_A^{u+d} = 0.44$ and $M_A = 1.2\,\text{GeV}$ \cite{Alexandrou_2021, Alexandrou:2023qbg}. The induced pseudoscalar form factors $G_P^{u-d}(Q^2)$ and $G_P^{u+d}(Q^2)$ can be written according
to the pole mass ansatz
\begin{equation}\label{eq:GPu_pm_d}
G_P^{u\pm d}(Q^2) =G_A^{u\pm d}(Q^2)\frac{C_{u\pm d}^2}{Q^2+M_{u\pm d}^2}
\end{equation}
where $C_{u-d} = 4m_N^2$, $C_{u+d} = 0.9\,\text{GeV}^2$, $M_{u-d} = m_\pi\simeq 0.137\,\text{GeV}$ and $M_{u+d} = 0.330\,\text{GeV}$ \cite{Alexandrou_2021, Alexandrou:2023qbg}, so that the pseudoscalar form factors
associated to the isovector and isoscalar currents are given by 
\begin{eqnarray}
G_5^{u-d}(Q^2) &=& \frac{g_A^{u-d}}{(1+Q^2/m_A^2)^2} \frac{m_\pi^2}{Q^2+m_\pi^2},\\
G_5^{u+d}(Q^2) &=& \frac{g_A^{u+d}}{(1+Q^2/m_A^2)^2}\left(1 - \frac{Q^2}{4m_N^2}\,\frac{C_{u+d}}{Q^2 + M_{u+d}^2}\right).
\end{eqnarray}
and the linear combinations of interest to us reads 
\begin{eqnarray}
G_{5, \,p}(Q^2) &\equiv& \sum_q g_{qa}G^q_{5,\,p}(Q^2) = \frac{g_{u a}}{2}[G_5^{u+d}+G_5^{u-d}] + \frac{g_{d a}}{2}[G_5^{u+d}-G_5^{u-d}]\\
G_{5, \,n}(Q^2) &\equiv& \sum_q g_{qa}G^q_{5,\,n}(Q^2) = \frac{g_{u a}}{2}[G_5^{u+d}-G_5^{u-d}] + \frac{g_{d a}}{2}[G_5^{u+d}+G_5^{u-d}].
\end{eqnarray}
We note that in this case the cross section 
vanishes for $Q^2 = 0$. 
Hence, we do not define any \virg{non-relativistic} cross section in this case. 
Then, we obtain 
\begin{equation}\label{eq:pseudoscalarelastic_app}
   \boxed{\frac{d\sigma^{\EL}_{a \chi N}}{dT_\chi d\mu_s} = \frac{
   g_{\chi a}^2 g_{N a }^2}{16 \pi s}
   \frac{(Q^2)^2}{(m_a^2+Q^2)^2}\widehat{G}_{5,\,N}^2(Q^2)\frac{\delta(\mu_s-\mu_s^{\EL}(T_N, T_\chi))}{T_\chi^{\max}(T_N)},}
\end{equation} 
where $g_{Na} \equiv m_N[g_{ua}(g_A^{u+d} \pm g_A^{u-d})+g_{da}(g_A^{u+d}\mp g_A^{u-d})]/(2\hat{m})$, with $+$ ($-$) holding for $N=p$ ($N=n$), and $\widehat{G}_{5,\,N} (Q^2)\equiv m_N G_{5,\,N}(Q^2)/(\hat{m}g_{Na})$.

\paragraph{\rule{1ex}{1ex} Vector mediator.} In the case of a vector mediator, the Feynman amplitude reads
\begin{equation}
i\mathcal{M}_{V \chi N} = g_{\chi V}\,\bar{u}(k_\chi)\gamma^\mu u(p_\chi)\frac{-g_{\mu\nu} + q_\mu q_\nu/m_{V}^2}{m_{V}^2+Q^2}\langle N(k_N)| v^\nu_{V, Q}|N(p_N)\rangle,
\end{equation}
where $q = p_\chi-k_\chi = k_N - p_N$ is the 4-momentum transfer and $v^\nu_{V, Q} = g_{u V} \bar{u}\gamma^\nu u + g_{d V} \bar{d}\gamma^\nu d$ is the quark vector current associated to the $V$ boson.
The hadronic matrix element can be decomposed as
\begin{equation}
    \langle N(k_N)| v^\mu_{V, Q}|N(p_N)\rangle = \bar{u}(k_N)\left[ \gamma^\mu F_1^{V,\,N}(Q^2) + \frac{i}{2m_p} \sigma^{\mu\nu}q_\nu F_2^{V,\,N}(Q^2)\right]u(p_N)
\end{equation}
where $F_{1,2}^{V,\,N}(Q^2)$ are the vector form factors associated to the nucleon $N$ \footnote{
Any term proportional to $Q^\mu$ has to vanish due to vector current conservation \cite{Bilenky:1994bs}.
}. 
The current $v_{V, Q}^\mu$ can be written as a linear combination of the electromagnetic current $j_{\EM, Q}^\mu \equiv (2/3)\bar{u}\gamma^\mu u - (1/3)\bar{d}\gamma^\mu d$ 
and the 3-isospin current $v^\mu_{3, Q} = (1/2)(\bar{u}\gamma^\mu u - \bar{d}\gamma^\mu d)$ via
\begin{equation}
v^\mu_{V, Q} = 3(g_{uV} +g_{dV})j_{\EM, Q}^\mu - 2 (g_{uV} + 2g_{dV}) v^\mu_{3, Q}.
\end{equation}
Consequently, assuming nuclear isospin symmetry, the vector form factors can be expressed in terms of the electromagnetic ones (see, e.g., \cite{Giunti:2007ry}) as
\begin{eqnarray}
F_{1,2}^{V,\,p}(Q^2) &=& g_{pV} F_{1,2}^{p}(Q^2) + g_{nV} F_{1,2}^{n}(Q^2),\\
F_{1,2}^{V,\,N}(Q^2) &=& g_{nV} F_{1,2}^{p}(Q^2) + g_{pV} F_{1,2}^{n}(Q^2)
\end{eqnarray}
where $F_{1,2}^{p,n}(Q^2)$ are the proton and neutron electromagnetic form factors, $g_{pV} \equiv 2g_{uV} + g_{dV}$
and $g_{nV} \equiv g_{uV} + 2g_{dV}$. 
The latter can be expressed in terms of the electric and magnetic form factors $G_{E,M}^{p,n}(Q^2)$,
which are defined as \cite{PhysRev.119.1105, RevModPhys.35.335, Perdrisat:2006hj}
\begin{eqnarray}
G_\text{E}^{p,n}(Q^2) &\equiv& F_1^{p,n}(Q^2) - \frac{Q^2}{4m_N^2}F_2^{p,n}(Q^2),\\
G_\text{M}^{p,n}(Q^2) &\equiv& F_1^{p,n}(Q^2) +F_2^{p,n}(Q^2),
\end{eqnarray}
satisfying $G_\text{E}^p(0) = 1$, $G_\text{E}^n(0) = 0$, $G_\text{M}^p(0) = \mu_p/\mu_N$ and $G_\text{M}^n(0) = \mu_n/\mu_N$,
with $\mu_N$ being the nuclear magneton, while $\mu_{p} \simeq 2.79\mu_N$ and $\mu_n\simeq -1.91\mu_N$ respectively the magnetic moments of the proton and of the neutron~\cite{Workman:2022ynf}.
We then approximate the from factors $G_{E,M}(Q^2)$ with a dipole-like function $G^{p,n}_{E,M}(Q^2) \simeq G^{p,n}_{E,M}(0)/(1+Q^2/\Lambda_p^2)^2$ \cite{Alarcon:2020kcz, Gao:2021sml}.
The differential cross section in this case reads
\begin{equation}\label{eq:vectorelastic_app}
\boxed{\frac{d\sigma^{\EL}_{V \chi N}}{dT_\chi d\mu_s} = \frac{\sigma^{\NR}_{V \chi N}}{s} \frac{m_N^4}{4\mu_{\chi N}^2} \left[ A^{V,\,N}(Q^2) +  \frac{(s-u)^2}{m_N^4} C^{V,\,N}(Q^2)+ \frac{m_\chi^2}{m_N^2}D^{V,\,N}(Q^2)\right]\frac{\delta(\mu_s-\mu_s^{\EL}(T_N, T_\chi))}{T_\chi^{\max}(T_N)(1+Q^2/m_{V}^2)^2},}
\end{equation}
where $\sigma^{\NR}_{V \chi N} \equiv g_{\chi V}^2 g_{NV}^2 \mu_{\chi N}^2/(\pi m_{V}^4)$ is total cross section in the non-relativistic limit and for $Q^2=0$ (which is non-zero only if $g_{dV} \neq -2g_{uV}$),
and $A^{V,\,N}(Q^2)$, $C^{V,\,N}(Q^2)$ and $D^{V,\,N}(Q^2)$ are the following functions of the normalised form factors $\widehat{F}_{1,2}^{V N}(Q^2) = F_{1,2}^{V,\,N}(Q^2)/g_{NV}$:
\begin{eqnarray}
A^{V,\,N}(Q^2) &=& -\frac{Q^2}{m_N^2}\,\Bigg\{\left(1-\frac{Q^2}{4m_N^2}\right)\left[\left(\widehat{F}_1^{V,\,N}(Q^2)\right)^2 - \frac{Q^2}{4m_N^2}\left(\widehat{F}_2^{V,\,N}(Q^2)\right)^2\right] \\
\nonumber &&~~~~~~~~~ -\,\frac{Q^2}{m_N^2}\widehat{F}_1^{V,\,N}(Q^2)\widehat{F}_2^{V,\,N}(Q^2)\Bigg\}\\
C^{V,\,N}(Q^2) &=& \frac{1}{4}\left[\left(\widehat{F}_1^{V,\,N}(Q^2)\right)^2 + \frac{Q^2}{4m_N^2}\left(\widehat{F}_2^{V,\,N}(Q^2)\right)^2 \right],\\
D^{V,\,N}(Q^2) &=& -\frac{Q^2}{m_N^2}\left[\widehat{F}_1^{V,\,N}(Q^2)+\widehat{F}_2^{V,\,N}(Q^2)\right]^2.
\end{eqnarray}
As can be checked, the above expressions agree with the corresponding ones associated to neutral current neutrino-proton elastic scattering
reported, e.g., in \cite{Giunti:2007ry} (see also \cite{Formaggio:2012cpf}) when the axial contribution is neglected,
but here we are keeping the contributions of order $\mathcal{O}(m_\chi^2/m_N^2)$. We note that terms $\mathcal{O}(Q^2/m_V^2)$ in the form factors cancel and thus do not appear in the expressions above.

\paragraph{\rule{1ex}{1ex} Axial mediator.} In the case of an axial mediator, the Feynman amplitude reads
\begin{equation}
i\mathcal{M}_{V' \chi N} = g_{\chi V'}\,\bar{u}(k_\chi)\gamma^\mu\gamma^5 u(p_\chi)\frac{-g_{\mu\nu} + q_\mu q_\nu/m_{V'}^2}{m_{V'}^2+Q^2}\langle N(k_N)| a^\nu_{V', Q}|N(p_N)\rangle,
\end{equation}
where $a^\mu_{V', Q} = g_{u V'} \bar{u}\gamma^\mu\gamma^5 u + g_{d V'}\bar{d}\gamma^\mu\gamma^5 d$ is the quark axial current associated to the $V'$ boson.
We decompose the hadronic matrix element as
\begin{equation}
    \langle N(k_N)| a^\mu_{V', Q}|N(p_N)\rangle = \bar{u}(k_N)\left[ \gamma^\mu G_A^{V'\,N}(Q^2) -  \frac{q^\mu}{2m_p}G_P^{V'\,N}(Q^2)\right]\gamma^5 u(p_N)
\end{equation}
with the functions $G_{A,P}^{V' N}(Q^2)$ being the associated axial and pseudoscalar form factors\footnote{
We neglect the contribution from any term proportional to $\sigma^{\mu\nu}Q_\nu$ as it would violate 
$\mathcal{G}$ conjugation, i.e. a combination of charge conjugation and a rotation by $\pi$ in isospin space \cite{Schindler_2007}.}.
The current $a^\mu_{V', Q}$ can be decomposed in terms of the isoscalar and isovector currents
\begin{equation}
a^\mu_{V', Q} = \frac{g_{uV'}}{2}(j^\mu_{u+d} + j^\mu_{u-d}) + \frac{g_{dV'}}{2}(j^\mu_{u+d} - j^\mu_{u-d}),
\end{equation}
and, correspondingly, the form factors can be expressed in terms of $G_{A,P}^{u\pm d}(Q^2)$ defined in Eqs.~\eqref{eq:GAu_pm_d} and \eqref{eq:GPu_pm_d} as
\begin{eqnarray}
G_{A,P}^{V',\,p}(Q^2) = \frac{g_{uV'}}{2}[G_{A,P}^{u+d}(Q^2) + G_{A,P}^{u-d}(Q^2)] + \frac{g_{dV'}}{2}[G_{A,P}^{u+d}(Q^2) - G_{A,P}^{u-d}(Q^2)],\\
G_{A,P}^{V',\,n}(Q^2) = \frac{g_{uV'}}{2}[G_{A,P}^{u+d}(Q^2) - G_{A,P}^{u-d}(Q^2)] + \frac{g_{dV'}}{2}[G_{A,P}^{u+d}(Q^2) + G_{A,P}^{u-d}(Q^2)],
\end{eqnarray}
The differential cross section in this case reads
\begin{equation}\label{eq:vectoraxialelastic_app}
\boxed{\frac{d\sigma^{\EL}_{V' \chi N}}{dT_\chi d\mu_s} = \frac{\sigma^{\NR}_{V' \chi N}}{s} \frac{m_N^4}{4\mu_{\chi N}^2}\left[ A^{V',\,N}(Q^2) +  \frac{(s-u)^2}{m_N^4} C^{V',\,N}(Q^2)+ \frac{m_\chi^2}{m_N^2}D^{V',\,N}(Q^2)\right]\frac{\delta(\mu_s-\mu_s^{\EL}(T_N, T_\chi))}{T_\chi^{\max}(T_N)(1+Q^2/m_{V'}^2)^2},}
\end{equation}
where $\sigma^{\NR}_{V' \chi N} \equiv g_{NV'}^2 {g^2_{\chi V'}} \mu_{\chi N}^2/(\pi m_{V'}^4)$ is the total cross section computed in the non-relativistic limit and for $Q^2=0$, with $g_{NV'}\equiv (\sqrt{3}/2)[g_{uV'}(g_{A}^{u+d} \pm g_{A}^{u-d}) + g_{dV'}(g_{A}^{u+d}\mp g_{A}^{u-d})]$, with $+$ ($-$) holding for $N=p$ ($N=n$),
while $A^{V',\,N}(Q^2)$, $C^{V',\,N}(Q^2)$ and $D^{V',\,N}(Q^2)$ are the following functions of the normalised form factors $\widehat{G}_{A,P}^{V',\,N}(Q^2)\equiv G_{A,P}^{V',\,N}(Q^2)/g_{NV'}$:
\begin{eqnarray}
A^{V',\,N}(Q^2) &=& \frac{Q^2}{m_N^2}\left(1+\frac{Q^2}{4m_N^2}
\right)\left(\widehat{G}_A^{V',\,N}(Q^2)\right)^2,\\
C^{V',\,N}(Q^2) &=& \frac{1}{4}\left(\widehat{G}_A^{V',\,N}(Q^2)\right)^2,\\
D^{V',\,N}(Q^2) &=& 8\left[1+\frac{Q^2}{8m_N^2} + \frac{Q^2}{m_V'^2} + \frac{1}{2}\left(\frac{Q^2}{m_V'^2}\right)^2\right]\left(\widehat{G}_A^{V',\,N}(Q^2)\right)^2 +\\
\nonumber && ~ + \, 2\,\frac{Q^2}{m_N^2}\left(1+\frac{Q^2}{m_V'^2}\right)^2 \widehat{G}_A^{V',\,N}(Q^2)\widehat{G}_P^{V',\,N}(Q^2) +\\
\nonumber && ~ + \,\frac{1}{4}\left(\frac{Q^2}{m_N^2}\right)^2\left(1+\frac{Q^2}{m_V'^2}\right)^2\left(\widehat{G}_P^{V',\,N}(Q^2)\right)^2.
\end{eqnarray}
It can be checked that the above expressions agree with the corresponding ones associated to neutral current neutrino-proton elastic scattering
reported, e.g., in \cite{Giunti:2007ry} (see also \cite{Formaggio:2012cpf}) when the vector contribution is neglected,
but here we are keeping the contributions of order $\mathcal{O}(m_\chi^2/m_N^2, Q^2/m_{V'}^2)$.

\section{Deep inelastic scattering cross sections}
\label{app:DIS}
We give here more details on the derivation of the DIS cross section that we reported in the main text. We consider a nucleon $N=p,n$ scattering inelastically off a DM particle.
If the momentum transfer is large enough, the DM particles interacts directly
with the quarks that constitute the nucleon via DIS. 
The process is diagrammatically represented below \cite{Ellis:2016jkw}
\begin{center}
\begin{tabular}{cc}
\begin{tikzpicture}[baseline=(current bounding box.center)]
\begin{feynman}
\vertex(V1);
\vertex(p1)[left=2cm of V1];
\vertex(k3)[right=2cm of V1];
\node[blob, below = 1.5cm of V1](V2);
\path (V2.+180) ++ (00:-1.58) node[vertex] (p2);
\path (V2.+145) ++ (00:-1.65) node[vertex] (p2up);
\path (V2.+220) ++ (00:-1.68) node[vertex] (p2down);
\vertex(kXup)[right=1.95cm of V2];
\path (V2.-45) ++ (00:1.68) node[vertex] (kXdown);
\path (V2.+45) ++ (45:1) node[vertex] (p2q);
\path (V2.+45) ++ (45:1) node[vertex] (p2q);
\vertex[right=1cm of p2q](k4q);
\diagram*{
(p1)--[fermion, edge label=\(\chi\), pos = .6](V1),
(p2)--[fermion](V2),
(p2up)--[with arrow=0.48](V2.+145),
(p2down)--[with arrow=0.47](V2.+220),
(V1)--[edge label' = \(Y\), color = nicepurple](p2q),
(V1) --[fermion, edge label=\(\chi\), pos = .6](k3),
(V2.+45) --[fermion, edge label' = \(q\,\text{, }\bar{q}\)](p2q)--[fermion, edge label = \(q\,\text{, }\bar{q}\)](k4q),
(V2) --[fermion](kXup),
(V2.-45) --[with arrow=0.53](kXdown),
};
\vertex(X1) [right=.2cm of k4q];
\vertex(X2) [right=.2cm of kXdown];
\draw [decoration={brace}, decorate] (X1) -- (X2) node [pos=0.5, right = .2em] {\(X\)};
\vertex(N1) [left=.1cm of p2up];
\vertex(N2) [left=.1cm of p2down];
\draw [decoration={brace, mirror}, decorate] (N1) -- (N2) node [pos=0.5, left = .2em] {\(N\)};
\end{feynman}
\end{tikzpicture}
\end{tabular}
\end{center}
where $Y=\phi, a,\,V,\,V'$ denotes either the scalar, pseudoscalar,
vector or axial mediator, $N$ the initial nucleon and $X$ the outgoing hadronic state.
We fix the momenta of the initial DM and quark respectively as $p_\chi =(E_\chi, \vec{p}_\chi)$ and $p_q = (E_q, \vec{p}_q)$, and, analogously, for the outgoing DM and quark $k_\chi = (E'_\chi, \vec{k}_\chi) $ and $k_q = (E'_q, \vec{k}_q
)$. We furthermore neglect the quark masses. The DIS is typically studied in the context of neutrino interactions, in the frame of reference in which the nucleon is at rest $\vec{p}_N = 0$ (LAB) and 
neglecting neutrino masses. A comprehensive discussion
of the SM neutrino-nucleon interaction can be found, e.g., in \cite{Giunti:2007ry}. Here, we will follow an approach similar to that outlined in \cite{DeMarchi:2024zer}, namely we derive
the DIS cross section in the LAB frame and write it in terms of Lorentz invariants, so to render any change of reference frame immediate. We will also keep
the DM masses non-zero as they are not necessarily negligible compared to the other mass and energy scales involved in the process.
It will prove useful to define also the initial nucleon momentum $p_N = p_q/a = (E_N, \vec{p}_N)$ and the 4-momentum transfer $q = p_\chi-k_\chi$.
We denote the squared centre-of-mass energy and momentum transfer respectively as $s = (p_N + p_\chi)^2$ and $Q^2 = -q^2 $. We also define the following Lorentz invariant quantities:
the \textit{energy transfer} $   \nu \equiv p_N\cdot q/m_N$, 
the \textit{inelasticity} parameter $ 
    y \equiv p_N\cdot q/(p_N\cdot p_\chi) = 2 \nu m_N/(s-m_N^2-m_\chi^2)$ 
and the \textit{Bjorken scaling variable} $
    x \equiv Q^2/(2 p_N\cdot q)= Q^2/[(s-m_N^2-m_\chi^2)\, y]$.
    Moreover, we have the following set of relations in the LAB frame:
\begin{eqnarray}
 Q^2 &\overset{\mathsmaller{\mathrm{LAB}}}{=}& -2m_\chi^2 + 2E_\chi E_\chi' - 2 |\vec{p}_\chi| |\vec{k}_\chi| \cos\theta,\\
 E_\chi &\overset{\mathsmaller{\mathrm{LAB}}}{=}& (s-m_N^2-m_\chi^2)/(2m_N),\\
 E_\chi' &\overset{\mathsmaller{\mathrm{LAB}}}{=}& E_\chi(1-y),\\
  v_\text{rel} &\overset{\mathsmaller{\mathrm{LAB}}}{=}& |\vec{p}_\chi|m_N/(E_\chi E_N),
\end{eqnarray}
where $\theta$ is the scattering angle in the LAB frame and $v_\text{rel}$ is the relative velocity between the DM and the nucleon. 
The master formula for the DIS differential cross section can be written as \cite{Peskin:1995ev}:
\begin{equation}\label{eq:xsec_parton}
    d\sigma^{\DIS}_{Y \chi N} =\, \frac{1}{4E_\chi E_N v_\rel}\frac{d^3k_\chi}{(2\pi)^32 E_\chi'}\sum_{\kappa = q,\bar{q}} \int_0^1 d\xi f_\kappa^{N}(a, Q^2)\frac{E_N}{E_\kappa}\int \frac{d^3k_a}{(2\pi)^32E_\kappa'}(2\pi)^4\delta^{(4)}(\xi p_N + q - k_\kappa)|\overline{\mathcal{M}_{Y\chi}^{\quark}}|^2,
\end{equation}
where $|\overline{\mathcal{M}_{Y\chi}^{\quark}}|^2$ is the squared Feynman amplitude of the
DM-quark scattering averaged over the initial spins, while
$f_{q(\bar{q})}^{N}$ is the Lorentz scalar \textit{parton distribution function} (PDF) 
for the (anti)quark $q=u,\,c,\,t,\,d,\,s,\,b$ ($\bar{q} = \bar{u},\,\bar{c},\,\bar{t},\,\bar{d},\,\bar{s},\,\bar{b}$).
\begin{figure}[t!]
    \centering
    
    \includegraphics[width=0.46\textwidth]{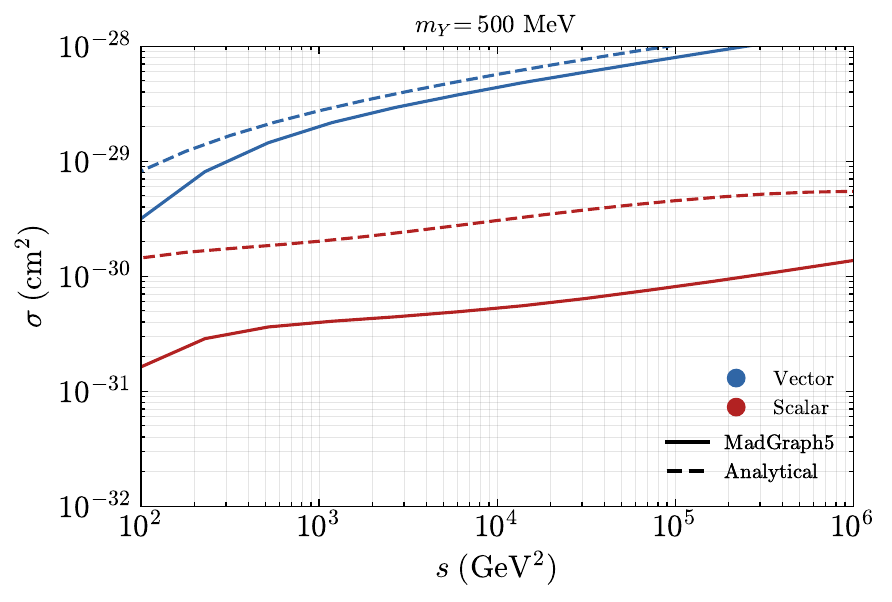}
    \includegraphics[width=0.46\textwidth]{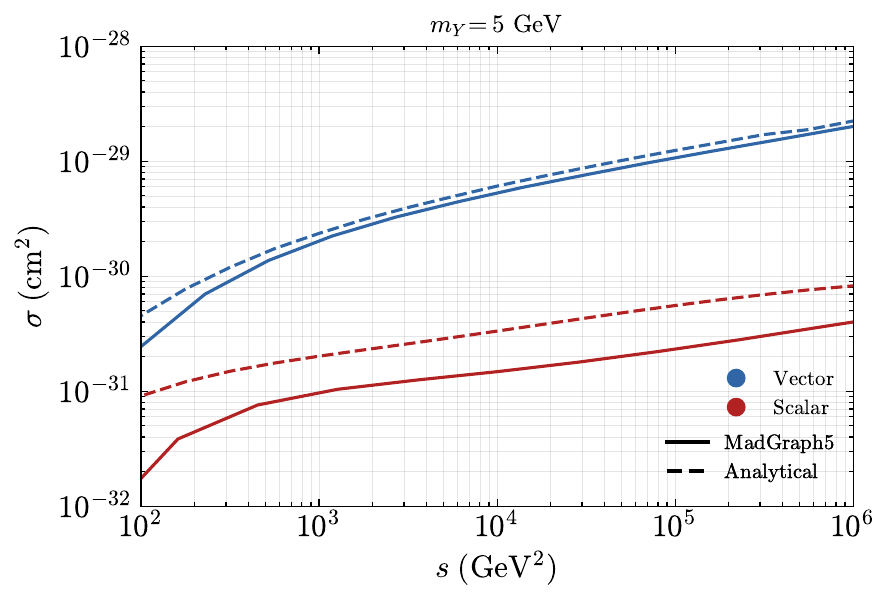}
    \includegraphics[width=0.46\textwidth]{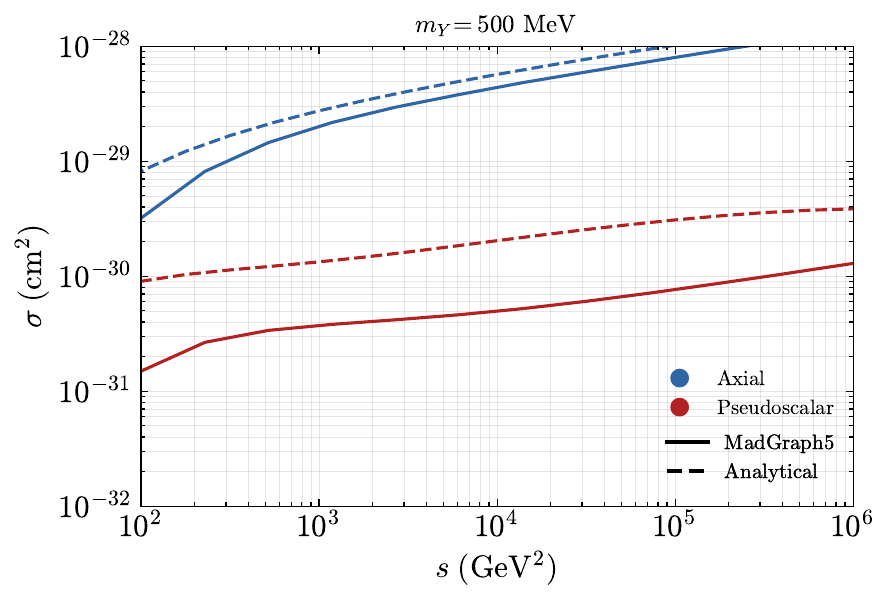}
    \includegraphics[width=0.46\textwidth]{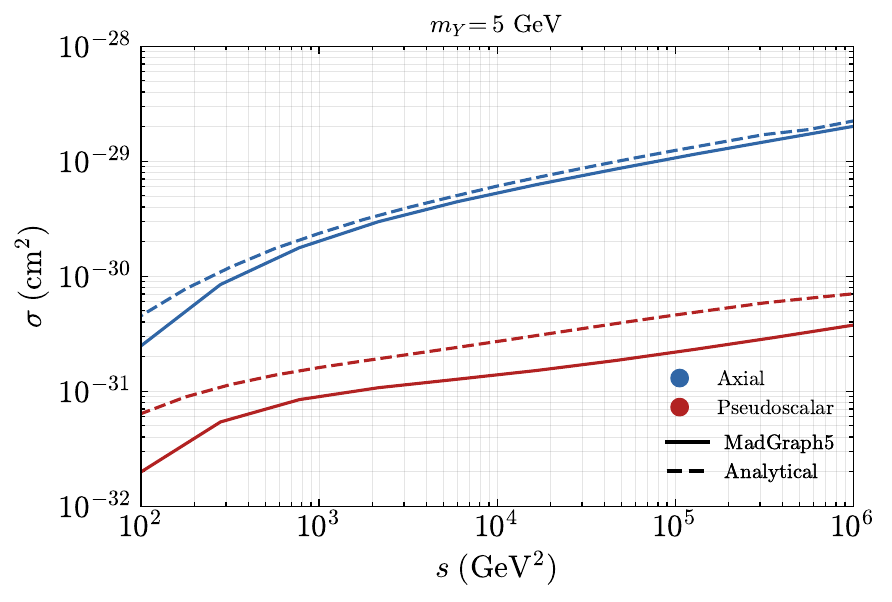}

    \caption{Comparison between the integrated inelastic cross sections (integration of the expressions in Eqs.~(\ref{eq: analytic scalar mediator cross section}-\ref{eq: analytic vector mediator cross section})) and the same quantities as evaluated by \textsc{Madgraph5}, for a mediator of mass $m_Y = 500$~MeV (5~GeV) in the left (right) panels. We have fixed $g_{\chi V} = g_{q V} = 1$. The upper (bottom) panels show the comparison for the case of vector and scalar (axial and pseudoscalar) mediators. 
    }
    \label{fig:cross section comparison}
\end{figure}
The expression above is Lorentz invariant except for the factor $E_\chi E_N v_\text{rel}$. This factor is, however, invariant in all the reference frames in which the particles are collinear (see, e.g., \cite{Peskin:1995ev}), so that
if we find a Lorentz invariant expression for the differential cross section in the rest frame of the proton, the same expression can 
be used in the rest frame in which the DM particle is at rest instead. We have that
\begin{equation}\label{eq:transformation}
    \frac{4\pi}{4E_\chi E_Nv_\text{rel}}\frac{d^3k_\chi}{(2\pi)^32E_\chi'} \overset{\mathsmaller{\mathrm{LAB}}}{=} \frac{d\nu dQ^2}{16\pi m_N (E_\chi^2-m_\chi^2)} = \frac{y}{8\pi}\,\frac{(Q^2)^2}{(Q^2)^2 - 4m_N^2 m_\chi^2 x^2 y^2}dxdy,
\end{equation}
where we have used the relations
\begin{equation}
d^3k_\chi \overset{\mathsmaller{\mathrm{LAB}}}{=} \frac{2\pi E_\chi'}{2|\vec{p}_\chi|}d\nu dQ^2\quad \text{and}\quad d\nu dQ^2 = \frac{(Q^2)^2}{2m_Nx^2y} dx dy.
\end{equation}
Then, the $\delta$-function in the DIS differential cross section can be integrated out via the four integrals over
the final quark's 3-momentum $\vec{k}_\kappa$ and the variable $\xi$ to give:
\begin{equation}
d\sigma^{\DIS}_{Y\chi N} \overset{\mathsmaller{\mathrm{LAB}}}{\simeq} \frac{\sum_\kappa f^N_\kappa(x,\, Q^2)|\overline{\mathcal{M}_{Y\chi}^{\quark}}|^2}{32\pi m_N Q^2 (E_\chi^2-m_\chi^2)} d\nu dQ^2
= \frac{y}{16 \pi}\frac{Q^2 \sum_{\kappa} f_\kappa^N(x,\,Q^2)|\overline{\mathcal{M}_{Y\chi}^{\quark}}|^2}{(Q^2)^2-4m_N^2m_\chi^2 x^2 y^2}  dx dy,
\end{equation}
where we have neglected quarks' masses, so that $\nu \overset{\mathsmaller{\mathrm{LAB}}}{=} E_\chi-E_\chi' \simeq E_\kappa'$, $\kappa = q,\,\bar{q}$. This last expression reproduces Eq.~\eqref{eq:Dismaster} in the main text for $N=p$.

We validate our analytical expressions for the inelastic scattering cross sections by comparing them with numerical results obtained using \textsc{Madgraph5}~\cite{Alwall:2014hca} for a mediator mass of $m_Y = 500\,\text{MeV}$ and $5~\mathrm{GeV}$, and show the result of such a comparison in Fig.~\ref{fig:cross section comparison}.
 We believe that the discrepancies between the numerical and analytical estimates of the cross section, which depend on the center-of-mass energy and mediator mass, are due to cuts implemented internally by \textsc{Madgraph5}, not captured by our analytical computation.

\section{A different approach for the Earth attenuation}
\label{app:EarthAtt}

We now investigate the effects of the Earth attenuation on DM in our toy models following the approaches of, e.g., \cite{Bringmann:2018cvk, Ema:2018bih, Ema:2020ulo, Das:2024ghw}. In particular, we approximate the DM energy loss, $dT_\chi^x$, during a single scattering event occurring within the infinitesimal distance interval $[x, x+dx]$, with its average value. We also assume the scatterings to occur only on protons and neutrons at rest. Then, the resulting DM kinetic energy at distance $x$, $T_\chi^x$, is controlled by the differential equation:
\begin{equation}
\label{eq:Attenuation}
\frac{d T^x_\chi}{dx}=-\sum_{N = p,\, n} n_N  \int_0^{T_N^{\max}(T_\chi^x)}dT_N\, T_N  \frac{d \sigma_{Y\chi N}^{\EL+ \DIS}}{d T_N }(T_N,T_\chi^x)\, ,
\end{equation}
where $n_N$ is the number density of the nucleon $N =p,n$ and 
\begin{equation}
T_N^{\max}(T_\chi^x) 
= \frac{2m_{N} T_\chi^x\left(T_\chi^x + 2m_\chi \right)}{2m_{N} T_\chi^x+ (m_\chi+m_{N})^2}\,,
\end{equation}
with $m_N$ the nucleon's mass. By solving numerically Eq.~\eqref{eq:Attenuation}, for given DM mass, couplings $g_{\chi Y}$, $g_{qY}$, and distance $x$, with the initial condition $T_\chi^{x=0} = T_\chi^{\zero}$, we extrapolate $T_\chi^x(T_\chi^{\zero})$ and $dT_\chi^x/dT_\chi^{\zero}$, as well as $T_\chi^{\zero}(T_\chi^x)$ and $dT_\chi^{\zero}/dT_\chi^x$. We also keep track of the average number of scatterings taking place over a distance $x$, $N^x_{\rm sct}$, by solving
\begin{equation}\label{eq:Nsct}
    \frac{dN_{\rm sct}^x}{dx} = \sum_{N = p,\, n} n_N \sigma_{Y\chi N}^{\EL + \DIS}(T_\chi^x).
\end{equation}
together with Eq.~\eqref{eq:Attenuation} and the initial condition $N_{\rm sct}^{x=0} = 0$.

To ensure the feasibility of the numerical evaluation, we adopt all the approximations as in our procedure described in the main text. Namely, we take the nucleon density $n_N \equiv n_p \simeq n_n \simeq \rho_{p+n}/(2m_N)$, with $\rho_{p+n} \simeq 2.7\,\text{g}/\text{cm}^3$ and $m_N \equiv m_p \simeq m_n$; we focus only the elastic scattering contribution with the form factor evaluated at zero momentum transfer; we consider equal DM-proton and DM-neutron cross sections $\sigma_{ Y\chi  N}^{\EL}\equiv \sigma_{ Y\chi  p}^{\EL} \simeq \sigma_{ Y \chi  n}^{\EL}$. Furthermore, we neglect the change of DM direction after each scattering and compute the differential DM flux at depth $x$ assuming flux conservation as
\begin{equation}
\frac{d\Phi_\chi^x}{dT_\chi^x}=\frac{dT_\chi^{\zero}}{dT_\chi^x}\frac{d\Phi_\chi}{dT_\chi^{\zero}}\Bigg|_{T_\chi^{\zero}=T_\chi^{\zero}(T_\chi^x)}\,.\label{eq:fluxattenuation}
\end{equation}

\begin{figure}[t!]
        \centering
        \includegraphics[width = 0.48\textwidth]{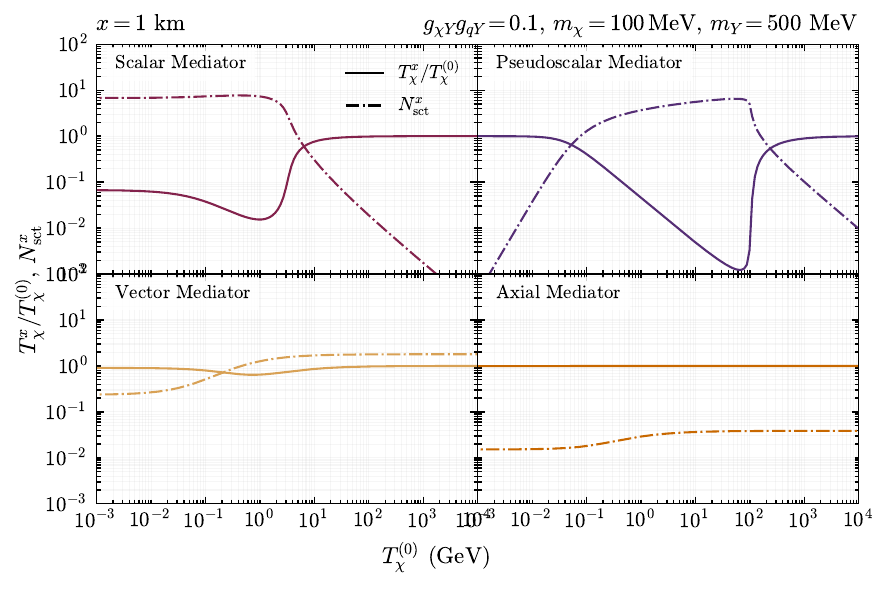}
        \includegraphics[width = 0.48\textwidth]{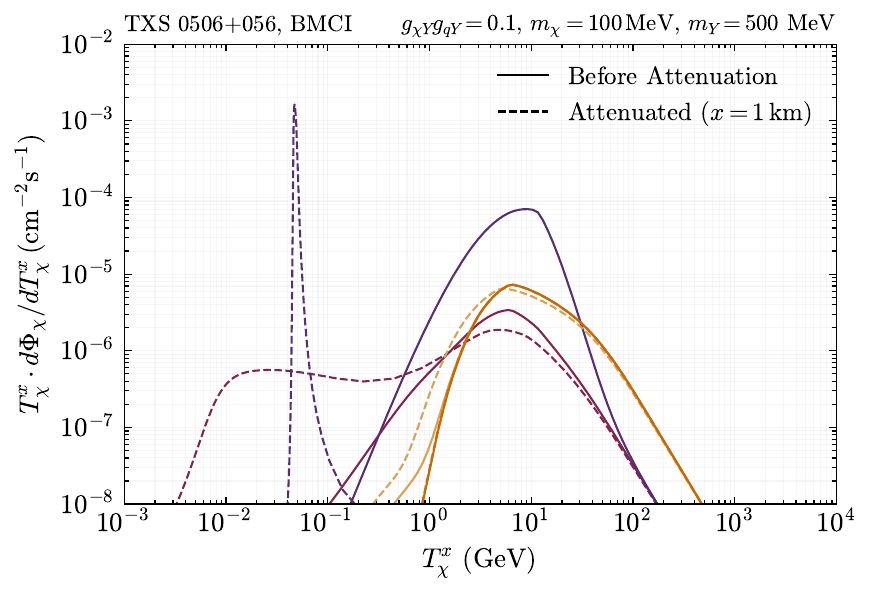}
        \caption{Effect of the Earth attenuation. Left panel: ratio of the attenuated DM kinetic energy to its value at the Earth surface  $T_\chi^x/T_\chi^{\zero}$ (solid) and the DM average number of scatterings $N_{\rm sct}^x$ (dashed) for a scalar (top left), pseudoscalar (top right), vector (bottom left) and axial (bottom right) mediator. Right panel: BBDM flux from TXS 0506+056 before (solid) and after (dot-dashed) Earth attenuation. We set $x=1\,\text{km}$, $m_Y=500\,\text{MeV}$, $m_\chi=100\,\text{MeV}$, $g_{\chi Y} g_{q Y} = 0.1$, $\Sigma_{\DM}^{\rm spike} = 6.9\times 10^{28}\,\text{GeV}\,\text{cm}^{-2}$ (BMCI).}
            \label{fig:Flux_Attenuation}
\end{figure}

In order to visualise the role played by the Earth attenuation within this approach, we show in the left panel of Fig.~\ref{fig:Flux_Attenuation} the ratio $T_\chi^x/T_\chi^{\zero}$ and $N_{\rm sct}^x$ (solid) obtained by solving the differential Eqs.~\eqref{eq:Attenuation} and \eqref{eq:Nsct} for our DM models, with mediator mass $m_Y = 500\,\text{MeV}$, DM mass $m_\chi = 100\,\text{MeV}$, couplings $g_{\chi Y} g_{q Y} = 0.1$ at depth $x=1$ km. In the right panel of the same figure, we plot the BBDM flux before attenuation (solid) and the attenuated flux (dashed) for the same DM model parameters and $\Sigma_{\DM}^{\rm spike} = 6.9\times 10^{28}\,\text{GeV}\,\text{cm}^{-2}$ as in BMCI.

From the left panel of Fig.~\ref{fig:Flux_Attenuation}  one can see that as long as $N_{\rm sct}^x<1$, then  $T_\chi^x\simeq T_\chi^{\zero}$, hence the effect of the Earth attenuation is negligible. In this situation, the average number of scatterings can be computed by integrating Eq.~\eqref{eq:Nsct}, which, under the adopted approximation, gives $N_{\rm sct}^x \simeq x\ell^{-1}$ with $ \ell^{-1}\equiv 2n_N \sigma_{Y \chi N}^{\EL}$. Earth attenuation becomes relevant when $N_{\rm sct}^x>\text{few}$, leading eventually to an exponential suppression of $T_\chi^x/T_\chi^{\zero}$.

The effect of the attenuation treated in this way on the BBDM flux is shown in the right panel of Fig.~\ref{fig:Flux_Attenuation}. For the pseudoscalar mediator case one has a sharp peak in the attenuated flux, which is produced by the DM particles whose $T_\chi^{\zero}$ is such that their average $N_{\rm sct}^x$ is larger than 1. This peak forms because $N_{\rm sct}^x$ eventually becomes smaller than 1 for sufficiently small $T_\chi^{\zero}$, due to the vanishing of the pseudoscalar elastic cross section, as it follows from Eq.~\eqref{eq:pseudoscalarelastic} in the $T_\chi \to 0$ limit. For the scalar case, instead, as the elastic cross section does not vanish for $T_\chi \to 0$, the effect of Earth attenuation is to smooth the peak of the unattenuated BBDM flux towards smaller kinetic energies.  Finally, for the vector and the axial mediators Earth attenuation is marginal for the adopted choice of parameters.

However, in our treatment of the Earth attenuation discussed in Sec.~\ref{subsec:attenuation} we have chosen not to follow the approach described in this Appendix, but rather to adopt the conservative one encoded in  Eq.~\eqref{eq:attenuationconservative}, where we set to zero the BBDM flux if the average number of DM scatterings $N_{\rm sct}^{x_\det} \simeq x_{\det}\ell^{-1}(T_\chi^{\zero})$ gets larger than 1. Two main considerations justify this choice:
i) our conservative treatment described in the main text -- which, by the way, makes the numerical implementation of the Earth attenuation much faster than the method outlined here -- is justified \textit{a posteriori} as the effects of the attenuation become relevant for values of the couplings much larger than the ones probed in our scenario for the scalar, vector and axial mediator cases; ii) in the pseudoscalar mediator scenario, where attenuation effects are more significant, the procedure described above leads to an artificial sharp peak in the flux (see right panel of Fig.~\ref{fig:Flux_Attenuation}). This is a consequence of the assumption that the DM particles maintain their direction and velocity across successive scatterings, which is unphysical. A more accurate treatment would require a dedicated simulation that tracks the change in DM direction and energy loss at each scattering event, lying beyond the goals of the present study.

\section{Results for BMCII}\label{app:BMCII}
\begin{figure}[t!]
        \centering
                        \includegraphics[width = 0.4\textwidth]{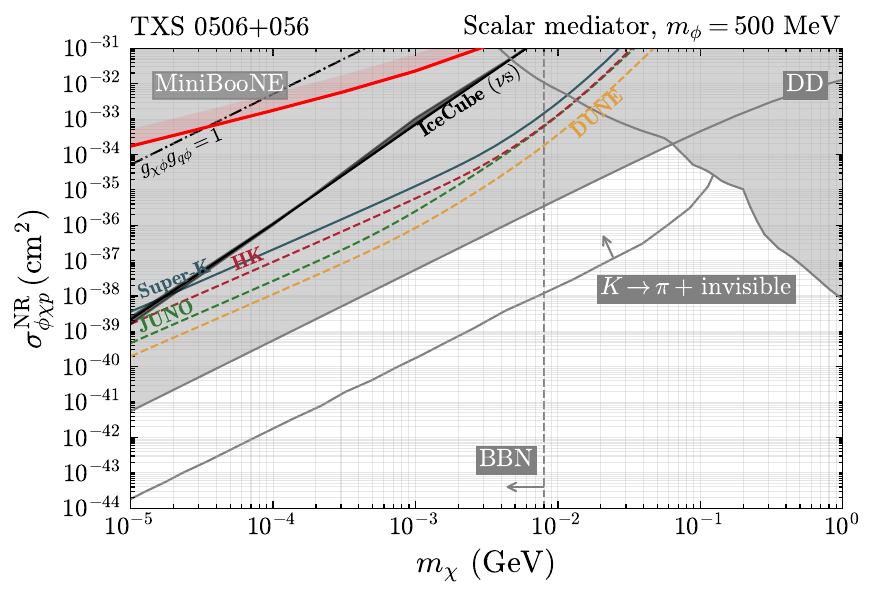}
                \includegraphics[width = 0.4\textwidth]{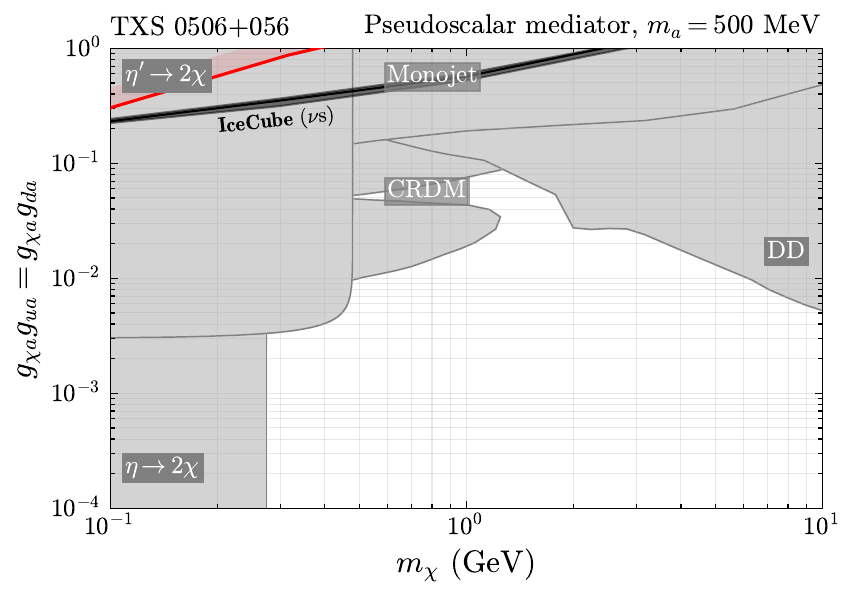}
                \includegraphics[width = 0.4\textwidth]{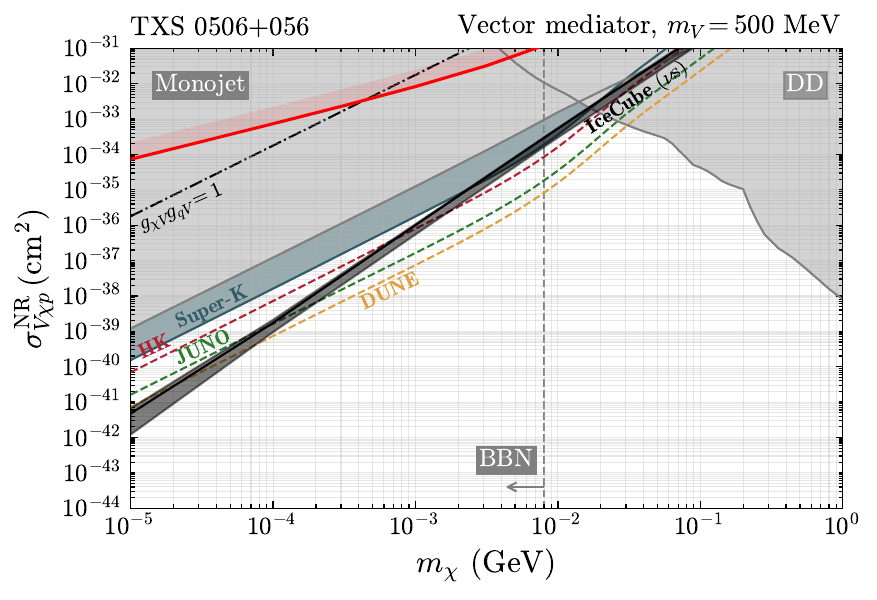}   
            \includegraphics[width = 0.4\textwidth]{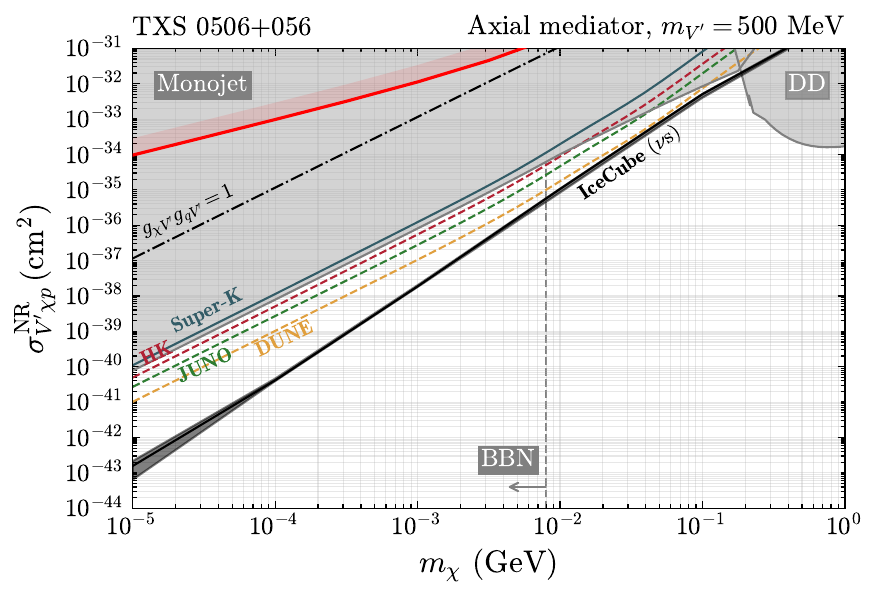}
                \includegraphics[width = 0.4\textwidth]{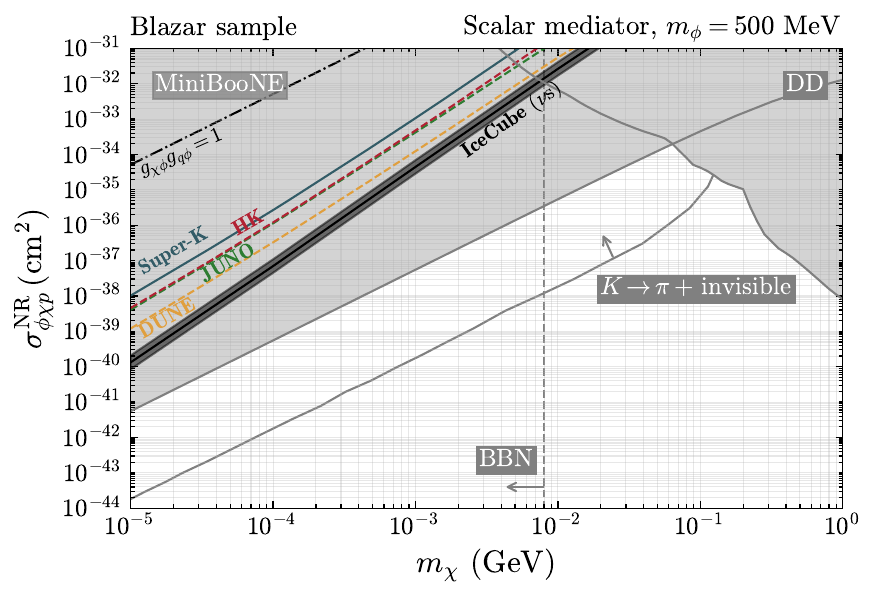}
                \includegraphics[width = 0.4\textwidth]{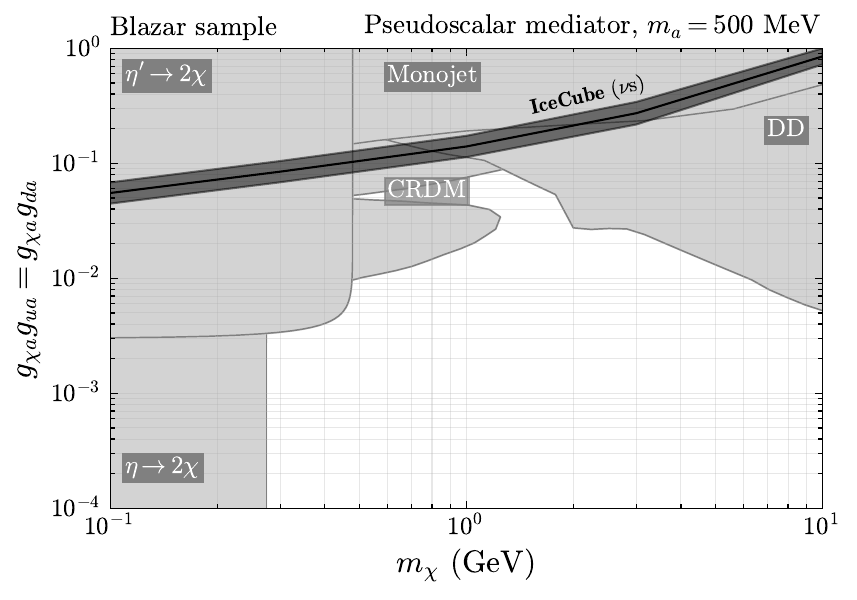}
                \includegraphics[width = 0.4\textwidth]{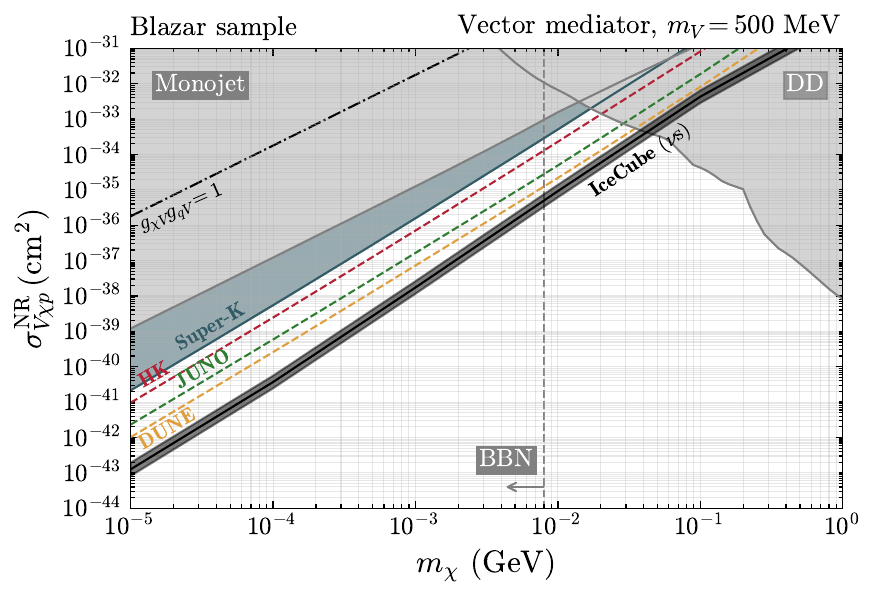}   
            \includegraphics[width = 0.4\textwidth]{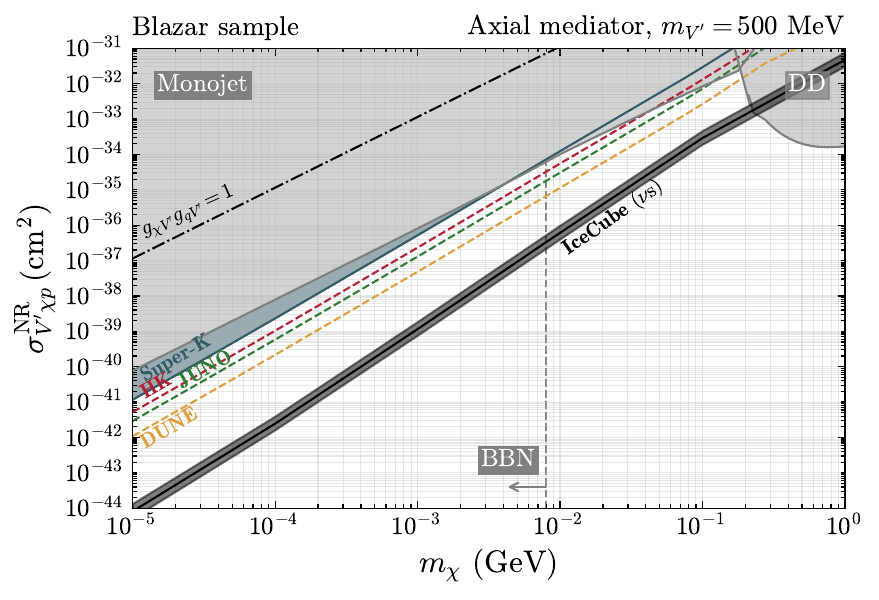}
    \caption{Limits and sensitivities to DM blazar signals from TXS 0506+056 (first four panels) and the considered blazar sample (last four panels) for $m_Y = 500\,\text{MeV}$ and $R_{\min} = 10^4\,R_S$ as in BMCII. Also shown in red for TXS 0506+056 are the spike depletion limit computed according to Eq.~\eqref{eq:bounddepletion} for this benchmark of the DM spike. All the other details are as in Fig.~\ref{fig:TXS_Sigma_vs_mx} and Fig.~\ref{fig:Sample_Sigma_vs_mx}}.
                \label{fig:Sigma_vs_mx_BMCII}
\end{figure}

We show in Fig.~\ref{fig:Sigma_vs_mx_BMCII} our limits and sensitivities to BBDM and neutrino signals for TXS 0506+056 and for the blazar sample, derived by assuming the conservative BMCII for the DM spike, namely $R_{\min} = 10^4\,R_S$, and choosing $m_Y = 500\,\text{MeV}$. 
For this benchmark choice, we find that the BBDM signal can test the DM hypothesis for the 2017 IceCube neutrino event from TXS 0506+056 in the vector mediator case in regions of the parameter space that are still allowed by current DM searches at terrestrial laboratories. This is also true in the scalar mediator case for DM masses approaching the $\mathcal{O}(10 \,\text{keV})$ scale (if the $K\to \pi+\text{invisible}$ bound does not apply), while otherwise the DM hypothesis for the neutrino signal at IceCube is incompatible with current DM searches. In the case of an axial mediator instead, a neutrino signal from DM around the blazar always comes before any BBDM signal at neutrino detectors. In the pseudoscalar case, the DM hypothesis is completely ruled out by current DM searches. We further notice that here, for the mass range considered in the plot, the selected neutrino detectors are not sensitive to any BBDM signal. This loss of sensitivity arises from the scaling of the BBDM signal with the LOS integral: both the projected limits and sensitivities obtained for BMCI are suppressed in BMCII by a factor of $\left(R_{\min}^{\rm BMCII}/R_{\min}^{\rm BMCI}\right)^{1/3} = 10^{2/3}$, resulting in a substantial reduction in the detectable signal strength.

Conversely, we find that the diffuse neutrino signal from DM-proton interaction in the blazar sample remains always dominant over the BBDM one. In the vector and axial mediator scenarios, both the diffuse BBDM signal at future neutrino detectors and the diffuse neutrino signal at IceCube from DM around blazar are currently allowed by existing DM constraints. However, the neutrino signal is expected to appear first, as the corresponding curve lies below the BBDM sensitivities. In contrast, the pseudoscalar case is entirely excluded by current bounds, while the scalar case remains largely, but not completely, ruled out.

We do not show the case of $m_Y = 5 \,\text{GeV}$ for the BMCII scenario, as the BBDM mechanism would already be excluded by current DM search constraints for all the considered toy models of DM-quark interactions. However, the DM interpretation for the IceCube neutrinos remains viable, see \cite{DeMarchi:2024riu, DeMarchi:2025xag}.

The spike depletion estimated according to Eq.~\eqref{eq:bounddepletion} is ineffective for BMCII for both the single source TXS 0506+056 in flare and the blazar sample.

\section{ Other processes that can alter the dark matter spike}\label{app:Spikedepletion}

\subsection{$4\chi \to 2\chi$ annihilations}
Throughout this work, we have assumed that DM is strongly coupled, i.e.~$g_{\chi Y} = 1$. This assumption, together with the very large densities found in the DM spike, could make $4\chi\rightarrow 2\chi$ annihilation processes efficient enough to deplete the DM spike itself, as pointed out in \cite{BetancourtKamenetskaia:2025ivl}. In this section, we estimate the size of this effect. We first approximate the \virg{cross section} (not really an area in this case) for this process as 
$\sigma_{4\chi\rightarrow2\chi} \sim g_{\chi Y}^8/m^8_Y$, and the energy density depletion rate as  $\dot{\rho}_{\DM} \sim m_\chi \sigma_{4\chi\rightarrow2\chi} (\rho_\DM/m_\chi)^4$.
Therefore, the timescale for the depletion is given by
\begin{equation}\tau_{4\chi\rightarrow 2\chi} = 
\frac{\rho_{\DM}}{\dot{\rho}_{\DM}} = \frac{1}{\sigma_{4\chi\rightarrow2\chi}}\left(\frac{m_\chi}{\rho_\DM}\right)^3.
\end{equation}
By inverting this relation we can find a maximal DM density that guarantees the spike is not depleted by these processes on a timescale $t_{\mathsmaller{\rm Accr.}}=10^8\,\text{yr}$:
\begin{equation}
    \rho_\max \approx 2\times 10^{16}\,\mathrm{GeV}\,\mathrm{cm}^{-3} \left(\frac{10^8\;\mathrm{yr}}{\tau_{4\chi\rightarrow 2\chi}}\right)^{1/3}\left(\frac{m_\chi}{10\;\mathrm{keV}}\right)
\left(\frac{m_Y}{1\;\mathrm{MeV}}\right)^{8/3}\left(\frac{1}{g_{\chi Y}}\right)^{8/3} \,.
\end{equation}
By comparing this result with Fig.~\ref{fig:DMDensitiyPlot} we note that the spike depletion due to $4\chi\to2\chi$ processes gets relevant in the inner part of the DM density profile only for benchmark values of $m_\chi$ and $m_Y$ well below those considered in our setup. We can thus safely neglect such effect in our calculations.

\subsection{$2\chi\to 2\chi$ scatterings}
The DM models outlined in the main text, besides describing DM interactions with the SM, induce interactions of DM with itself, 
thus realising a self-interacting DM (SIDM) model. This class of models 
can help alleviating small-scale tensions in the 
standard cosmological model \cite{Spergel:1999mh} and generally predicts a suppression of the power spectrum at small scales and the formation of isothermal cores in galactic halos. Dwarf galaxies \cite{Ando:2025qtz} and galaxy clusters \cite{Andrade:2020lqq} put bounds on the size of the SIDM cross section of the order $\sigma_{\mathsmaller{\text{SIDM}}}/m_\chi  \lesssim \mathcal{O}(1)\,\mathrm{cm}^2/\text{g}$.
For comparison, we estimate the SIDM cross section for the $2\chi \to 2 \chi$ processes in the scenarios considered by us as
$\sigma_{\mathsmaller{\text{SIDM}}} \sim g_\chi^4 m_\chi^2/m_Y^4, 
$ and thus
\begin{equation}
\frac{\sigma_{\mathsmaller{\text{SIDM}}}}{m_\chi}  
\simeq 2\times 10^{-4} \,\text{cm}^2\,\text{g}^{-1}\left(\frac{g_\chi}{1}\right)^4\left(\frac{m_\chi}{\mathrm{GeV}}\right)\left(\frac{\mathrm{GeV}}{m_Y}\right)^4,
\end{equation}
which is well below the maximum allowed value.

Additionally, SIDM leads to cores in galactic haloes, 
and the same should occur around supermassive BHs like those that power blazars. In particular, the effect of $2\chi \to 2 \chi$ processes 
is to suppress the spike \cite{BetancourtKamenetskaia:2025ivl, Shapiro:2014oha} to 
\begin{equation}
\rho_{\mathsmaller{\rm SIDM}}(r) \sim r^{-(3+k)/4}
\end{equation}
for a SIDM cross section with velocity dependence $\sigma_{\mathsmaller{\text{SIDM}}} \sim v^{-k}$. We estimate the DM density required for this process to be efficient in altering the spike profile according to Eq.~(3.5) of \cite{BetancourtKamenetskaia:2025ivl}. This happens approximately when each DM particle undergoes at least one scattering during the age of the halo, that is $(\rho_{\mathsmaller{\text{SIDM}}} /m_\chi)\langle\sigma_\mathsmaller{\text{SIDM}} v\rangle\,t_\mathsmaller{\rm Accr.}\sim 1$
where we once again compare to the accretion time as our reference timescale. Then this condition is satisfied when
\begin{equation}\label{eq:SIDMestimate}
\rho_{\mathsmaller{\text{SIDM}}}\sim 3\times 10^3\; \mathrm{GeV}\;\mathrm{cm}^{-3}\left(\frac{10^8 \mathrm{yr}}{t_\mathsmaller{\rm Accr.}}\right)\left(\frac{10^{-2}}{v}\right)\left(\frac{1}{g_\chi}\right)^4\left(\frac{\mathrm{GeV}}{m_\chi}\right)\left(\frac{m_Y}{\mathrm{GeV}}\right)^4,
\end{equation}
assuming $k=0$.
Qualitatively, as long as the DM density stays below the maximal value in Eq.~\eqref{eq:SIDMestimate}, the SIDM effects of spike relaxation can be neglected.  
By tweaking $m_\chi$ and $m_Y$ alone, we find that BMCII can still be consistent, while BMCI requires $g_\chi \sim 10^{-2}$, thus strengthening all monojet and DD constraints. We point out, however, that we could easily circumvent such limitation by considering a model of inelastic DM, i.e.~taking two DM states separated in mass by a relatively small mass splitting. If the mass splitting among the two DM states is large enough, $2\rightarrow2$ scattering processes are kinematically forbidden at tree-level and are only achievable at 1-loop. Furthermore, if the mass splitting is sufficiently smaller than the typical momentum exchanged in $p-\chi$ scatterings, then BBDM and neutrino signals would be unaffected by the presence of the two DM states, whereas DD bounds would be even weaker due to an additional loop suppression.

\small

\providecommand{\href}[2]{#2}\begingroup\raggedright\endgroup

\end{document}